\documentclass[11pt,a4paper]{article}
\usepackage{alltt}
\usepackage{amscd}
\usepackage{amsfonts}
\usepackage{amsmath}
\usepackage{amssymb,latexsym,cite}
\usepackage{amsthm} 
\usepackage[english]{babel}
\usepackage{booktabs}
\usepackage{bbm}
\usepackage{bibentry}
\usepackage[makeroom]{cancel}
\usepackage{caption}
\usepackage{chngcntr}

\usepackage[pdfencoding=auto, psdextra]{hyperref}
\hypersetup{urlcolor=blue, colorlinks=false} 
\usepackage{cleveref}
    \crefformat{section}{\S#2#1#3}
    \crefformat{subsection}{\S#2#1#3}
    \crefformat{subsubsection}{\S#2#1#3}
    \crefrangeformat{section}{\S\S#3#1#4 to~#5#2#6}
    \crefmultiformat{section}{\S\S#2#1#3}{ and~#2#1#3}{, #2#1#3}{ and~#2#1#3}

\usepackage{changepage}
\usepackage{chngcntr}
\usepackage{color}
\usepackage{enumitem}
\usepackage{etoolbox}
\usepackage{faktor}
\usepackage{fancyhdr}
\usepackage[T1]{fontenc}
\usepackage{geometry}
\usepackage{graphicx}
\graphicspath{ {.} } 

\usepackage{imakeidx}
\usepackage[utf8]{inputenc}
\usepackage{lmodern}
\usepackage{mathdots}
\usepackage{mathrsfs} 
\usepackage{mathtools}
\usepackage{microtype}
\usepackage{multirow}

\usepackage{pgffor}  
\usepackage{pdflscape}
\usepackage{pgfplots}
    \pgfplotsset{compat=1.16}
\usepackage{siunitx}
\usepackage{shuffle}
\usepackage{slashed}
\usepackage{stmaryrd} 
\usepackage{tabu}
\usepackage{tabularx}
\usepackage[most]{tcolorbox}
\usepackage{tensor}
\usepackage[normalem]{ulem}
\usepackage{verbatim}  
\usepackage{wrapfig} 
\usepackage{xcolor}
\usepackage{xfrac}
\usepackage{xspace}
\usepackage[matrix,arrow,color]{xy}

\usepackage{tikz}
\pgfdeclarelayer{nodelayer}
\pgfdeclarelayer{edgelayer}
\pgfsetlayers{edgelayer,nodelayer,main}
\tikzstyle{dashed edge}=[<->, dashed]
\tikzstyle{blue pointer}=[->, draw=blue]
\tikzstyle{squiggle}=[->, decorate, decoration={snake,amplitude=.4mm,segment length=2mm, post length=1mm,pre length=0mm}]
\tikzstyle{dashed ->}=[->, dashed]
\tikzstyle{block}= [rounded corners, draw, thick, rectangle, minimum height = 1em, minimum width = 3em]

\tikzstyle{none}=[inner sep=0pt]
\usepackage{tikz}
\usepackage{tikz-cd}
\usetikzlibrary{decorations.markings}
\usepackage{tkz-euclide}
\usepackage{tikz-feynman} 
\usepackage[tikz]{bclogo}                   

\usetikzlibrary{arrows,shapes}
\usetikzlibrary{trees}
\usetikzlibrary{decorations}
\usetikzlibrary{matrix,arrows}              
\usetikzlibrary{positioning}                
\usetikzlibrary{calc,through}               
\usetikzlibrary{decorations.pathreplacing}  

\usetikzlibrary{decorations.pathmorphing}   
\usetikzlibrary{decorations.markings}
\usetikzlibrary{intersections}              

\tikzset{
    vector/.style={decorate, decoration={snake}, draw},
    fermion/.style={postaction={decorate},
        decoration={markings,mark=at position .55 with {\arrow{>}}}},
    fermionbar/.style={draw, postaction={decorate},
        decoration={markings,mark=at position .55 with {\arrow{<}}}},
    fermionnoarrow/.style={},
    gluon/.style={decorate,
        decoration={coil,amplitude=4pt, segment length=5pt}},
    scalar/.style={dashed},
    scalarbar/.style={dashed, postaction={decorate},
        decoration={markings,mark=at position .55 with {\arrow{<}}}},
    scalarnoarrow/.style={dashed,draw},
    vectorscalar/.style={loosely dotted,draw=black, postaction={decorate}},
}

\def\mathcolor#1#{\@mathcolor{#1}}
\def\@mathcolor#1#2#3{%
  \protect\leavevmode
  \begingroup
    \color#1{#2}#3%
  \endgroup
}

\def\slasha#1{\setbox0=\hbox{$#1$}#1\hskip-\wd0\hbox to\wd0{\hss\sl/\/\hss}}

\def\periodb#1{\setbox0=\hbox{$#1$}#1\hskip-\wd0\hbox to\wd0{-}}

\newcommand{\ident}{\mathbbm{1}}             
\newcommand{\ii}{\mathrm{i}}            
\newcommand{\e}{\mathrm{e}}             

\newcommand{\CA}{\mathcal{A}}               

\newcommand{\CCA}{\mathscr{A}}

\newcommand{\CCL}{\mathscr{L}}

\newcommand{\CCF}{\mathscr{F}}

\newcommand{\CJ}{\mathcal{J}}

\newcommand{\CL}{\mathcal{L}}
\newcommand{\CM}{\mathcal{M}}

\newcommand{\CN}{\mathcal{N}}

\newcommand{\CS}{\mathcal{S}}

\newcommand{\CCT}{\mathscr{T}}

\newcommand{\CCW}{\mathscr{W}}

\newcommand{\lvpa}{\overset{\leftarrow}{\partial}}
\newcommand{\rvpa}{\overset{\rightarrow}{\partial}}
\newcommand{\lvlie}{\overset{\leftarrow}{\pounds}}
\newcommand{\rvlie}{\overset{\rightarrow}{\pounds}}

\newcommand{\frg}{\mathfrak{g}}             

\newcommand{\frR}{\mathfrak{R}}

\newcommand{\fru}{\mathfrak{u}}

\newcommand{\frC}{\mathfrak{C}}
\newcommand{\frL}{\mathfrak{L}}

\def\hodge{{\textrm{\tiny H}}}

\def\CS{{\textrm{\tiny CS}}}
\def\ym{{\textrm{\tiny YM}}}
\def\sw{{\textrm{\tiny SW}}}
\def\hodge{{\textrm{\tiny H}}}

\def\tv{{\textrm{\tiny $V$}}}
\def\tv1{{\textrm{\tiny $V[1]$}}}

\newcommand{\mbf}[1]{{\boldsymbol {#1} }}

\newcommand{\FR}{\mathbbm{R}}               
\newcommand{\FC}{\mathbbm{C}}               
\newcommand{\RZ}{\mathbbm{Z}}               

\newcommand{\Ch}{\mathsf{Ch}}

\newcommand{\dd}{\mathrm{d}}                

\newcommand{\sU}{\mathsf{U}}                

\newcommand{\sG}{\mathsf{G}}

\newcommand{\sSL}{\mathsf{SL}}

\newcommand{\sEnd}{\mathsf{End}}

\renewcommand{\comment}[1]{}                  
\def\tyng(#1){\hbox{\tiny$\yng(#1)$}}           
\def\tyoung(#1){\hbox{\tiny$\young(#1)$}}           

\newcommand{\beq}{\begin{eqnarray}}
\newcommand{\eeq}{\end{eqnarray}}

\newcommand{\sfp}{{\sf p}}
\newcommand{\sfH}{{\sf H}}

\newcommand{\sfh}{{\sf h}}

\newcommand{\sfGamma}{{\sf \Gamma}}

\newcommand{\sfe}{\mathsf{e}}

\definecolor{outrageousorange}{rgb}{1.0, 0.43, 0.29}

\newcommand{\Tr}{\mathrm{Tr}}

\newcommand{\midwedge}{\text{\Large$\wedge$}}

\def\ds{\stackrel{\star}{,}}

\newcommand{\Scal}{\mathfrak{Scal}}
\newcommand{\Kin}{\mathfrak{Kin}}
\newcommand{\BAS}{\mathfrak{BAS}}
\newcommand{\AS}{\mathfrak{AS}}
\newcommand{\ChS}{\mathfrak{CS}}
\newcommand{\YM}{\mathfrak{YM}}

\def\beq{\begin{equation}}
\def\bee{\begin{equation}}
\def\eeq{\end{equation}}
\def\bea{\begin{eqnarray}}
\def\eea{\end{eqnarray}}
\def\ba{\begin{align}}
\def\ea{\end{align}}

\renewcommand{\thefootnote}{\fnsymbol{footnote}}
\theoremstyle{plain}

\theoremstyle{definition}

\newtheorem{remark}[equation]{Remark}

\numberwithin{equation}{section}
\catcode`@=12

\begin{document}
\begin{titlepage}
    
    \begin{flushright}
        \small
        {\sf EMPG--23--10}
    \end{flushright}
    
    \begin{center}
        
        \vspace{1cm}
        
        \baselineskip=24pt
        
        {\Large\bf Homotopy double copy of noncommutative gauge theories}
        
        \baselineskip=14pt
        
        \vspace{1cm}
        
{\bf Richard
            J. Szabo}${}^{\,(a),(b),}$\footnote{Email: \ {\tt R.J.Szabo@hw.ac.uk}} \ \ and \ \ {\bf Guillaume Trojani}${}^{\,(a),}$\footnote{Email: \ {\tt gt43@hw.ac.uk}}
        \\[6mm]

\noindent  {${}^{(a)}$ {\it Department of Mathematics, Heriot-Watt University\\ Colin Maclaurin Building,
            Riccarton, Edinburgh EH14 4AS, U.K.}}\\ and {\it Maxwell Institute for
            Mathematical Sciences, Edinburgh, U.K.} \\[3mm]
\noindent{${}^{(b)}$ {\it Higgs Centre
            for Theoretical Physics, Edinburgh, U.K.}}
        \\[30mm]
        
    \end{center}
    
    \begin{abstract}
        \noindent
We discuss the double copy formulation of Moyal--Weyl type noncommutative gauge theories from the homotopy algebraic perspective of factorisations of $L_\infty$-algebras. 
    We define new noncommutative scalar field theories with rigid colour symmetries taking the role of the zeroth copy, where the deformed colour algebra plays the role of a kinematic algebra; some of these theories have a trivial classical limit but exhibit colour-kinematics duality, from which we construct the double copy theory explicitly. We show that noncommutative gauge theories exhibit a twisted form of colour-kinematics duality, which we use to show that their double copies match with the commutative case. We illustrate this explicitly for Chern--Simons theory, and also for Yang--Mills theory where we obtain a modified Kawai–Lewellen–Tye relation whose momentum kernel is linked to a binoncommutative biadjoint scalar theory. 
We reinterpret rank one noncommutative gauge theories as double copy theories, and discuss how our findings tie in with recent discussions of Moyal--Weyl deformations of self-dual Yang--Mills theory and gravity.
    \end{abstract}
    
    \vspace{1cm}
    
\begin{center}
{\sl\small Contribution to the Special Issue of Symmetry on `Quantum Geometry and Symmetries of String Theory'}
\end{center}
    
\end{titlepage}

{\baselineskip=12pt
    \tableofcontents}
    
\bigskip

\setcounter{page}{1}

\setcounter{footnote}{0}
\renewcommand{\thefootnote}{\arabic{footnote}}

\section{Introduction} 
\label{sec:introduction}

The \emph{double copy}  provides a novel perspective on observables in gravitational theories and related field theories by extracting information from gauge theory, offering an efficient tool for the calculation of perturbative gravity amplitudes while at the same time simplifying calculations in gauge theories. It has led to new insights into the geometric and algebraic structures underlying scattering amplitudes, and suggests profound relationships between a wide range of theories with various disparate properties. Very loosely put, it is a map between theories which follows the slogan 
\begin{align}\label{eq:slogan}
 \text{Gravity} \ = \ (\text{Gauge~Theory})^2
\end{align}
though both sides have broad meanings and generalisations. There has been an explosion of intense activity on the subject in recent years from all sorts of directions as well as with different goals and applications in mind. The literature is by now vast and extensive; see e.g.~\cite{Bern:2019prr,Bern:2022wqg,Adamo:2022dcm} for reviews and more complete lists of references. 

The aim of this paper is to understand how  standard noncommutative gauge theories, such as those which arise naturally from string theory, fit into the paradigm of colour-kinematics duality and the double copy of gauge theory to gravity. As an offspring of our investigations, we  shall encounter some novel
noncommutative scalar field theories with rigid colour symmetries that
have no commutative counterparts. These have not appeared in the literature before and are worthy of further studies in their own right. They form the building blocks of the double copy relations considered in the present paper.

In this section we give a short and informal introduction to the subject, providing some  motivation and background to what we set out to do in the present paper. We start by briefly sketching the key ideas behind three double copy relations that will play a prominent role in our treatment. We then summarise the main ideas and results of this paper.

\paragraph{KLT double copy.}

The Kawai--Lewellen--Tye (KLT) relations~\cite{Kawai:1985xq} lie historically at the origins of the double copy. Symbolically, they prescribe the squaring operation \eqref{eq:slogan} in the form
\begin{align}\label{eq:KLTintro}
\CCA_n^{\textrm{\tiny L}\otimes\textrm{\tiny R}} = \sum_{w,w'} \, \CA_n^{\textrm{\tiny L}}(w) \ \boldsymbol{S}_n(w|w') \ \CA_n^{\textrm{\tiny R}}(w') \ .
\end{align}
The left-hand side involves the $n$-point tree amplitudes in the double copy theory labelled by $^{\textrm{\tiny L}\otimes\textrm{\tiny R}}$, while the right-hand side involves colour-stripped $n$-point amplitudes  of potentially different ``left'' and ``right'' theories called single copies, and respectively labelled by $^{\textrm{\tiny L}}$ and $^{\textrm{\tiny R}}$. The sum runs over two choices from $(n-3)!$ of the $n!$ possible single-trace colour orderings of $n$ external particles. The KLT momentum kernel $\boldsymbol{S}_n(w|w')$ is a polynomial of degree $n-3$ in $n$-point Mandelstam variables encoding the kinematic invariants of the scattering process.

A simple example is the four-graviton tree amplitude in Einstein gravity. In the evident notation, it can be expressed in terms of the product of colour ordered four-gluon scattering amplitudes of Yang--Mills theory as
\begin{align}\label{eq:4graviton}
\CCA_4^{\textrm{\tiny GR}} = \boldsymbol{S}_4(1234|1243) \ \CA_4^{\textrm{\tiny YM}}(1234) \, \CA_4^{\textrm{\tiny YM}}(1243) \qquad \mbox{with} \quad \boldsymbol{S}_4(1234|1243) = s \ ,
\end{align}
where $(s,t,u) = (s_{12},s_{23},s_{13})$ and $s_{ij}=(p_i+p_j)^2$ are Mandelstam variables.

The KLT relations originate from string theory, where closed string states factorise into products of left-moving and right-moving open string states. In this case $\CA_n^{\textrm{\tiny L/R}}$ represent the colour-stripped left/right open string disk amplitudes, while $\CCA_n^{\textrm{\tiny L}\otimes\textrm{\tiny R}}$ represents the closed string sphere amplitude. At a topological level it can be understood from the gluing of two disks, regarded as northern and southern hemispheres, into a sphere. Its low-energy limit gives the field theory relations \eqref{eq:KLTintro}.

\paragraph{BCJ double copy.}

The Bern--Carrasco--Johansson (BCJ) relations~\cite{Bern:2008qj} appeared over 20~years after the KLT relations. The starting point is to reorganise the full (colour-dressed) left/right tree amplitudes in the form
\begin{align} \label{eq:CKintro}
\CCA_n^{\textrm{\tiny L/R}} = \sum_\sfGamma \, \frac{c_\sfGamma \, n_\sfGamma^{\textrm{\tiny L/R}}}{D_\sfGamma} \ ,
\end{align}
where the sum runs through all $(2n-5)!!$ trivalent graphs $\sfGamma$ with $n$ external legs. Here \smash{$c_\sfGamma$} are colour factors, i.e. combinations of generators of the gauge algebra, while the numerator factors \smash{$n_\sfGamma^{\textrm{\tiny L/R}}$} are kinematic weights made from Lorentz-invariant contractions of external momenta, polarisations, flavour, fermion wavefunctions, and so on. The denominators $D_\sfGamma=\prod_{e\in \sfGamma}\,s_e$ come from propagators, where $s_e=p^2_e$ are Mandelstam invariants of the momenta $p_e$ flowing through internal lines $e\in \sfGamma$.

The key requirement of the choice of decomposition \eqref{eq:CKintro} is \emph{colour-kinematics duality}: On each subgraph, a generalised Jacobi identity among colour numerators \smash{$c_{\sfGamma_s} + c_{\sfGamma_t} + c_{\sfGamma_u}=0$} inherited from the gauge algebra  is  obeyed identically among kinematic numerators \smash{$n_{\sfGamma_u}^{\textrm{\tiny L/R}} + n_{\sfGamma_t}^{\textrm{\tiny L/R}} + n_{\sfGamma_u}^{\textrm{\tiny L/R}}=0$} under particle permutations, and individual factors are antisymmetric. 

Consider, for example, the tree-level on-shell four-gluon amplitude in non-abelian gauge theory whose $s$-channel cubic graph is
\begin{equation}\label{eq:gluon4pt}
\begin{split}
\begin{tikzpicture}[scale=0.8]
{\small
    \draw[gluon] (-140:1.5)--(0,0);
    \draw[gluon] (140:1.5)--(0,0);
    \draw[gluon] (0:1.5)--(0,0);
    \node at (-145:1.6) {$1 \ \ $};
    \node at (145:1.6) {$2 \ \ $};
    \node at (.5,.3) {$$};    
\begin{scope}[shift={(1.5,0)}]
    \draw[gluon] (-40:1.5)--(0,0);
    \draw[gluon] (40:1.5)--(0,0);
    \node at (-35:1.6) {$ \ \ 4$};
    \node at (35:1.6) {$ \ \ 3$};    
\end{scope} } \normalsize
\end{tikzpicture}
\end{split}
\end{equation}    
The associated colour factor is $c_s=f^{a_1a_2b}\,f^{a_3a_4b}$ where $f^{abc}$ are structure constants for a gauge Lie algebra $\frg$. When summed over the three Mandelstam channels, obtained by degree~$3$ cyclic permutations of the colour labels, they satisfy $c_{s}+c_{t}+c_{u} = 0$ as a consequence of the Jacobi identity for $\frg$. Then the same Jacobi identity must be satisfied by the kinematic numerator $n_{s}$ of the graph \eqref{eq:gluon4pt}, obtained by contracting the Feynman rules with incoming on-shell polarisation vectors and using crossing symmetry.

When colour-kinematics duality holds for at least one of the left and right theories, then the replacement of colour factors $c_\sfGamma$ with kinematic factors \smash{$n_\sfGamma^{\textrm{\tiny R/L}}$} in \smash{$\CCA_n^{\textrm{\tiny L/R}}$} results in the tree amplitudes of another field theory, called the ``double copy'' theory. This prescribes the squaring operation \eqref{eq:slogan} in the form
\begin{align} \label{eq:BCJintro}
\CCA_n^{\textrm{\tiny L}\otimes\textrm{\tiny R}} = \sum_\sfGamma \, \frac{n_\sfGamma^{\textrm{\tiny L}} \, n_\sfGamma^{\textrm{\tiny R}}}{D_\sfGamma} \ .
\end{align}
This theory is invariant under diffeomorphisms. It is a gravitational theory with a dynamical spin two field if there are dynamical spin one gauge fields in each numerator. On the other hand, it may not be readily identifiable with a gravity theory when no on-shell spin two states exist. 

For the double copy of Yang--Mills theory with itself, \cref{eq:BCJintro} expresses the four-graviton amplitude \eqref{eq:4graviton} as
\begin{align}
\CCA_4^{\textrm{\tiny GR}} = \frac{\big(n_s^{\textrm{\tiny YM}}\big)^2}s + \frac{\big(n_t^{\textrm{\tiny YM}}\big)^2}t + \frac{\big(n_u^{\textrm{\tiny YM}}\big)^2}u \ .
\end{align}
For amplitudes of $\CN=4$ supergravity, one may double copy amplitudes of $\CN=4$ supersymmetric Yang--Mills theory with amplitudes of pure Yang--Mills theory, and so on.

The statement of colour-kinematics duality amounts to replacing the colour structure constants $f^{abc}$ by momentum-dependent factors that represent structure constants of an infinite-dimensional kinematic algebra~\cite{Monteiro2011,Bjerrum-Bohr:2012kaa,Fu:2012uy}. From this perspective, gravity is a gauge theory for which colour is substituted by kinematics, thus reminiscent of old ideas of viewing gravity as a gauge theory whose gauge algebra is the infinite-dimensional Lie algebra of diffeomorphisms. One of the main problems in the double copy relation between gauge theory and gravity is a detailed understanding of this kinematic algebra.

Colour ordered amplitudes that satisfy the {BCJ amplitude relations}~\cite{Bern:2008qj} along with the Kleiss--Kuijf relations~\cite{Kleiss:1988ne} ensure that the double copy amplitude \eqref{eq:BCJintro} does not depend on the choice of bases for the decomposition of \cref{eq:CKintro}. These
can be used to directly obtain the KLT formula \eqref{eq:KLTintro}.  The KLT momentum kernel can be understood in this setting as the inverse of a matrix of bi-coloured scalar amplitudes~\cite{Cachazo:2013iea}, whose role  is to ensure the correct propagator structure of the double copy amplitudes. These arise as the tree amplitudes of a cubic biadjoint scalar field theory, which result instead from the replacement of kinematic factors $n_\sfGamma^{\textrm{\tiny L/R}}$ in \smash{$\CCA_n^{\textrm{\tiny L/R}}$} with a second set of colour factors $\bar c_\sfGamma$ for another gauge algebra. They are labelled by $^{\textrm{\tiny BAS}} $ and take the form
\begin{align} \label{eq:BASintro}
\CCA_n^{\textrm{\tiny BAS}} = \sum_\sfGamma \, \frac{c_\sfGamma \, \bar c_\sfGamma}{D_\sfGamma} \ .
\end{align}

The amplitude \eqref{eq:BASintro} has manifest colour-kinematics duality, with either of the two colour factors $c_\sfGamma$ or $\bar c_\sfGamma$ regarded as kinematical numerators.
The biadjoint scalar theory thus serves as the ``identity model'' in the double copy operations: its doubly colour ordered amplitudes behave like identity matrices in KLT double copies with any other single copy theory, while replacing $c_\sfGamma$  in  \smash{$\CCA_n^{\textrm{\tiny L/R}}$} by one of the two biadjoint scalar colour factors  returns the same amplitude. As such, it is called the ``zeroth copy'' theory.

Altogether, the chain of relations amongst the theories involved in the double copy construction can be depicted symbolically as
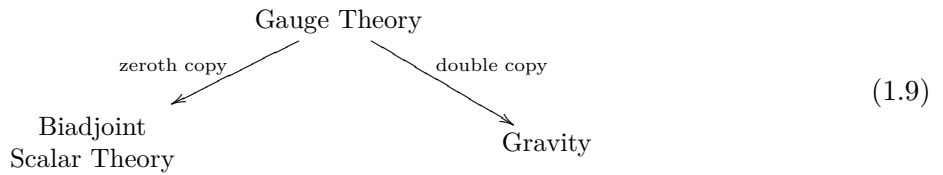
\begin{align} \label{eq:symred}
\small
\begin{split}
\xymatrix{ & \text{Gauge Theory} \ar[dl]_{\text{zeroth copy \ }} \ar[dr]^{\text{ \ double copy}}& \\
{\begin{matrix} \text{Biadjoint} \\ \text{Scalar Theory} \end{matrix}} & & \textrm{Gravity}
}
\end{split}
\normalsize
\end{align}
The double copy procedure can also be applied to loop amplitudes at the level of momentum space integrands, provided that colour-kinematics duality holds on all unitarity cuts, thereby enabling the double copy construction of gravitational amplitudes~\cite{Bern:2010ue}.

\paragraph{Homotopy double copy.}

Moving forward another 13~years, the more recent remarkable perspective of Borsten, Jur\v{c}o, Kim, Macrelli, S\"amann and Wolf~\cite{Borsten:2021hua} reformulates the double copy relations by exploiting the fact that the kinematical and dynamical data of any field theory are organised by homotopy algebras. By the strictification theorem for $L_\infty$-algebras, any theory that can be quantized in the Batalin--Vilkovisky (BV) formalism is perturbatively equivalent, via addition of suitable auxiliary fields, to  a theory with only cubic interaction vertices. The underlying $L_\infty$-algebras are quasi-isomorphic.

Let $\frL^{\textrm{\tiny L/R}}$ denote the strict $L_\infty$-algebras of left/right  theories with only cubic interactions based on a colour Lie algebra $\frg$ of internal symmetries. Suppose that they can be factorised into tensor products
\begin{align} \label{eq:factintro}
\frL^{\textrm{\tiny L/R}} = \frg\otimes\big(\Kin^{\textrm{\tiny L/R}}\otimes_{\tau^{\textrm{\tiny L/R}}}\Scal\big) \ .
\end{align}
Here $\Kin^{\textrm{\tiny L/R}} $ are finite-dimensional `kinematical' vector spaces encoding the kinematic degrees of freedom of the field theories, $\Scal$ is the strict $L_\infty$-algebra of a cubic scalar field theory encoding the trivalent interactions, and \smash{$\Kin^{\textrm{\tiny L/R}} \otimes_{\tau^{\textrm{\tiny L/R}}} \Scal$} denote the twisted tensor products with twist data \smash{$\tau^{\textrm{\tiny L/R}} $} which define kinematical strict $C_\infty$-algebras.

If at least one of the factorisations \eqref{eq:factintro} is compatible with colour-kinematics duality, then one can double copy the field theory by replacing the colour factor $\frg$ with the kinematic factor $\Kin^{\textrm{\tiny R/L}}$ in $\frL^{\textrm{\tiny L/R}} $ and twist the tensor product by $\tau^{\textrm{\tiny R/L}}$. This results in a field theory organised by the strict $L_\infty$-algebra
\begin{align}
\frL^{\textrm{\tiny L}\otimes\textrm{\tiny R}} = \Kin^{\textrm{\tiny L}}\otimes_{\tau^{\textrm{\tiny L}}}\big(\Kin^{\textrm{\tiny R}}\otimes_{\tau^{\textrm{\tiny R}}}\Scal\big) \ .
\end{align}
This realises the squaring operation \eqref{eq:slogan} as the ``homotopy double copy''. It provides a precise mathematical description of the bilinear multiplicative structure underlying the double copy operation on the space of certain classes of field theories.

In this framework the KLT momentum kernels are constructed from the strict $L_\infty$-algebra of a biadjoint scalar theory, which is obtained by instead replacing the kinematic vector spaces $\Kin^{\textrm{\tiny L/R}}$ by a second colour Lie algebra $\bar\frg$ in the factorisation \eqref{eq:factintro}  to get the strict $L_\infty$-algebra
\begin{align}
\BAS = \frg\otimes\big(\bar\frg\otimes\Scal\big)  \ .
\end{align}
This is called the ``homotopy zeroth copy''. Note that the same scalar $L_\infty$-algebra $\Scal$ is common to all three factorisations. 

In this setting, the chain of relations \eqref{eq:symred} is depicted by
\begin{align} \label{eq:symredhomotopy}
\begin{split}
\xymatrix{ & \frg\otimes\big(\Kin^{\textrm{\tiny L/R}}\otimes_{\tau^{\textrm{\tiny L/R}}}\Scal\big) \ar[dl]_{\text{zeroth copy \ }} \ar[dr]^{\text{ \ double copy}}& \\
\frg\otimes\big(\bar\frg\otimes\Scal\big) & & \Kin^{\textrm{\tiny L}}\otimes_{\tau^{\textrm{\tiny L}}}\big(\Kin^{\textrm{\tiny R}}\otimes_{\tau^{\textrm{\tiny R}}}\Scal\big)
}
\end{split}
\end{align}
Altogether this gives a complete and elegant algebraic formulation of the double copy operations at all levels, which moreover has the potential to elucidate the algebraic origins of colour-kinematics duality~\cite{Borsten:2022vtg,Bonezzi:2022bse,Bonezzi:2023pox}, and to
give an off-shell non-perturbative definition of the double copy. 

\paragraph{Noncommutative gauge theories and gravity.}

In this paper we are interested in how  noncommutative gauge theories fit into the double copy paradigm. We focus on noncommutative field theories of the type which arise as low-energy limits of open string theory with stacks of D-branes in $B$-fields, such as noncommutative Yang--Mills theory with the Moyal--Weyl star-product; see e.g.~\cite{Douglas:2001ba,Szabo:2001kg} for early reviews of the subject. Although in this paper we deal only with the simplest Moyal--Weyl deformations to clearly illustrate the main ideas and constructions, many aspects of our constructions such as planar equivalence and colour-stripping should work also for more general (twisted) Poisson structures induced by non-constant $B$-fields, and in particular for more general twist deformations.

To avoid undue suspense, let us right away state the main message of this paper: the double copy operation \eqref{eq:slogan} in the noncommutative world is replaced by
\begin{align} \label{eq:NCslogan}
\text{Ordinary Gravity} \ = \ (\text{Noncommutative Gauge~Theory})^2
\end{align}
where we use the adjective `ordinary' to distinguish commutative theories from their deformed counterparts. In other words, we find that the double copy of a noncommutative gauge theory is \emph{not} a deformed gravitational theory, but rather coincides with the result of the ordinary double copy. While this somewhat mundane conclusion may seem obvious to experts, it raises some interesting conceptual and theoretical questions whose answers lead to novel perspectives on the double copy.  We discuss in detail how these expectations are borne out from the perspective of the homotopy double copy, drawing on the homotopy algebras organising the relevant noncommutative gauge theories that are discussed in~\cite{Blumenhagen:2018kwq,Giotopoulos:2021ieg}. We believe that the homotopy algebraic manipulations used to arrive at the slogan \eqref{eq:NCslogan} are interesting and useful, and further point towards various generalisations of our work. 

The standard examples of double copy relations mostly involve  renormalizable field theories, but higher derivative operators are also expected to take part in some form of the double copy. Noncommutative gauge theories provide such an example: they are non-local because they involve an infinite tower of higher derivative operators, and they are generally non-renormalizable because they are plagued by the famous UV/IR mixing problem~\cite{Minwalla:1999px}. As such, in this paper we work only with noncommutative field theories at tree-level: when we speak of `double copy' in this context we mean at the level of tree amplitudes or homotopy algebras at the classical  (Lagrangian) level. UV/IR mixing becomes problematic only beyond one-loop, and the double copy operation \eqref{eq:NCslogan} perhaps offers a different perspective on the non-renormalizability of certain gravitational theories.

Another question that can be addressed by our result \eqref{eq:NCslogan} is the extent to which color-kinematics duality depends on Poincar\'e invariance. If colour-kinematics duality and the double copy are truly intrinsic properties of certain quantum field theories, they should exist in some form in backgrounds that break spacetime symmetries. Our results show that this is the case for noncommutative deformations, which explicitly break Lorentz symmetry. 
This may explain why ordinary gravity is recovered: in certain Lorentz-violating scenarios, field theory scattering amplitudes are still severely constrained by unitarity and locality. The existence of massless spin two particles forces three-particle amplitudes (including gravitons) to be Lorentz invariant, and hence cubic graviton interactions in Minkowski space must be those of general relativity up to certain unique higher derivative corrections; this is conjectured to be true for all $n$-particle amplitudes in~\cite{Pajer:2020wnj}. This fits in nicely with our perspective on ordinary Einstein gravity as the double copy of noncommutative Yang--Mills theory.

A somewhat mysterious issue is the precise meaning of colour-kinematics duality in noncommutative gauge theories, which are invariant under a deformed gauge symmetry that ensures consistency of the theory: a unitary colour algebra $\frg=\fru(N)$ combines intricately with kinematical degrees of freedom of the fields into an infinite-dimensional Lie algebra $\fru_\star(N)$ parametrizing noncommutative gauge transformations. This leads to the well-known colour-kinematics \emph{mixing} in noncommutative theories, which in turn spoils standard colour-stripping and colour-kinematics duality. 

In noncommutative field theories, the color-kinematics approach to the double copy is  extended by considering generalized numerators $n_\sfGamma$ that simultaneously depend on both colour and kinematics, while still satisfying Jacobi-like identities. This provides a novel and precise realisation of the kinematic Lie algebras underlying some theories. It is similar in spirit to the double copy of effective field theories which involve (finitely-many) higher derivative operators, see e.g.~\cite{Carrasco:2019yyn,Low:2019wuv,Carrasco:2021ptp}, while providing an explicit realisation of the complementary construction of colour-kinematics duality for algebraic relations satisfied by the symmetric structure constants $d^{abc}$ of the unitary Lie algebra $\frg=\fru(N)$~\cite{Carrasco:2022jxn}. We show that this modification is most naturally explained through the lens of twisted tensor products of homotopy algebras, leading to deformations of the standard constructions that we call `twisted homotopy factorisation' and `twisted colour-kinematics duality'. 

Our modifications of colour-kinematics duality, and ultimately the operation \eqref{eq:NCslogan}, are most naturally explained by the ultraviolet completion of noncommutative field theory in tree-level open bosonic string amplitudes with $B$-fields. In particular, we examine their connections with the BCJ and KLT amplitude relations. Higher derivative generalizations of the KLT relations are discussed in~\cite{Mizera:2016jhj,CEHJP21}; our KLT approach to the double copy is taken by generalizing the momentum kernel by an infinite tower of higher derivative operators. The double copy amplitudes obtained with our modified KLT kernel can be equivalently achieved by the traditional kernel. 

The $B$-field modified momentum kernel in the KLT relations thus also involves infinitely-many higher derivative corrections. It satisfies the requisite minimal rank condition of~\cite{CEHJP21} and is  given by the inverse matrix of the colour ordered amplitudes in a `binoncommutative biadjoint scalar theory'. This new scalar theory plays the role of the zeroth copy theory for double copies of noncommutative gauge theories. It reduces to various interesting noncommutative scalar field theories by taking rank one specialisations; in a certain sense these theories are the most novel and interesting aspects of the story we tell in the following. One of these theories, which we call the `adjoint scalar theory', is a simple example with rigid colour symmetry where all of our considerations can be made very explicit. It is an unusual example of a noncommutative scalar field theory, in the sense that its commutative limit is a free theory. 

The adjoint scalar theory has an interesting homotopy double copy which is a noncommutative scalar field theory  that we construct and describe explicitly. While the classical limit is trivial, its semi-classical limit reproduces some known topological theories such as the special galileon theory
in two dimensions and self-dual gravity in four dimensions. In these cases the semi-classical colour-kinematics replacements involve replacing gauge algebras by algebras of area-preserving diffeomorphisms, which are naturally explained as mutual isomorphisms between the (noncommutative) theories. These relate the biadjoint scalar theory to the Zakharaov--Mikhailov and special galileon theories in two dimensions, and to self-dual gauge and gravity theories in four dimensions. The non-perturbative double copy proposals in this case~\cite{Cheung:2022mix,Armstrong-Williams:2022apo} are thus naturally explained via the elegant formulation of the double copy prescription
using homotopy algebras, which enables an off-shell Lagrangian-level formulation and allows for replacement of colour algebras by kinematic algebras beyond amplitude level.

As is well-known, the  rank one specialisations of noncommutative theories with gauge symmetries are interesting interacting field theories of photons, unlike their commutative counterparts; like the adjoint scalar theory, in the classical limit they become free theories. These are low-energy effective theories on a single D-brane in a background $B$-field, and they have long been believed to contain gravitation~\cite{Ishibashi:2000hh}. Superficial evidence for this comes from the interplay between noncommutative gauge transformations and spacetime diffeomorphisms~\cite{Gross:2000ba,Lizzi:2001nd}: spatial translations are equivalent to gauge transformations (up to global symmetry transformations), and thus these theories are at least toy models of general relativity, the only other theory that shares this property. More substantial arguments revolve around emergent gravity phenomena in noncommutative gauge theories; see e.g.~\cite{Langmann:2001yr,Rivelles:2002ez,Yang:2004vd,Szabo:2006wx, Steinacker:2007dq,Steinacker:2010rh} and references therein for a partial list of early works in this direction.

The rank one noncommutative gauge theories have no colour symmetries, but they are invariant under a kinematic Lie algebra $\fru_\star(1)$ of infinitesimal noncommutative $\sU(1)$ gauge transformations. Here the adjoint scalar theory discussed above plays a decisive role which offers a new perspective on the emergence of gravitation in these theories: $\fru_\star(1)$ gauge theories can be constructed via the homotopy double copy of the adjoint scalar theory with the corresponding commutative gauge theory for any colour Lie algebra $\frg$. This double copy operation is written symbolically as
\begin{align} \label{eq:DCU1}
\begin{matrix}
\text{Noncommutative} \\ \text{$\sU(1)$ Gauge Theory}
\end{matrix}
 \ \ = \ \ \begin{matrix}
 \text{Adjoint} \\ \text{Scalar Theory} 
 \end{matrix}
 \ \ \otimes \ \ \begin{matrix}
 \text{Ordinary} \\ \text{Gauge Theory}
 \end{matrix}
\end{align}

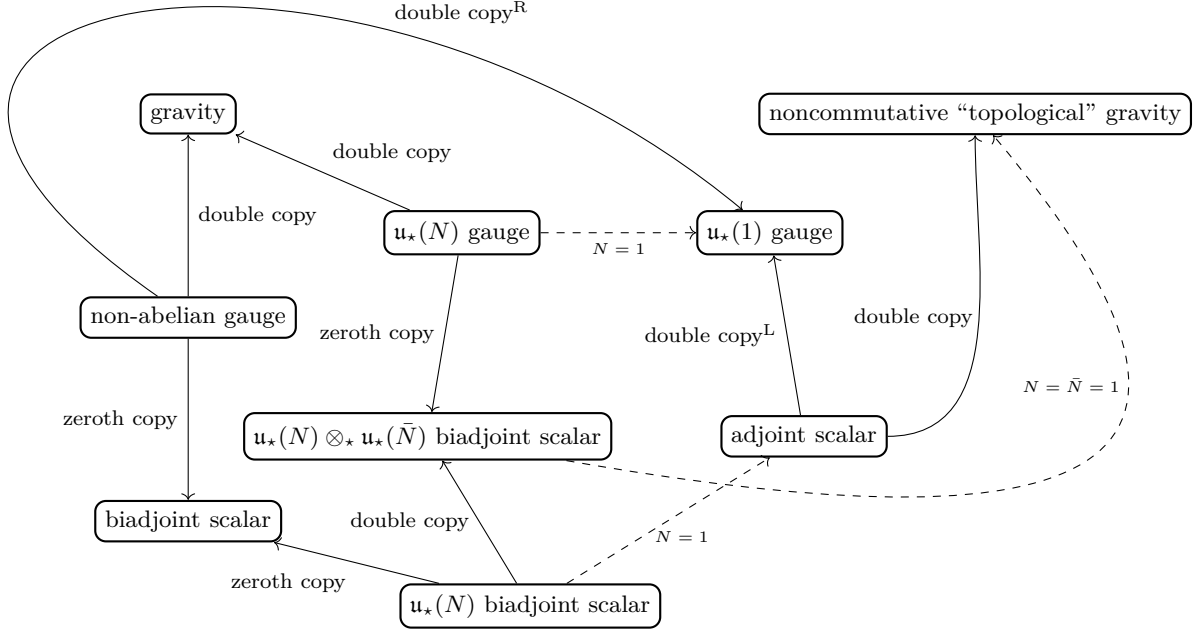
\begin{figure}
\hspace{-.3in}
\begin{center}
{
\begin{tikzpicture}[ yscale=0.9, xscale = 0.9]
\clip (-3,-1) rectangle + (18,10);
        \node [style=block] (0) at (0, 1.25) {\footnotesize  biadjoint scalar};
        \node [style=block] (1a) at (5, 0) {\footnotesize $\fru_\star(N)$ biadjoint scalar};
        \node [style=block] (1) at (3.5, 2.5) {\footnotesize $\fru_\star( N)\otimes_\star\fru_\star(\bar N)$ biadjoint scalar};
        \node [style=block] (2) at (0, 4.25) {\footnotesize non-abelian gauge};
        \node [style=block] (3) at (4, 5.5) {\footnotesize $\fru_\star(N)$ gauge};
        \node [style=block] (5) at (8.5, 5.5) {\footnotesize $\mathfrak{u}_\star(1)$ gauge};
        \node [style=none] (5p) at (8.47, 5.85) {};
        \node [style=block] (6) at (9, 2.5) {\footnotesize adjoint scalar};
        \node [style=block] (7) at (0, 7.25) {\footnotesize gravity};
        \node [style=block] (9) at (11.5, 7.25) {\footnotesize noncommutative ``topological'' gravity};
        \node (10) at (10.6, 4.25) {\scriptsize double copy};
        \node (10a) at (12.9, 3.25) {\tiny $ N=\bar N=1$};
        \node (11) at (4, 8.75) {\scriptsize double copy$^{\textrm{\tiny R}}$};
        \draw [style=dashed ->] (1a) -- (6) node[midway,below]{\tiny \ \ \ \ $N=1$};
        \draw [style=dashed ->] (3) -- (5) node[midway,below]{\tiny $N=1$};
        \draw [style=->] (1a) -- (1)node[midway,left]{\scriptsize double copy};
        \draw [style=->] (1a) -- (0)node[midway,below left]{\scriptsize zeroth copy};
        \draw [style=->] (2) -- (7) node[midway, right]{\scriptsize double copy};
        \draw [style=->] (3) -- (7) node[midway, above right]{\scriptsize double copy};
        \draw [style=->] (2) -- (0) node[midway, left]{\scriptsize zeroth copy};
        \draw [style=->] (3) -- (1) node[midway, left]{\scriptsize zeroth copy};
        \draw [style=->] (6) -- (5) node[midway, left]{\scriptsize double copy$^{\textrm{\tiny L}}$};
        \draw [style=->, in=270, out=0] (6) to (9);
        \draw [style=->, in=140, out=145, looseness=2.3] (2) to (5);
        \draw [style=dashed ->, in=310, out=-10, looseness=2.3] (1) to (9);
\end{tikzpicture}
}
\caption{\small The different commutative and noncommutative field theories encountered in this paper, and the relations between them. Dashed lines relate theories through rank one specialisations of the colour algebra, while solid lines relate theories through double or zeroth copy operations. Double copy operations without superscripts indicate a double copy of a theory with itself, while those with superscripts indicate the double copy of a ``left'' theory with a ``right'' theory.
\label{fig:web}}
\end{center}  
\end{figure}

In the case of noncommutative Chern--Simons theory, we corroborate the relation \eqref{eq:DCU1} by showing explicitly that the left-hand side realises a subalgebra of diffeomorphisms of spacetime. But the most compelling evidence comes from the self-dual sector of noncommutative Yang--Mills theory in four dimensions: rank one noncommutative self-dual Yang--Mills theory is equivalent to a known noncommutative deformation of self-dual gravity, in which case the double copy operation \eqref{eq:DCU1} becomes
\begin{align} \label{eq:DCSD}
\begin{matrix}
\text{Noncommutative} \\ \text{Self-Dual Gravity}
\end{matrix}
 \ \ = \ \ \begin{matrix}
 \text{Adjoint} \\ \text{Scalar Theory} 
 \end{matrix}
 \ \ \otimes \ \ \begin{matrix}
 \text{Ordinary Self-Dual} \\ \text{Yang--Mills Theory}
 \end{matrix}
\end{align}
Altogether, we arrive at an intricate web of relations among various commutative and noncommutative theories, which we display symbolically in fig.~\ref{fig:web}.

\paragraph{Outline.} 
\label{par:outline_}

This paper works through the double copy operations for noncommutative field theories in a detailed and expository manner. Since the standard noncommutative field theories are organised by homotopy algebras largely in parallel to their commutative counterparts, some parts of this work can be read as a review upon taking classical limits. However, some of our homotopy double copy constructions, for example in the cases of the adjoint scalar theory, Chern--Simons theory and the first order formalism of Yang--Mills theory in four dimensions, have not appeared before in the literature as far as we are aware. On the other hand, we carefully analyse the modifications due to noncommutativity of colour-kinematics duality and the precise forms of the kinematic Lie algebras underlying certain theories. The outline of the remainder of this paper is as follows.

In  \cref{sec:building}, we study two scalar field theories which form the `commutative skeletons' of all theories studied in this paper. We formulate them in the language of $L_\infty$-algebras,  and use this to illustrate a simple application of  homotopy factorisation techniques. 

In  \cref{sec:NCscal} we begin by reviewing the Moyal--Weyl deformation of scalar field theories with and without rigid colour symmetry. We apply the homotopy factorisation procedure to a noncommutative version of the biadjoint scalar theory (designated as `$\fru_\star(N)$ biadjoint scalar' in fig.~\ref{fig:web}) and its rank one limit, the adjoint scalar theory.
Most importantly, we observe that colour-stripping in the deformation to noncommutative theories fits into the twisted homotopy factorisation introduced in \cite{Borsten:2021hua}.
We show explicitly that the adjoint scalar theory exhibits colour-kinematics duality at the amplitude level involving the  kinematic Lie algebra $\fru_\star(1)$, from which we construct its homotopy double copy with kinematic factors based on the twisted tensor product of $\mathfrak{u}_\star(1)$ with itself in Feynman rules. 
Guided by this result, we define the binoncommutative biadjoint scalar field theory (designated as `$\mathfrak{u}_\star(N) \otimes_\star \mathfrak{u}_\star(\bar{N})$ biadjoint scalar' in fig.~\ref{fig:web}), seen as the double copy of the noncommutative biadjoint scalar theory. We discuss how our constructions encompass already known double copies in the literature, including the special galileon theory in two dimensions as well as (noncommutative) self-dual gravity in four dimensions.

In  \cref{sec:cs} we start by reviewing the Moyal--Weyl deformation of theories involving differential forms.
We then apply the twisted homotopy factorisation to noncommutative Chern--Simons gauge theory, and explicitly identify a twisted form of colour-kinematics duality at the amplitude level involving the  kinematic Lie algebra of volume-preserving diffeomorphisms (in the Lorenz gauge). 
This allows for the homotopy double copy of Chern--Simons theory with itself to be constructed, which matches with results from \cite{Ben-Shahar:2021zww}.
We furthermore realise the rank one limit of noncommutative Chern--Simons theory as a double copy of the adjoint scalar theory with commutative Chern--Simons theory, together with an explicit diffeomorphism invariance identified at the amplitude level.

In  \cref{sec:klt_with_b_field} we first review the homotopy algebraic structure of noncommutative Yang--Mills theory. 
We then show that the twisted homotopy factorisation procedure is compatible with strictification in the second order formalism, inspired by the construction of~\cite{Borsten:2021hua} in the commutative case.
We also directly exhibit the twisted homotopy factorisation of the first order formalism in four dimensions.
We then proceed to consider the ultraviolet completion of noncommutative Yang--Mills theory to open string theory with background $B$-fields. We introduce a modification of the KLT relations to account for the $B$-field, making use of the factorisation of noncommutativity phase factors in open string amplitudes.
In the low-energy limit, this validates our understanding that gravitational amplitudes can be built from two copies of noncommutative Yang--Mills amplitudes with a modified momentum kernel; this is also used to exhibit the corresponding modifications of the BCJ amplitude relations.
We show that this modified kernel is exactly sourced by the binoncommutative biadjoint scalar theory. The rank one limit is further shown to reproduce noncommutative self-dual gravity as the double copy of the adjoint scalar theory with commutative self-dual Yang--Mills theory; this points towards a more precise gravitational interpretation of rank one noncommutative Yang--Mills theory beyond the self-dual sector.

In \cref{sec:final_remarks} we conclude with some final remarks summarising the main points of this paper and briefly discuss some future directions.

Supplementing this paper is an extensive appendix \S\ref{sec:linf}, in which we provide a pedagogical review of various aspects of homotopical algebra and their applications in quantum field theory. We discuss $L_\infty$-algebra methods for quantum field theory, twisted tensor products of homotopy algebras, as well as the construction of scattering amplitudes from Berends--Giele and minimal model recursion relations.


\paragraph{Acknowledgements.}

We thank Marija Dimitrijevi\'c \'Ciri\'c, Igor Prlina and Christian S\"amann for helpful discussions. {\sc R.J.S.} would like to thank the editor Edvard Musaev for the invitation to contribute to this special issue. R.J.S. thanks the Centro de
Matem\'atica, Computa\c{c}\"ao e Cogni\c{c}\"ao of the Universidade de Federal do ABC (S\~ao Paulo,
Brazil) for hospitality and support during part of this work. The work of {\sc R.J.S.} was supported in part by
the STFC Consolidated Grant ST/P000363/1 and by the FAPESP Grant 2021/09313-8. The work of {\sc G.T.} is supported by the STFC Doctoral Training Partnership Award ST/T506114/1. 


\section{Building blocks: Scalar field theories}
\label{sec:building}
 
\subsection{Cubic scalar field theory} 
\label{sub:nc_strict_scalar_field_theory}

An important theory in our discussions of the various factorizations of $L_\infty$-algebras that we will encounter is the massless  scalar field theory with $g\, \phi^3$ interaction on $d$-dimensional Minkowski spacetime $\mathbbm{R}^{1,d-1}$; it forms the skeleton theory encoding trivalent interactions on which the propagating degrees of freedom of all field theories considered in this paper will be built. 
We use standard coordinates $x^0,x^1,\dots,x^{d-1}$ on $\FR^{1,d-1}$, where $x^0$ represents time, and use the shorthand notation $\partial_\mu$ for the partial derivative $\frac\partial{\partial x^\mu}$. The standard Minkowski metric of signature $(+-\cdots -)$ is denoted $\eta_{\mu\nu}$, and the inner product of two vectors $v=(v^\mu)$ and $v'=(v'{}^\mu)$ in $\FR^{1,d-1}$ is $v\cdot v'  = \eta_{\mu\nu}\,v^\mu\,v'{}^\nu= v_\mu\,v'{}^\mu$.\footnote{Unless otherwise explicitly stated, implicit summation over repeated indices is always understood throughout this paper.} The standard volume form on $\FR^{1,d-1}$ is
\begin{align}
\dd^dx = \dd x^0\wedge\dd x^1\wedge\cdots\wedge\dd x^{d-1} \ ,
\end{align}
and $\square=\eta^{\mu\nu} \, \partial_\mu \, \partial_\nu=\partial_\mu\,\partial^\mu$ is the wave operator.

    The action functional is given by 
    \begin{equation}\label{eq:scalar functional}
        S_{\rm Scal}[\phi] = \int\,\dd^dx \ \frac12\, \phi \, \square \,\phi - \frac g{3!} \, \phi^3 \ ,
    \end{equation}
for a function (regarded as a zero-form) $\phi\in\Omega^0(\FR^{1,d-1})$.
    The three-point Feynman vertex in momentum space is 
    \begin{equation}\label{eq:feynman vertex ncscal}
\begin{split}
\begin{tikzpicture}[scale=0.7]
{\small
    \draw[fermionnoarrow] (140:1.8)node[above]{}--(0,0);
    \draw[fermionnoarrow] (-140:1.8)node[below]{}--(0,0);
    \draw[fermionnoarrow] (0:1.8)node[right]{} --(0,0);
    \draw[->, shift={(-0.2,0.45)}] (140:.5) node[above]{$p_1$} -- (0,0);
    \draw[->, shift={(-0.2,-0.45)}] (-140:.5) node[below]{$p_2$} -- (0,0);
    \draw[->, shift={(0.2,0.2)}] (0:.5) node[above]{$p_3$} -- (0,0);
    }\normalsize
    \node at (5, 0) {$ \ = \ -\ii\,g \ \delta(p_1 + p_2 +p_3) \ .$};
\end{tikzpicture}
\end{split}
\end{equation}
    
    This theory is organised into a cyclic strict $L_\infty$-algebra $\mathfrak{Scal}$ whose underlying cochain complex $\Ch_{\rm Scal} := \Ch(\mathfrak{Scal})$ is
    \begin{align}\label{eq:scalarcochaincomplex}
     \Ch_{\rm Scal} = \Big(  \Omega^0(\FR^{1,d-1})[-1] \xrightarrow{ \ \square \ } \Omega^0(\FR^{1,d-1})[-2]\Big)
    \end{align}
    with differential $\mu_1^{\rm Scal} = \square$; there are no gauge symmetries so the degree~$0$ and~$3$ subspaces are trivial. 
    The only non-trivial higher bracket is the $2$-bracket $\mu_2^{\rm Scal}:\Omega^0(\FR^{1,d-1})[-1]^{\otimes2} \longrightarrow \Omega^0(\FR^{1,d-1})[-2]$ that scales pointwise multiplication of functions as\footnote{This can be regarded as the pointwise multiplication in momentum space, or equivalently the convolution product in position space, and \emph{vice versa}.}
    \begin{equation}
    \begin{split}\label{eq:ncscalarmu2}
\mu_2^{\rm Scal}(\phi_1 ,\phi_2) = -g\,\phi_1\,\phi_2 \ ,
    \end{split}
    \end{equation}
    which is  symmetric, or graded antisymmetric with our degree conventions, as required.
    The Maurer--Cartan equation \eqref{eq:Maurer Cartan equation} yields the equation of motion
    \begin{align}\label{eq:scalarMCeq}
        \mu_1^{\rm Scal}(\phi) +\tfrac12\,\mu_2^{\rm Scal}(\phi,\phi) = \square\,\phi - \tfrac{ g}{2} \, \phi^2 = 0 \ .
    \end{align}
    
    The cyclic inner product of degree~$-3$ is given by the single non-vanishing pairing between fields and antifields
    \begin{align}\label{eq:ncscalarpairing}
        \langle \phi,\phi^+\rangle_{\rm Scal} = \int \, \dd^d x \  \phi \, \phi^+
    \end{align}
    for $\phi\in\Omega^0(\FR^{1,d-1})[-1]$ and $\phi^+\in\Omega^0(\FR^{1,d-1})[-2]$.
    With this inner product, the Maurer--Cartan functional \eqref{eq:MC functional} for the cubic scalar field theory is
    \begin{align}\label{eq:scalarMCaction}
    \begin{split}
        S_{\rm Scal}[\phi] &= \tfrac12\,\langle\phi,\mu_1^{\rm Scal}(\phi)\rangle_{\rm Scal}
        + \tfrac1{3!} \, \langle\phi,\mu_2^{\rm Scal}(\phi,\phi)\rangle_{\rm Scal} \ ,
    \end{split}
    \end{align}
    which is just the action functional of \cref{eq:scalar functional}.

    In summary, the cubic scalar field theory is organised into the cyclic strict $L_\infty$-algebra
    \begin{equation}\label{eq:linf scal star}
        \mathfrak{Scal} = \big(\mathsf{Ch}_{\rm Scal}, \mu_2^{\rm Scal}, \langle - , - \rangle_{\rm Scal}\big) \ .
    \end{equation}
In this paper we will build other field theories by adding extra data to the $L_\infty$-algebra $\Scal$ which preserve the $L_\infty$-structure. 
    
\subsection{Biadjoint scalar theory} 
\label{sub:biadjoint}

A theory that will become very useful in our discussion of the double copy is the biadjoint scalar theory  on $\FR^{1,d-1}$, which has a rigid $\sG\times \bar\sG$ symmetry under a pair of compact Lie groups $\sG$ and $\bar\sG$ equipped with bi-invariant metrics. Let $\frg$ and $\bar\frg$ be corresponding quadratic Lie algebras with brackets $[-,-]_\frg$ and $[-,-]_{\bar\frg}$, and invariant bilinear forms $\Tr_\frg$ and $\Tr_{\bar\frg}$.
Choose a basis of generators $\{T^a\}$ for $\frg$ with structure constants $f^{abc}$, i.e. $[T^a,T^b]_\frg=f^{abc}\,T^c$, and normalisation $\Tr_\frg(T^a\otimes T^b)=\delta^{ab}$.
Similarly, choose a basis $\{\bar T^{\bar a}\}$ for $\bar\frg$ with $[\bar T^{\bar a},\bar T^{\bar b}]_{\bar\frg}=f^{\bar a\bar b\bar c}\,\bar T^{\bar c}$, and normalisation $\Tr_{\bar\frg}(\bar T^{\bar a}\otimes \bar T^{\bar b})=\delta^{\bar a\bar b}$. 

Let $\Omega^0(\FR^{1,d-1},\frg\otimes\bar\frg)$ be the space of zero-forms on $\FR^{1,d-1}$ in the biadjoint representation of $\sG\times\bar\sG$ on $\frg\otimes\bar\frg$; we stress that $\mathfrak{g} \otimes \bar{\mathfrak{g}}$ is \textit{not} a Lie algebra.
An element $\phi\in\Omega^0(\FR^{1,d-1},\frg\otimes\bar\frg)$ can be written as $\phi=\phi_{a\bar a}\, T^a \otimes\bar T^{\bar a}$ with $\phi_{a\bar a}\in\Omega^0(\FR^{1,d-1})$.
The biadjoint scalar theory on $\FR^{1,d-1}$ is defined by the action functional 
\begin{equation}\label{eq:biadactioncomp}
\begin{split}
  S_{\textrm{\tiny BAS}}[\phi] = \int\, \mathrm{d}^dx \ \frac{1}{2}\, \phi^{a \bar a} \, \square\, \phi_{a \bar a} - \frac{g}{3!}\, f^{abc}\, \bar{f}^{\bar a \bar b \bar c}\, \phi^{a \bar a} \, \phi^{b \bar b} \, \phi^{c \bar c} \ .
\end{split}
\end{equation}
The equation of motion derived from \cref{eq:biadactioncomp} has the form
\begin{equation}\label{eq:biMCeq}
\begin{split}
 \square\,\phi^{a\bar a} - \tfrac{g}{2} \, f^{abc} \, \bar f^{\bar a\bar b\bar c} \, \phi^{b\bar b} \, \phi^{c\bar c} = 0 \ .
\end{split}
\end{equation}

This theory is organised by the cyclic strict $L_\infty$-algebra $\BAS$ whose 
 underlying two-term cochain complex $\Ch_{\textrm{\tiny BAS}}:=\Ch(\BAS)$ is simply formed with the differential $\mu_1^{\textrm{\tiny BAS}} = \square$, acting as $\mu_1^{{\textrm{\tiny BAS}}}(\phi) = \square\,\phi_{a\bar a}\, T^a \otimes\bar T^{\bar a}$:
\begin{align}\label{eq:nc biadjoint cochain complex}
    \mathsf{Ch}_{\textrm{\tiny BAS}} = \Big(
    \Omega^0(\mathbbm{R}^{1,d-1}, \mathfrak{g} \otimes \bar{\mathfrak{g}})[-1] \xrightarrow{ \ \square \ } \Omega^0(\mathbbm{R}^{1,d-1}, \mathfrak{g} \otimes \bar{\mathfrak{g}})[-2]
    \Big)\ .
\end{align}
To write the remaining structure maps, we note that there is a symmetric bilinear operation on $\Omega^0(\FR^{1,d-1},\frg\otimes\bar\frg)$ given by
\begin{align}\label{eq:ncbiadbracket}
    \llbracket \phi_1,\phi_2\rrbracket_{\frg\otimes\bar\frg} 
    = 
    f^{abc} \, \bar f^{\bar a\bar b\bar c} \, \phi_1^{a\bar a} \,  \phi_2^{b\bar b}  \, T^c \otimes\bar T^{\bar c} \ ,
\end{align}
and a local bilinear form $\Tr_{\frg\otimes\bar\frg}:\Omega^0(\FR^{1,d-1},\frg\otimes\bar\frg)^{\otimes 2}\longrightarrow \Omega^0(\FR^{1,d-1})$ defined as
\begin{align}\label{eq:double trace bilinear form}
    \Tr_{\frg\otimes\bar\frg}(\phi_1\otimes\phi_2) = \delta_{ab} \, \delta_{\bar a\bar b} \,\phi_1^{a\bar a} \, \phi_2^{b\bar b} \ .
\end{align}
The single non-vanishing higher bracket $\mu_2^{{\textrm{\tiny BAS}}}:\Omega^0(\FR^{1,d-1},\frg\otimes\bar \frg)[-1]^{\otimes2} \longrightarrow \Omega^0(\FR^{1,d-1},\frg\otimes\bar \frg)[-2]$ is then given by
\begin{equation}\label{eq:nc biad bracket}
    \mu_2^{\textrm{\tiny BAS}}(\phi_1, \phi_2) = -g\, \llbracket \phi_1 , \phi_2 \rrbracket_{\mathfrak{g} \otimes \bar{\mathfrak{g}}}
     =- g\, f^{abc}\, \bar{f}^{\bar a \bar b \bar c}\, \phi_1^{a\bar a} \, \phi_2^{b \bar b} \, T^c \otimes \bar T^{\bar c} \ .
\end{equation}
This is symmetric because the bracket \eqref{eq:ncbiadbracket} is symmetric.  The cyclic inner product of degree~$-3$ is given by the non-zero pairing
\begin{align}\label{eq:bipairing}
\langle\phi,\phi^+\rangle_{\textrm{\tiny BAS}} = \int\,\dd^d x \ \Tr_{\frg\otimes\bar\frg}(\phi\otimes\phi^+) = \int\,\dd^dx \ \phi^{a\bar a}\,\phi^+_{a\bar a} \ ,
\end{align}
for $\phi\in\Omega^0(\FR^{1,d-1},\frg\otimes\bar\frg)[-1]$ and $\phi^+\in\Omega^0(\FR^{1,d-1},\frg\otimes\bar\frg)[-2]$. 

Altogether this defines  a cyclic strict $L_\infty$-algebra
\begin{equation}\label{eq:linf for biadjoint}
    \BAS = \big( \mathsf{Ch}_{\textrm{\tiny BAS}} , \mu_2^{\textrm{\tiny BAS}}, \langle -,- \rangle_{\textrm{\tiny BAS}}\big) \ .
\end{equation}
This $L_\infty$-algebra is well-known and is very useful for computations in the double copy.
From the Maurer--Cartan equation \eqref{eq:Maurer Cartan equation} we obtain the equation of motion
\begin{equation}
\begin{split}
  \mu_1^{{\textrm{\tiny BAS}}}(\phi) + \tfrac12\,\mu_2^{{\textrm{\tiny BAS}}}(\phi,\phi) =  \square\, \phi -  \tfrac{g}{2}\, \llbracket \phi , \phi\rrbracket_{\mathfrak{g} \otimes \bar{\mathfrak{g}}} = 0 \ ,
\end{split}
\end{equation}
which in components coincides with \cref{eq:biMCeq}.
The Maurer--Cartan functional \eqref{eq:MC functional} gives the action functional
\begin{equation}
\begin{split}
  S_{\textrm{\tiny BAS}}[\phi] &= \tfrac{1}{2}\, \langle \phi, \mu_1^{\textrm{\tiny BAS}}(\phi) \rangle_{\textrm{\tiny BAS}} +
    \tfrac{1}{3!}\, \langle \phi, \mu_2^{\textrm{\tiny BAS}}(\phi,\phi) \rangle_{\textrm{\tiny BAS}} \\[4pt]
    &= \int\,\dd^dx \ \Tr_{\frg\otimes\bar\frg}\Big( \frac12\,\phi\,\square\,\phi - \frac g{3!} \, \phi\, [\![\phi,\phi]\!]_{\frg\otimes\bar\frg}\Big) \ ,
\end{split}
\end{equation}
whose component form coincides with \cref{eq:biadactioncomp}.

\paragraph{Homotopy factorisation.} 
\label{par:factorisation}
{
    The three-point Feynman vertex of the biadjoint scalar theory is given by 
    \begin{equation}\label{eq:feynman vertex ncbiad}
\begin{split}
\begin{tikzpicture}[scale=0.8]
{\small
    \draw[fermionnoarrow] (140:1.8)node[above]{$a,\bar a$}--(0,0);
    \draw[fermionnoarrow] (-140:1.8)node[below]{$b ,\bar b$}--(0,0);
    \draw[fermionnoarrow] (0:1.8)node[right]{$c,\bar c$} --(0,0);
    \draw[->, shift={(-0.2,0.45)}] (140:.5) node[above]{$p_1$} -- (0,0);
    \draw[->, shift={(-0.2,-0.45)}] (-140:.5) node[below]{$p_2$} -- (0,0);
    \draw[->, shift={(0.2,0.2)}] (0:.5) node[above]{$p_3$} -- (0,0);
    }\normalsize
    \node at (6.4, 0) {$ = \ -\ii\,g\, f^{abc}\,\bar f^{\bar a \bar b \bar c} \ \delta(p_1 + p_2 +p_3) \ .$};
\end{tikzpicture}
\end{split}
\end{equation}
We observe the factorisation of \cref{eq:feynman vertex ncbiad} in terms of the colour structure carried by the pair of Lie algebras $(\mathfrak{g}, \bar{\mathfrak{g}})$ and the cubic interaction vertex for the scalar field theory in \cref{eq:feynman vertex ncscal}, which manifests the double copy structure between the colour group $\sG$ and its `dual' colour group $\bar\sG$.\footnote{Since the degrees of freedom here are scalars rather than gluons, the notion of `colour'  should really be referred to as `flavour'. We prefer to uniformly use the term `colour' for internal symmetries throughout this paper, always understood as a rigid symmetry for scalar fields.}
This motivates the idea of factorisation of $L_\infty$-structures in terms of a Lie structure and a $C_\infty$-structure.

It is easy to demonstrate that the cyclic  strict biadjoint scalar $L_\infty$-algebra $\BAS$ factorizes in terms of the cyclic  strict scalar $L_\infty$-algebra $\mathfrak{Scal}$ from \cref{sub:nc_strict_scalar_field_theory} as
\begin{align}\label{eq:ncbiadjointfact}
    \BAS= \frg\otimes(\bar\frg\otimes \mathfrak{Scal}) \ ,
\end{align}
with each Lie algebra regarded as a cyclic differential graded (dg) Lie algebra sitting in degree~$0$ with the zero differential. For this, we identify $\Omega^0(\FR^{1,d-1},\frg\otimes\bar\frg)$ with $\frg\otimes\bar\frg\otimes\Omega^0(\FR^{1,d-1})$ and write each element $\phi$ of $\Omega^0(\FR^{1,d-1},\frg\otimes\bar\frg)$ in the form $\phi=T^a\otimes \bar T^{\bar a}\otimes\phi_{a\bar a}$ with $\phi_{a\bar a}\in\Omega^0(\FR^{1,d-1})$. 
With this identification it follows that \eqref{eq:ncbiadjointfact} holds at the level of the underlying graded vector spaces.

Recalling the differential $\mu_1^{\rm Scal} = \square $ on $\mathfrak{Scal}$, we see that the differential on the biadjoint complex splits as
\begin{align}
    \mu_1^{{\textrm{\tiny BAS}}}(T^a\otimes \bar T^{\bar a}\otimes\phi_{a\bar a}) 
    = T^a\otimes \bar T^{\bar a}\otimes\square\,\phi_{a\bar a} 
    = T^a\otimes \bar T^{\bar a}\otimes \mu_1^{\rm Scal}(\phi_{a\bar a}) \ .
\end{align}
On the other hand, the bracket  \eqref{eq:nc biad bracket} factorises in terms of the bracket \eqref{eq:ncscalarmu2} on $\mathfrak{Scal}$ as
\begin{align}
\begin{split}
    \mu_2^{{\textrm{\tiny BAS}}}(T^a\otimes \bar T^{\bar a}\otimes\phi^{a\bar a}_1,T^b\otimes \bar T^{\bar b}\otimes\phi^{b\bar b}_2) 
    &= T^c\otimes\bar T^{\bar c} \otimes \big( - g \, f^{abc} \, \bar f^{\bar a\bar b\bar c} \, \phi_1^{a\bar a} \,  \phi_2^{b\bar b} \big) \\[4pt]
    &= [T^a,T^b]_\frg \otimes [\bar T^{\bar a},\bar T^{\bar b}]_{\bar \frg} \otimes  \big( - g\,\phi_1^{a\bar a} \,  \phi_2^{b\bar b}\big) \\[4pt]
    &= [T^a,T^b]_\frg \otimes [\bar T^{\bar a},\bar T^{\bar b}]_{\bar \frg} \otimes \mu_2^{\rm Scal}(\phi_1^{a\bar a},\phi_2^{b\bar b}) \ .
\end{split}
\end{align}
This establishes \eqref{eq:ncbiadjointfact} at the level of dg-Lie algebras. 

Finally, we factorise the cyclic inner product \eqref{eq:bipairing} with respect to the inner product \eqref{eq:ncscalarpairing} on $\mathfrak{Scal}$ as
\begin{align}
\begin{split}
\langle T^a\otimes\bar T^{\bar a}\otimes \phi_{a\bar a} , T^b\otimes\bar T^{\bar b}\otimes\phi_{b\bar b}^+\rangle_{\textrm{\tiny BAS}} &= \delta^{ab} \, \delta^{\bar a\bar b} \, \int\,\dd^dx \ \phi_{a\bar a}\, \phi_{b\bar b}^+ \\[4pt]
&= \Tr_\frg(T^a\otimes T^b) \ \Tr_{\bar \frg}(\bar T^{\bar a}\otimes \bar T^{\bar b}) \ \langle\phi_{a\bar a},\phi_{b\bar b}^+\rangle_{\rm Scal} \ ,
\end{split}
\end{align}
which shows \eqref{eq:ncbiadjointfact} at the full level of cyclic strict $L_\infty$-algebras.

\paragraph{Amplitudes.}
Scattering amplitudes for the biadjoint scalar theory can be computed efficiently using the minimal model of the $L_\infty$-algebra $\BAS$, which also encodes its perturbiner expansion, as we review in \S\ref{sub:scattering_in_the_textlinf_formalism}. For example, the tree-level four-point off-shell amplitude is found in this way to be
\begin{equation}\label{eq:BAS4pt}
\begin{split}
\begin{tikzpicture}[line width=1. pt, scale=0.8]
    \draw[fermionnoarrow] (-140:1)--(0,0);
    \draw[fermionnoarrow] (140:1)--(0,0);
    \draw[fermionnoarrow] (0:1)--(0,0);
    \node at (-140:1.2) {${}_2$};
    \node at (140:1.2) {${}_1$};
    \node at (.5,.3) {$$};    
\begin{scope}[shift={(1,0)}]
    \draw[fermionnoarrow] (-40:1)--(0,0);
    \draw[fermionnoarrow] (40:1)--(0,0);
    \node at (-40:1.2) {${}_4$};
    \node at (40:1.2) {${}_3$};    
\end{scope}
\node at ((2.7,0) {$+$};
\begin{scope}[shift={(4.2,-.5)}]
    \begin{scope}[rotate=90]
            \draw[fermionnoarrow] (-140:1)--(0,0);
            \draw[fermionnoarrow] (140:1)--(0,0);
            \draw[fermionnoarrow] (0:1)--(0,0);
            \node at (-140:1.2) {${}_4$};
            \node at (140:1.2) {${}_2$};
            \node at (.5,.3) {$$};    
        \begin{scope}[shift={(1,0)}]
            \draw[fermionnoarrow] (-40:1)--(0,0);
            \draw[fermionnoarrow] (40:1)--(0,0);
            \node at (-40:1.2) {${}_3$};
            \node at (40:1.2) {${}_1$};    
        \end{scope}
    \end{scope}
\end{scope}
\node at ((6,0) {$+$};
\begin{scope}[shift={(8,-.5)}]
    \begin{scope}[rotate=90]
            \draw[fermionnoarrow] (-140:1)--(1,0);
            \draw[fermionnoarrow] (140:1)--(0,0);
            \draw[fermionnoarrow] (0:1)--(0,0);
            \node at (-140:1.2) {${}_4$};
            \node at (140:1.2) {${}_2$};
            \node at (.5,.3) {$$};    
        \begin{scope}[shift={(1,0)}]
            \draw[fermionnoarrow] (-40:1)--(-1,0);
            \draw[fermionnoarrow] (40:1)--(0,0);
            \node at (-40:1.2) {${}_3$};
            \node at (40:1.2) {${}_1$};    
        \end{scope}
    \end{scope}
\end{scope}
\node at (12, 0) {$\displaystyle = \ \frac{c_s\,\bar c_s}s + \frac{c_t\,\bar c_t}t + \frac{c_u\,\bar c_u}u \ ,$};
\end{tikzpicture}
\end{split}
\end{equation}    
where we introduced the Mandelstam variables
\begin{align}
s = (p_1+p_2)^2 \ , \quad t = (p_2+p_3)^2 \qquad \mbox{and} \qquad u = (p_1+p_3)^2 \ ,
\end{align}
along with the colour factors
\begin{align}\label{eq:colournum}
c_s = g\,f^{a_1a_2b}\,f^{a_3a_4b} \ , \quad c_t = g\,f^{a_3a_1b}\,f^{a_2a_4b} \qquad \mbox{and} \qquad c_u = g\,f^{a_2a_3b}\,f^{a_1a_4b} \ ,
\end{align}
and similarly for the barred colour factors. See~\cite{Lopez-Arcos:2019hvg} for further details. Note that $c_s+c_u+c_t=0$ as a consequence of the Jacobi identity for $\frg$, and similarly for $\bar\frg$.

\paragraph{Colour ordering and decomposition.}

Tree-level amplitudes of the biadjoint scalar theory, such as the four-point amplitude $\CCA_4(p,a,\bar a)$ of \cref{eq:BAS4pt}, are organised in terms of bi-invariant partial amplitudes as~\cite{Cachazo:2013iea}
\begin{align}
\begin{split}
\CCA^{\textrm{\tiny BAS}}_n(p,a,\bar a) &= \sum_{\sigma,\sigma'\in S_n/\RZ_n} \,  \mathrm{Tr}_{\frg}(T^{a_{\sigma(1)}} \cdots T^{ a_{\sigma(n)}}) \, \mathrm{Tr}_{\bar\frg}(\bar T^{\bar a_{\sigma'(1)}} \cdots \bar T^{\bar a_{\sigma'(n)}}) \\
& \hspace{5cm} \times \CA^{\textrm{\tiny BAS}}_n\big( \sigma(1), \dots, \sigma(n)\big|\sigma'(1), \dots, \sigma'(n)\big) \ ,
    \end{split}
\end{align}
where the sums run over non-cyclic orderings.
This defines the colour decomposition of biadjoint scalar amplitudes. The partial amplitudes $ \CA^{\textrm{\tiny BAS}}_n$ each appear four times.

For a permutation $\sigma\in S_n$ on $n$ letters, a colour ordering is an equivalence class of $n$-tuples $[\sigma(1),\dots,\sigma(n)]$ with respect to the equivalence relation 
\begin{align}
\big(\sigma(1),\dots,\sigma(n-1),\sigma(n)\big)\sim \big(\sigma(n),\sigma(1),\dots,\sigma(n-1)\big)\sim \big(\sigma(n),\sigma(n-1),\dots,\sigma(1)\big) 
\end{align}
on $S_n$ generated by the action of the subgroup $\RZ_n\rtimes\RZ_2\subset S_n$; the first identification takes care of cyclic ordering symmetry while the second implements the Kleiss--Kuijf relations.
A canonical representative  of a colour ordering has $\sigma(1)=1$ and $\sigma(2)<\sigma(n)$. Given a colour ordering $[\sigma(1),\dots,\sigma(n)]$ in canonical form, the associated colour factor is
\begin{align}
C(\sigma) := \mathrm{Tr}_{\frg}(T^{a_{\sigma(1)}} \cdots T^{ a_{\sigma(n)}}) + (-1)^n \, \mathrm{Tr}_{\frg}(T^{a_{\sigma(n)}} \cdots T^{ a_{\sigma(1)}}) \ ,
\end{align}
and similarly we define the colour factor $\bar C(\sigma)$ for the dual Lie algebra $\bar\frg$. There are $\frac12\,(n-1)!$ colour factors $C(\sigma)$ and $\bar C(\sigma)$, and $\sigma$ is called their planar orderings. 

In terms of these the $n$-point amplitudes become
\begin{align}
\CCA^{\textrm{\tiny BAS}}_n(p,a,\bar a) = \sum_{\sigma,\sigma'\in S_n/\RZ_n\rtimes\RZ_2} \,  C(\sigma) \, \bar C(\sigma') \ \CA^{\textrm{\tiny BAS}}_n\big( \sigma(1), \dots, \sigma(n)\big|\sigma'(1), \dots, \sigma'(n)\big) \ .
\end{align}
The partial amplitudes $ \CA^{\textrm{\tiny BAS}}_n$ have a simple expression as a sum over tree-level Feynman diagrams of the cubic scalar field theory from \cref{sub:nc_strict_scalar_field_theory} which are planar with respect to both orderings~\cite{Cachazo:2013iea,Mafra:2016ltu}. This is of course the amplitude incarnation of the homotopy factorisation \eqref{eq:ncbiadjointfact}. For example, the bi-colour ordered four-point amplitude is given by
\begin{align}
\CA^{\textrm{\tiny BAS}}_4(1,2,3,4|1,2,3,4) = \frac{1}t - \frac{1}s \ .
\end{align}
Conversely, the cubic scalar tree-level amplitudes can be computed by summing over all colour orderings of $\CA^{\textrm{\tiny BAS}}_n$~\cite{Dolan:2014ega}.

\section{Noncommutative scalar theories with rigid colour symmetries} 
\label{sec:NCscal}

\subsection{Moyal--Weyl deformation of scalar fields}
\label{sub:MWdeformation}

In the following we will describe natural noncommutative deformations of the biadjoint scalar theory from \cref{sub:biadjoint}. While the formalism can be quite generally applied using any Drinfel'd twist deformation in the universal enveloping algebra of the Lie algebra of vector fields on $\FR^{1,d-1}$, for definiteness we work with the simplest and best known example of the Moyal--Weyl twist. This is the case which is best understood both algebraically and geometrically, and it is the natural one which will arise in our later discussions of the relation between scattering amplitudes of noncommutative Yang--Mills theory and open string theory. It is also the case which has appeared in recent discussions of the double copy.

The Moyal--Weyl deformation of the algebra of functions $\Omega^0(\FR^{1,d-1})$ is parametrized by a constant Poisson bivector $\theta=\frac12\,\theta^{\mu\nu}\,\partial_\mu\wedge\partial_\nu$ on $\FR^{1,d-1}$, and is defined by replacing the commutative pointwise multiplication of functions $\phi_1\,\phi_2$ with the associative but noncommutative star-product
\begin{align}\label{eq:starproduct}
\phi_1\star\phi_2 = \phi_1 \, \exp\Big(-\frac{\ii\,t}2 \ \lvpa_\mu \, \theta^{\mu\nu}\, \rvpa_\nu \Big) \, \phi_2 \ ,
\end{align}
which is understood by expanding the exponential of the bidifferential operator as a formal power series in the deformation parameter $t\in\FR$.\footnote{There is also a convergent integral convolution formula for the Moyal--Weyl product~\cite{Szabo:2001kg}, whose asymptotic expansion in $t$ coincides with \eqref{eq:starproduct}.} The product \eqref{eq:starproduct} is associative and the commutative limit $t=0$ returns the usual pointwise product of $\phi_1$ and $\phi_2$, i.e. $\phi_1\star\phi_2 = \phi_1\,\phi_2 + O(t)$. It is not real, but it is \emph{Hermitian}:
\begin{align}
\overline{\phi_1\star\phi_2 } = \phi_2\star\phi_1 \ ,
\end{align}
for $\phi_1,\phi_2\in\Omega^0(\FR^{1,d-1})$. 

If we restrict to Schwartz functions, then the star-product obeys the well-known integration by parts identity
\begin{align} \label{eq:MWintparts}
\int\,\dd^d x \ \phi_1\star\phi_2 = \int\,\dd^d x \ \phi_2\star \phi_1 = \int\,\dd^d x \ \phi_1 \, \phi_2 \ ,
\end{align}
which will translate to the statement that free field theories are unaffected by the Moyal--Weyl deformation. This identity implies that integration defines a trace on the deformed algebra of functions, i.e. it is cyclic with respect to the star-product \eqref{eq:starproduct}.

\paragraph{The kinematical Lie algebra $\boldsymbol{\fru_\star(1)}$.}
Using the star-product we define the \emph{star-commutator} of two functions $\phi_1,\phi_2\in\Omega^0(\FR^{1,d-1})$ by
\begin{align}\label{eq:commutatorfunctionsstar_1}
[\phi_1\ds\phi_2]_{\fru(1)} := \phi_1\star\phi_2 - \phi_2\star\phi_1 = -2\,\ii\,\phi_1\,\sin\Big(\frac{t}2 \ \lvpa_\mu \, \theta^{\mu\nu}\, \rvpa_\nu \Big)\,\phi_2 \ .
\end{align}
By construction the star-commutator defines a Lie bracket, and it is a deformation of the Poisson bracket $\{\phi_1,\phi_2\}_\theta = \theta^{\mu\nu}\,\partial_\mu\phi_1\,\partial_\nu\phi_2$ in the sense that its semi-classical limit is given by
\begin{align}
[\phi_1\ds\phi_2]_{\fru(1)} = -\ii\,t \, \{\phi_1,\phi_2\}_\theta + O(t^2) \ .
\end{align}
We write $\fru_\star(1)$ for the infinite-dimensional Lie algebra $\big(\Omega^0(\FR^{1,d-1}),[-\ds-]_{\fru(1)}\big)$. In the following this will appear naturally as a kinematical Lie algebra in our considerations of the double copy duality. 

A basis for the kinematical Lie algebra $\fru_\star(1)$ is given via Fourier transformation by the plane waves
\begin{align}
e_k(x) := \e^{\,\ii\,k\cdot x}
\end{align}
for $x\in\FR^{1,d-1}$ and $k\in(\FR^{1,d-1})^*$. They obey the star-commutation relations
\begin{align}\label{eq:starcommrels}
[e_k\ds e_p]_{\fru(1)} = \int\,\dd^dq \ \bar F(k,p,q) \ e_q \qquad \mbox{with} \quad \bar F(k,p,q) = 2\,\ii\,\sin\big(\tfrac{t}2\,k\cdot\theta\,p\big) \ \delta(k+p-q) \ .
\end{align}

A detailed description of the infinite-dimensional Lie algebra $\fru_\star(1)$ is found in~\cite{Lizzi:2001nd}, including its realization as a deformation of the Poisson--Lie algebra $\mathfrak{sdiff}(\FR^{1,d-1})$ of symplectic diffeomorphisms of $\FR^{1,d-1}$. In~\cite{Hoppe1989,Lizzi:2001nd} its relation to the infinite unitary Lie algebra~$\fru(\infty)$ is also described. By using an alternative monomial basis of generators for $\fru_\star(1)$ in $d=2$ dimensions, it is shown in~\cite{Pope:1989sr} that the deformation of the Poisson bracket to the Moyal--Weyl bracket corresponds to a deformation of the $w_{1+\infty}$ algebra of Poisson diffeomorphisms of the plane to a certain type of $W_{1+\infty}$ algebra called the symplecton algebra; see also~\cite{Lizzi:2001nd} for other features of $\fru_\star(1)$ in monomial and other bases as well as in arbitrary dimensionalities.

\paragraph{The colour Lie algebra $\boldsymbol{\fru_\star(\bar N)}$.}
The construction above can be generalized to include colour degrees of freedom in the following way. Let $\bar\frg$ be a matrix Lie algebra which is closed under both commutators and anticommutators; for definiteness we can take $\bar\frg=\fru(\bar N)$. Denote the Lie bracket of $\fru(\bar N)$ by $[-,-]_{\fru(\bar N)}$, and choose a basis $\{\bar T^{\bar a}\}$ of anti-Hermitian matrices for $\fru(\bar N)$ with structure constants \smash{$\bar f^{\bar a\bar b\bar c}$}, i.e. \smash{$[\bar T^{\bar a},T^{\bar b}]_{\fru(\bar N)}=\bar f^{\bar a\bar b\bar c}\,\bar T^{\bar c}$}. 

Let $\Omega^0(\FR^{1,d-1},\fru(\bar N))$ be the space of functions with values in $\fru(\bar N)$. This is an associative algebra under the composition of matrix multiplication with the pointwise product of functions, which as usual we make into a Lie algebra under the commutator bracket:
\begin{align}\label{eq:commutator0forms}
[\phi_1,\phi_2]_{\fru(\bar N)} = \phi_1\,\phi_2 - \phi_2\,\phi_1 \ .
\end{align}
There is a vector space isomorphism over $\FR$
\begin{align}\label{eq:vectorspaceiso0}
\Omega^0(\FR^{1,d-1},\fru(\bar N)) \simeq \fru(\bar N)\otimes\Omega^0(\FR^{1,d-1}) \ ,
\end{align}
and this is also true at the level of Lie algebras: the Lie algebra $\Omega^0(\FR^{1,d-1},\fru(\bar N))$ factorizes into the tensor product of the Lie algebra $\fru(\bar N)$ with the commutative algebra $\Omega^0(\FR^{1,d-1})$:
\begin{align}
\begin{split}
[\phi_1,\phi_2]_{\fru(\bar N)}=[\bar T^{\bar a}\otimes \phi_1^{\bar a},\bar T^{\bar b}\otimes\phi_2^{\bar b}]_{\fru(\bar N)} = [\bar T^{\bar a},\bar T^{\bar b}]_{\fru(\bar N)}\otimes (\phi_1^{\bar a}\,\phi_2^{\bar b}) \ ,
\end{split}
\end{align}
where $\phi_1^{\bar a},\phi_2^{\bar b}\in\Omega^0(\FR^{1,d-1})$ and we used commutativity of the pointwise multiplication of functions.

Let us now pass to the Moyal--Weyl deformation and define the star-commutator of functions in $\Omega^0(\FR^{1,d-1},\fru(\bar N))$ by
\begin{align}\label{eq:commutatorfunctionsstar}
[\phi_1\ds\phi_2]_{\fru(\bar N)} = \phi_1\star\phi_2 - \phi_2\star\phi_1 \ ,
\end{align}
where here the associative operation $\star$ means the composition of matrix multiplication with the star-product \eqref{eq:starproduct}.
This bracket again makes $\Omega^0(\FR^{1,d-1},\fru(\bar N))$ into an infinite-dimensional Lie algebra, which we denote by $\fru_\star(\bar N)$. However, while the vector space isomorphism \eqref{eq:vectorspaceiso0} still holds, it is no longer an isomorphism of Lie algebras, as now noncommutativity implies
\begin{align}\label{eq:factlack}
\begin{split}
[\phi_1\ds\phi_2]_{\fru(\bar N)}=[\bar T^{\bar a}\otimes \phi_1^{\bar a}\ds \bar T^{\bar b}\otimes\phi_2^{\bar b}]_{\fru(\bar N)} &= [\bar T^{\bar a},\bar T^{\bar b}]_{\fru(\bar N)}\otimes \tfrac12\,\big(\phi_1^{\bar a}\star\phi_2^{\bar b} + \phi_2^{\bar b} \star\phi_1^{\bar a}\big) \\
& \quad \, + \{\bar T^{\bar a},\bar T^{\bar b}\}_{\fru(\bar N)}\otimes \tfrac12\,\big(\phi_1^{\bar a}\star\phi_2^{\bar b} - \phi_2^{\bar b}\star\phi_1^{\bar a}\big) \ .
\end{split}
\end{align}

This lack of factorisation, which is the well-known intertwining between  colour and kinematical degrees of freedom in noncommutative field theories, will make colour-kinematics duality somewhat subtle in these instances. Indeed, as noted by e.g.~\cite{Borsten:2021hua}, colour-stripping is only possible in theories whose interactions are constructed exclusively from Lie algebra commutators \smash{$[\bar T^{\bar a},\bar T^{\bar b}]_{\fru(\bar N)}$}, even if all fields are valued in the adjoint representation of $\fru(\bar N)$. This is evidently not the case in noncommutative field theories, whose interactions also involve the anticommutators \smash{$\{\bar T^{\bar a},\bar T^{\bar b}\}_{\fru(\bar N)}$}. A related theory which violates the criterion is the non-abelian Dirac--Born--Infeld theory, whose fields are all ${\rm ad}(\fru(\bar N))$-valued but whose interactions also involve \smash{$\{\bar T^{\bar a},\bar T^{\bar b}\}_{\fru(\bar N)}$}. 
We shall discuss  in detail below how to handle this difficulty in our theories: it is precisely the \emph{mixing} of kinematics with the various colour weights which will permit a double copy construction. This is analogous to the extensions of colour-kinematics duality discussed in~\cite{Carrasco:2019yyn,Low:2019wuv,Carrasco:2021ptp,Carrasco:2022jxn} for theories involving higher derivative operators and gauge algebra anticommutators.

Let us rewrite the star-commutator in terms of the symmetric d-coefficients $\bar d^{\bar a\bar b\bar c}$ defined by $\{\bar T^{\bar a},\bar T^{\bar b}\}_{\fru(\bar N)} = \ii\,\bar d^{\bar a\bar b\bar c}\,\bar T^{\bar c}$ to get
\begin{align}\label{eq:scalcommanticomm}
[\phi_1\ds\phi_2]_{\fru(\bar N)}=\tfrac12\,\big(\bar f^{\bar a\bar b\bar c}\,\{\phi_1^{\bar a}\ds\phi_2^{\bar b}\}_{\fru(1)} + \ii\,\bar d^{\bar a\bar b\bar c}\,[\phi_1^{\bar a}\ds\phi_2^{\bar b}]_{\fru(1)}\big) \, \bar T^{\bar c} \ ,
\end{align}
where we introduced the star-anticommutator of functions 
\begin{align}
\big\{\phi_1^{\bar a}\ds\phi_2^{\bar b}\big\}_{\fru(1)} := \phi_1^{\bar a}\star\phi_2^{\bar b} + \phi_2^{\bar b} \star\phi_1^{\bar a} = 2\,\phi_1^{\bar a}\, \cos\Big(\frac{t}2 \ \lvpa_\mu \, \theta^{\mu\nu}\, \rvpa_\nu \Big)\,\phi_2^{\bar b} \ .
\end{align}
A momentum space basis for $\fru_\star(\bar N)$ is given by the $\fru(\bar N)$-valued plane waves
\begin{equation}\label{eq:ustarnbasis}
    e^{\bar a}_k(x) := \e^{\,\ii\,k \cdot x} \, \bar T^{\bar a} \ \in \  \Omega^0(\mathbbm{R}^{1,d-1}, \mathfrak{u}(\bar N))\ ,
\end{equation}
whose star-commutation relations
\begin{equation}
\label{eq:star commutator with colour}
    \big[e_k^{\bar a} \ds e_p^{\bar b}\big]_{\fru(\bar N)} = \int\, \mathrm{d}^dq \ \bar F^{\bar a\bar b\bar c}(k,p,q) \ e^{\bar c}_q 
\end{equation}
can be expressed in terms of structure constants
\begin{equation}\label{eq:u(n) star structure constant}
\begin{split}
\bar F^{\bar a\bar b\bar c}(k,p,q) &= \Big(\bar f^{\bar a\bar b\bar c} \cos \big(\tfrac{t}{2} \, k \cdot \theta\, p\big) 
    +\ii\,\bar d^{\bar a\bar b\bar c}\sin\big( \tfrac{t}{2}\, k \cdot \theta\, p\big)\Big) \ \delta(k+p-q) \ .
\end{split}
\end{equation}

These relations again highlight the mixing of colour and kinematics in the noncommutative algebra of $\fru(\bar N)$-valued functions, which allows the incorporation of noncommutative field theories into the standard $L_\infty$-algebra formalism~\cite{Blumenhagen:2018kwq,Giotopoulos:2021ieg}. See~\cite{Monteiro:2022xwq} for a description of the infinite-dimensional Lie algebra $\fru_\star(\bar N)$ as a deformation of a Kac--Moody algebra in $d=2$ dimensions, where the colour-kinematics mixing in the context of the double copy was also noted.

\subsection{Noncommutative biadjoint scalar theory}
\label{sub:NCbiadjoint}

A natural noncommutative deformation of the biadjoint scalar theory of \S\ref{sub:biadjoint} is defined by replacing the second Lie algebra $\bar\frg$ by the infinite-dimensional Lie algebra $\fru_\star(\bar N)$, using the Lie bracket \eqref{eq:scalcommanticomm}. It is organised by a cyclic strict $L_\infty$-algebra $\BAS_\star$ whose cochain complex $\Ch(\BAS_\star)=\Ch_{\textrm{\tiny BAS}}$ coincides with that of the commutative case \eqref{eq:nc biadjoint cochain complex} with $\bar\frg=\fru(\bar N)$, which again identifies the differential $\mu_1^{\star\textrm{\tiny BAS}}=\mu_1^{\textrm{\tiny BAS}}=\square$. 

The higher bracket $\mu_2^{\star\textrm{\tiny BAS}}: \Omega^0(\FR^{1,d-1},\frg\otimes\fru(\bar N))[-1]^{\otimes 2} \longrightarrow\Omega^0(\FR^{1,d-1},\frg\otimes\fru(\bar N))[-2]$ is given by
\begin{align}
\begin{split}
\mu_2^{\star\textrm{\tiny BAS}}(\phi_1,\phi_2) =&\, -g\,\llbracket\phi_1,\phi_2\rrbracket_{\frg\otimes\fru_\star(\bar N)} \\[4pt]
 :=&\, -\tfrac g2 \, f^{abc} \,\big(\bar f^{\bar a\bar b\bar c} \, \{\phi_1^{a\bar a}\ds\phi_2^{b\bar b}\}_{\fru(1)} + \ii\,\bar d^{\bar a\bar b\bar c} \, [\phi_1^{a\bar a}\ds\phi_2^{b\bar b}]_{\fru(1)} \big) \, T^c\otimes\bar T^{\bar c} \ .
\end{split}
\end{align}
Note that
\begin{align}\label{eq:NCbiadell2fact}
[\![T^a\otimes\bar T^{\bar a}\otimes \phi_1^{a\bar a},T^b\otimes \bar T^{\bar b}\otimes \phi_2^{b\bar b}]\!]_{\frg\otimes\fru_\star(\bar N)} = [T^a,T^b]_\frg\otimes [\bar T^{\bar a}\otimes \phi_1^{a\bar a}\ds\bar T^{\bar b}\otimes\phi_2^{b\bar b}]_{\fru(\bar N)} \ ,
\end{align}
and hence the $2$-bracket is symmetric. 

The cyclic inner product of degree $-3$ is given by the non-zero pairing
\begin{align}
\langle\phi,\phi^+\rangle_{\textrm{\tiny BAS}}^\star = \int\,\dd^dx \ \Tr_{\frg\otimes\fru(\bar N)}(\phi\star\phi^+) = \int\,\dd^dx \ \phi^{a\bar a}\star\phi_{a\bar a}^+= \langle\phi,\phi^+\rangle_{\textrm{\tiny BAS}} \ ,
\end{align}
for $\phi\in\Omega^0(\FR^{1,d-1},\frg\otimes\fru(\bar N))[-1]$ and $\phi^+\in\Omega^0(\FR^{1,d-1},\frg\otimes\fru(\bar N))[-2]$, where we used \cref{eq:MWintparts}. Note that
\begin{align}\label{eq:NCbiadcyclicfact}
\langle T^a\otimes\bar T^{\bar a}\otimes\phi_{a\bar a},T^b\otimes\bar T^{\bar b}\otimes\phi_{b\bar b}^+\rangle_{\textrm{\tiny BAS}}^\star = \Tr_{\frg}(T^a\otimes T^b) \ \Tr_{\fru_\star(\bar N)}(\bar T^{\bar a}\,\phi_{a\bar a}\star\bar T^{\bar b}\,\phi_{b\bar b}^+) \ ,
\end{align}
where $\Tr_{\fru_\star(\bar N)}=\int\,\circ\,\Tr_{\fru(\bar N)}$ defines a cyclic trace on the noncommutative algebra of functions on $\FR^{1,d-1}$ with values in the matrix Lie algebra $\fru(\bar N)$.

Altogether, we obtain a cyclic strict $L_\infty$-algebra
\begin{align}
\BAS_\star = \big( \mathsf{Ch}_{\textrm{\tiny BAS}} , \mu_2^{\star\textrm{\tiny BAS}}, \langle -,- \rangle_{\textrm{\tiny BAS}}\big) \ .
\end{align}
The underlying free field theory from \S\ref{sub:biadjoint} is unchanged and only the interactions undergo noncommutative deformation. Nevertheless, for the purposes of presentation, we continue to delineate all operations with the symbol `$\star$' to emphasise that we are working with a noncommutative field theory. 

The Maurer--Cartan equation \eqref{eq:Maurer Cartan equation} gives the equation of motion for $\phi\in\Omega^0(\FR^{1,d-1},\frg\otimes\fru(\bar N))$ as
\begin{align}
\mu_1^{\star\textrm{\tiny BAS}}(\phi) + \tfrac12\,\mu_2^{\star\textrm{\tiny BAS}}(\phi,\phi) = \square\,\phi - \tfrac g2\,\llbracket\phi,\phi\rrbracket_{\frg\otimes\fru_\star(\bar N)} = 0 \ ,
\end{align}
which on component functions in $\Omega^0(\FR^{1,d-1})$ reads as
\begin{align}
\square\,\phi^{a\bar a} -\tfrac g2\,f^{abc}\,\bar\lambda^{\bar a\bar b\bar c}\, \phi^{b\bar b}\star\phi^{c\bar c} = 0 \ ,
\end{align}
where
\begin{align}
\bar\lambda^{\bar a\bar b\bar c} = \bar f^{\bar a\bar b\bar c} + \ii\,\bar d^{\bar a\bar b\bar c} \ .
\end{align}
The Maurer--Cartan functional \eqref{eq:MC functional} is
\begin{align}\label{eq:NCbiadMCaction}
\begin{split}
S_{\textrm{\tiny BAS}}^\star[\phi] &= \tfrac12\,\langle\phi,\mu_1^{\star{\textrm{\tiny BAS}}}(\phi)\rangle^\star_{\textrm{\tiny BAS}} + \tfrac1{3!} \, \langle\phi,\mu_2^{\star{\textrm{\tiny BAS}}}(\phi,\phi)\rangle^\star_{\textrm{\tiny BAS}} \\[4pt]
&= \int\,\dd^dx \ \Tr_{\frg\otimes\fru(\bar N)}\Big( \frac12\,\phi\star\square\,\phi - \frac g{3!} \, \phi\star [\![\phi,\phi]\!]_{\frg\otimes\fru_\star(\bar N)}\Big)  \\[4pt]
&= \int\,\dd^dx \ \frac12\,\phi^{a\bar a}\star\square\,\phi_{a\bar a} - \frac g{3!} \,  f^{abc} \, \bar\lambda^{\bar a\bar b\bar c} \, \phi^{a\bar a} \star \phi^{b\bar b} \star \phi^{c\bar c} \ .
\end{split}
\end{align}

\remark{ \label{rem:NCbiadjointfamily}
We stress that the noncommutative biadjoint scalar theory possesses a \emph{rigid} symmetry under $\frg\oplus\fru(\bar N)$, because the ${\rm ad}(\fru(\bar N))$-action preserves independently the $\fru(\bar N)$ commutator and anticommutator in \cref{eq:factlack}. In particular, we may define different noncommutative field theories with the same rigid symmetry by replacing $\bar\lambda^{\bar a\bar b\bar c}$ by $\bar f^{\bar a \bar b\bar c}$, $\ii\,\bar d^{\bar a\bar b\bar c}$ or any linear combination of the two sets of structure constants. That is, there exists in fact a two-parameter family of noncommutative deformations of the biadjoint scalar theory defined by the more general structure constants
\begin{align}
\bar\lambda_{\bar u,\bar v}^{\bar a \bar b\bar c} := \bar u\,\bar f^{\bar a \bar b\bar c} + \ii\,\bar v\,\bar d^{\bar a\bar b\bar c} \ , \quad \bar u,\bar v\in \FR \ .
\end{align}
In the following we will mostly concentrate on the natural member of this family induced by the structure constants of the Lie algebra $\fru_\star(\bar N)$, with $\bar u=\bar v=1$, as above.
}

\subsection{Twisted homotopy factorisation}
\label{sub:factNCbiadjoint}

Let us now consider the problem of factorizing the cyclic strict $L_\infty$-algebra $\BAS_\star$. The three-point Feynman vertex of the noncommutative biadjoint scalar theory is given by
\begin{equation}\label{eq:2pointncbiad}
\begin{split}
\begin{tikzpicture}[scale=0.8]
{\small
    \draw[fermionnoarrow] (140:1.8)node[above]{$a,\bar a$}--(0,0);
    \draw[fermionnoarrow] (-140:1.8)node[below]{$b ,\bar b$}--(0,0);
    \draw[fermionnoarrow] (0:1.8)node[right]{$c,\bar c$} --(0,0);
    \draw[->, shift={(-0.2,0.45)}] (140:.5) node[above]{$p_1$} -- (0,0);
    \draw[->, shift={(-0.2,-0.45)}] (-140:.5) node[below]{$p_2$} -- (0,0);
    \draw[->, shift={(0.2,0.2)}] (0:.5) node[above]{$p_3$} -- (0,0);
    }\normalsize
    \node at (5.9, 0) {$= \ - \ii\,g\, f^{abc}\,\bar F^{\bar a \bar b \bar c}(p_1,p_2,p_3) \ .$};
\end{tikzpicture}
\end{split}
\end{equation}
The lack of factorisation of the star-commutator \smash{$[-\ds-]_{\fru(\bar N)}$}, discussed in \S\ref{sub:MWdeformation}, means that 
we cannot simply strip off the rigid $\fru(\bar N)$ symmetry, like we did with $\bar\frg$ in the biadjoint theory  from \S\ref{sub:biadjoint}. Instead, we show that the infinite-dimensional Lie algebra $\fru_\star(\bar N)$ admits a `twisted factorisation', with the twist implementing the intertwining of colour and kinematical degrees of freedom of the scalar fields. 

For this, we factorize $\BAS_\star$ into three parts: a colour part, a twisted colour part and a cyclic strict $L_\infty$-algebra which fully describes the trivalent interactions. There are two steps: In the first step, we show that $\BAS_\star$ admits a factorization into the colour Lie algebra $\frg$ and a kinematical cyclic strict $C_\infty$-algebra $\frC_{\fru_\star(\bar N)}$, which corresponds to colour-stripping the cyclic strict $L_\infty$-algebra~$\BAS_\star$. 

The cochain complex underlying $\frC_{\fru_\star(\bar N)}$ is
 \begin{align}\label{eq:biadcochaincomplex}
    \mathsf{Ch}(\frC_{\fru_\star(\bar N)}) =
    \Big(
    \Omega^0(\FR^{1,d-1},\fru(\bar N))[-1] \xrightarrow{ \ \square \ } \Omega^0(\FR^{1,d-1},\fru(\bar N))[-2] \Big)\ ,
\end{align}
which again identifies the differential ${m}_1^{\fru_\star(\bar N)} = \square$ acting as ${m}_1^{\fru_\star(\bar N)}(\phi) = \square\,\phi_{\bar a}\, \bar T^{\bar a} $. 
The product \smash{$m_2^{\fru_\star(\bar N)}:\Omega^0(\FR^{1,d-1},\fru(\bar N))[-1]^{\otimes2} \longrightarrow \Omega^0(\FR^{1,d-1},\fru(\bar N))[-2]$} is defined by
\begin{align}
m_2^{\fru_\star(\bar N)}(\phi_1,\phi_2) =  -g\,[\phi_1\ds\phi_2]_{\fru(\bar N)} \ .
\end{align}
This operation is clearly graded commutative, and for degree reasons it satisfies the Leibniz rule as well as the graded associativity relation. The cyclic structure is defined by  $\Tr_{\fru_\star(\bar N)}=\int\,\circ\,\Tr_{\fru(\bar N)}$. 

This cyclic strict $C_\infty$-algebra allows for a factorization
\begin{align}
\BAS_\star = \frg \otimes \frC_{\fru_\star(\bar N)} \ .
\end{align}
For this, we identify the vector space $\Omega^0(\FR^{1,d-1},\frg\otimes\fru(\bar N))$ in the usual way with the tensor product $\frg\otimes\Omega^0(\FR^{1,d-1},\fru(\bar N))$, recall the definition of the differential $\mu_1^{\star{\textrm {\tiny BAS}}}$ and the $2$-bracket $\mu_2^{\star{\textrm {\tiny BAS}}}$, and use \cref{eq:NCbiadell2fact} as well as \cref{eq:NCbiadcyclicfact}. It then easily follows that the structure maps of $\BAS_\star$ factor through \smash{$\mu_1^{\star\textrm{\tiny BAS}} = \mathbbm{1}\otimes m_1^{\fru_\star(\bar N)}$}, \smash{$\mu_2^{\star\textrm{\tiny BAS}} = [-,-]_\frg\otimes m_2^{\fru_\star(\bar N)}$} and \smash{$\langle-,-\rangle_{\textrm {\tiny BAS}}^\star = \Tr_\frg\otimes\Tr_{\fru_\star(\bar N)}$}.

The second step consists in stripping off the kinematical factor encoding the trivalent interactions. This uses the notion of twisted tensor product of homotopy algebras from~\cite{Borsten:2021hua},\footnote{The general notion of `twist' in this context should not be confused with Drinfel'd twist deformation in the context of noncommutative geometry, though in our setting they both play a similar role in defining a noncommutative deformation.} which we review in~\S\ref{sub:preliminaries}. Regarding $\fru(\bar N)$ as a graded vector space sitting in degree~$0$, we wish to compute the twisted tensor product between $\fru(\bar N)$ and the cyclic strict $L_\infty$-algebra $\Scal$ of the commutative $ g\,\phi^3$-theory on $\FR^{1,d-1}$ from \S\ref{sub:nc_strict_scalar_field_theory}. The twist datum \smash{$\tau^{\fru_\star(\bar N)} = \big(\tau_1^{\fru_\star(\bar N)},\tau_2^{\fru_\star(\bar N)}\big)$} consists of a pair of maps \smash{$\tau_1^{\fru_\star(\bar N)}:\fru(\bar N)\longrightarrow \fru(\bar N)\otimes\sEnd(L)$} and \smash{$\tau_2^{\fru_\star(\bar N)}:\fru(\bar N)\otimes\fru(\bar N)\longrightarrow \fru(\bar N)\otimes\sEnd(L)\otimes\sEnd(L)$}, where 
\begin{align} \label{eq:scalarvectorspace}
L=\Omega^0(\FR^{1,d-1})[-1]\oplus\Omega^0(\FR^{1,d-1})[-2]
\end{align}
is the graded vector space underlying the cochain complex $\Ch_{\rm Scal}$ from \cref{eq:scalarcochaincomplex}. 

We set
\begin{align}
\tau^{\fru_\star(\bar N)}_1(\bar T^{\bar a}) = \bar T^{\bar a}\otimes \mathbbm{1}
\end{align}
and
\begin{align}
\tau_2^{\fru_\star(\bar N)}(\bar T^{\bar a}, \bar T^{\bar b}) = [\bar T^{\bar a},\bar T^{\bar b}]_{\fru(\bar N)}\otimes\cos\big(\tfrac t2\,\theta^{\mu\nu} \, \partial_\mu \otimes \partial_\nu\big) + \ii\,\{\bar T^{\bar a},\bar T^{\bar b}\}_{\fru(\bar N)} \otimes \sin\big(\tfrac t2\,\theta^{\mu\nu} \, \partial_\mu \otimes \partial_\nu\big) \ .
\end{align}

Following the prescription of~\cite{Borsten:2021hua}, the tensor product $\fru(\bar N)\otimes L$ now carries the structure of a cyclic strict $C_\infty$-algebra with differential
\begin{align}
m_1^{\tau^{\fru_\star(\bar N)}}(\bar T^{\bar a}\otimes\phi_{\bar a}) = \bar T^{\bar a}\otimes\mu_1^{\rm Scal}(\phi_{\bar a})= \bar T^{\bar a}\otimes\square\,\phi_{\bar a} \ ,
\end{align}
and product
\begin{align}
\begin{split}
m_2^{\tau^{\fru_\star(\bar N)}}(\bar T^{\bar a}\otimes\phi^{\bar a}_1,\bar T^{\bar b}\otimes \phi^{\bar b}_2) &= [\bar T^{\bar a},\bar T^{\bar b}]_{\fru(\bar N)}\otimes\mu_2^{\rm Scal}\circ\cos\big(\tfrac t2\,\theta^{\mu\nu} \, \partial_\mu \otimes \partial_\nu\big) (\phi_1^{\bar a}\otimes\phi_2^{\bar b}) \\
& \quad \, + \ii\,\{\bar T^{\bar a},\bar T^{\bar b}\}_{\fru(\bar N)} \otimes \mu_2^{\rm Scal}\circ\sin\big(\tfrac t2\,\theta^{\mu\nu} \, \partial_\mu \otimes \partial_\nu\big)(\phi_1^{\bar a}\otimes\phi_2^{\bar b}) \\[4pt]
&=  -g\,[\bar T^{\bar a}\otimes\phi^{\bar a}_1\ds\bar T^{\bar b}\otimes\phi^{\bar b}_2]_{\fru(\bar N)} \ ,
\end{split}
\end{align}
where we used \cref{eq:twisted 2-brackets,eq:ncscalarmu2}. The cyclic structure does not twist.
This cyclic strict $C_\infty$-algebra is the twisted tensor product of $\fru(\bar N)$ and $\Scal$, denoted \smash{$\fru(\bar N)\otimes_{\tau^{\fru_\star(\bar N)}}\Scal$}. From the definitions it immediately follows that \smash{$m_1^{\fru_\star(\bar N)}=m_1^{\tau^{\fru_\star(\bar N)}}$} and \smash{$m_2^{\fru_\star(\bar N)}=m_2^{\tau^{\fru_\star(\bar N)}}$}, and hence the cyclic strict $C_\infty$-algebra $\frC_{\fru_\star(\bar N)}$ factorizes as
\begin{align}\label{eq:CCCbiadstarfact}
\frC_{\fru_\star(\bar N)} = \fru(\bar N)\otimes_{\tau^{\fru_\star(\bar N)}}\Scal \ .
\end{align}

Altogether, we have shown that the cyclic strict $L_\infty$-algebra of the noncommutative biadjoint scalar theory on $\FR^{1,d-1}$ admits the factorization
\begin{align}\label{eq:NCbiadfact}
\BAS_\star = \frg\otimes\big(\fru(\bar N)\otimes_{\tau^{\fru_\star(\bar N)}}\Scal\big) \ .
\end{align}
This is one of the main messages of this paper: colour-stripping is achievable in noncommutative field theories through a suitable notion of twisted factorisation, which enables one to disentangle colour and kinematical degrees of freedom. From this perspective noncommutativity is completely absorbed into the twist maps $\tau^{\fru_\star(\bar N)}$. This will also be applicable to the noncommutative theories with local gauge symmetry that we consider later on.
\subsection{Adjoint scalar theory } 
\label{sub:adjoint_scalar_theory_as_a_strict_linf}

The limiting rank one case $\bar N=1$ of the noncommutative biadjoint scalar theory is an interesting noncommutative scalar field theory that does not have a non-trivial commutative counterpart, in the sense that it becomes non-interacting in the commutative limit. We therefore refrain from using the adjective `noncommutative' and simply refer to it as the adjoint scalar theory on $\FR^{1,d-1}$. We begin by briefly summarising its main features, which follow by setting $\bar N=1$ everywhere in \S\ref{sub:NCbiadjoint}.

The cyclic strict $L_\infty$-algebra $\AS$ of this theory has cochain complex
\begin{align}\label{eq:adcochaincomplex}
\Ch(\AS) = \Big(\Omega^0(\FR^{1,d-1},\frg)[-1] \xrightarrow{ \ \square \ } \Omega^0(\FR^{1,d-1},\frg)[-2] \Big) \ ,
\end{align}
and the single non-vanishing higher bracket 
\begin{align}\label{eq:adell2fact}
\mu_2^{\textrm{\tiny AS}}(\phi_1,\phi_2) = -g\, [\![T^a\otimes \phi_1^a,T^b\otimes \phi_2^b]\!]_{\textrm{\tiny AS}} = -\ii\,g\, [T^a,T^b]_\frg\otimes [\phi_1^a\ds\phi_2^b]_{\fru(1)} \ ,
\end{align}
where we abbreviate $\llbracket-,-\rrbracket_{\textrm{\tiny AS}} := \llbracket-,-\rrbracket_{\frg\otimes\fru_\star(1)}$.
Note that this binary operation is not the antisymmetric Lie bracket $[\phi_1\ds\phi_2]_\frg$; in particular, it vanishes in the commutative limit $t=0$. The cyclic structure is given by
\begin{align}
\langle\phi,\phi^+\rangle_{\textrm{\tiny AS}} = \int\,\dd^dx \ \Tr_\frg(\phi\star\phi^+) = \int\,\dd^dx \ \phi^a\star\phi_a^+ \ .
\end{align}

The equations of motion
\begin{align}\label{eq:adMCeq}
\square\,\phi^{a} - \ii\, g \, f^{abc} \, \phi^{b} \star \phi^{c} = 0
\end{align}
of the adjoint scalar theory are the variational equations of the action functional
\begin{align}\label{eq:adMCaction}
\begin{split}
    S_{\textrm{\tiny AS}}[\phi] &= \int\,\dd^dx \ \frac12\,\phi^{a}\star\square\,\phi_{a} - \frac {\ii\,g}{3} \,  f^{abc} \, \phi^{a} \star \phi^{b} \star \phi^{c} \ .
\end{split}
\end{align}
The three-point Feynman vertex defined by \cref{eq:adMCaction} is
\begin{equation}\label{eq:3pointvertexadjoint}
\begin{split}
\begin{tikzpicture}[scale=0.8]
{\small
   \draw[fermionnoarrow] (140:1.8)node[above]{$a$}--(0,0);
    \draw[fermionnoarrow] (-140:1.8)node[below]{$b$}--(0,0);
    \draw[fermionnoarrow] (0:1.8)node[right]{$c$} --(0,0);
    \draw[->, shift={(-0.2,0.45)}] (140:.5) node[above]{$p_1$} -- (0,0);
    \draw[->, shift={(-0.2,-0.45)}] (-140:.5) node[below]{$p_2$} -- (0,0);
    \draw[->, shift={(0.2,0.2)}] (0:.5) node[above]{$p_3$} -- (0,0);
    }\normalsize
    \node at (5.2, 0) {$= \ -\ii\,g\,f^{abc} \, \bar F(p_1,p_2,p_3) \ . $};
\end{tikzpicture}
\end{split}
\end{equation}
This is completely antisymmetric due to symmetry of the structure constants of $\fru_\star(1)$ from \cref{eq:starcommrels} under cyclic permutations of the momenta, which follows from  momentum conservation.

\remark{
The adjoint scalar theory is formally obtained from the standard biadjoint scalar theory of \S\ref{sub:biadjoint} by replacing the second Lie algebra $\bar\frg$ with the kinematical Lie algebra $\fru_\star(1)$, via the prescription
\begin{align}\label{eq:adprescr}
    \phi^{a\bar a}\longrightarrow \phi^a \ , \quad \bar f^{\bar a\bar b\bar c} \longrightarrow \bar F(k,p,q)  \qquad \mbox{and} \qquad \Tr_{\frg\otimes\bar\frg} \longrightarrow \textrm{\small$\int$}\,\circ\,\Tr_\frg
\end{align}
in the Lagrangian of the commutative biadjoint scalar theory. This line of reasoning was used in \cite{Cheung:2022mix} to discuss aspects of a non-perturbative Lagrangian-level double copy between these field theories.
}

\paragraph{Twisted factorisation.}

The factorisation of the adjoint scalar $L_\infty$-algebra $\AS$ follows from the $\bar N=1$ limit of the twisted factorisation \eqref{eq:NCbiadfact}, but with a crucial simplification. We treat the Lie algebra $\fru_\star(1)$ as a kinematical factor by identifying $\fru(1)\simeq\FR$ and regarding it as the kinematical vector space of the adjoint scalar theory: 
\begin{align}
\Kin_{\wedge^0} := \FR \ .
\end{align}
The tensor products (over $\FR$) with the one-dimensional vector space $\FR$ are trivial, so that we may identify $\FR\otimes W\simeq W$ for any real vector space $W$. The definition of the twist datum $\tau^{\fru_\star(1)}$ correspondingly simplifies to
\begin{align}
\tau^{\fru_\star(1)}_1(1) = \mathbbm{1} \qquad \mbox{and} \qquad 
\tau_2^{\fru_\star(1)}(1) = 2\,\ii\,\sin\big(\tfrac t2\,\theta^{\mu\nu} \, \partial_\mu \otimes \partial_\nu\big) \ ,
\end{align}
from which one recovers the star-commutator bracket $[- \stackrel{\star}{,} -]_{\mathfrak{u}(1)}$.

Thus the cyclic strict $L_\infty$-algebra of the adjoint scalar theory on $\FR^{1,d-1}$ factorises as
\begin{align}\label{eq:adfact}
\AS = \frg\otimes(\Kin_{\wedge^0}\otimes_{\tau^{\fru_\star(1)}}\Scal) \ .
\end{align}
This agrees with the prescription in \eqref{eq:adprescr}, and it identifies the commutative biadjoint scalar theory as the zeroth copy of the adjoint scalar theory: replacing the kinematic factor $\Kin_{\wedge^0} = \FR$ in the factorisation \eqref{eq:adfact} with the second colour factor $\bar\frg$ yields the cyclic strict $L_\infty$-algebra \eqref{eq:ncbiadjointfact}, 
which is the correct homotopy algebraic structure in the zeroth copy prescription. 

\subsection{Colour-kinematics duality}
\label{sub:adjointCKduality}

In terms of the factorization \eqref{eq:adfact}, the double copy prescription for the adjoint scalar theory is immediately apparent. 
However, we must first verify that the factorisation is compatible with colour-kinematics duality. 
To do this, we will demonstrate how colour-kinematics duality is manifest in the Berends--Giele currents of the adjoint scalar theory.  This framework realises the double copy as a generalisation of the colour-kinematics dual formulation (relying on kinematic Jacobi identities), rather than the KLT formulation (relying on relations amongst colour ordered amplitudes) that we will encounter in \cref{sec:klt_with_b_field}.

\paragraph{Perturbative calculations.} 
\label{par:perturbative_calculations_scalar}

We first give the perturbiner expansion, reviewed in \S\ref{sub:scattering_in_the_textlinf_formalism}, for the adjoint scalar theory.
The contracting homotopy satisfying the Hodge--Kodaira decomposition for the projection to the minimal model of the $L_\infty$-algebra $\AS$ is simply the massless Feynman propagator
\begin{equation}
\begin{split}
    (\mu_1^{\textrm{\tiny AS}}\circ G_{\textrm{F}})(x,y) = \ii \, \delta(x-y) \qquad \mbox{with} \quad
    G_{\textrm{F}}(\e^{\,\ii\,p\cdot x})= -\frac{\ii}{p^2} \, \e^{\,\ii\,p\cdot x} \ ,
\end{split}
\end{equation}
where $\mu_1^{\textrm{\tiny AS}}=\square$.
From a perturbiner element $\phi^\circ \in \sfH^1(\AS)=\ker(\square)$, we look for a quasi-isomorphism $\psi_n: \sfH^1(\AS)^{\otimes n} \longrightarrow L^1$ which is given by the recursion relations \eqref{eq:minimal model recursion relations}.
Given $\phi^\circ$, a perturbiner expansion for the interacting theory is nothing but the Maurer--Cartan field under this quasi-isomorphism, namely
\begin{equation}\label{eq:MCadjoint}
    \phi = \sum_{n=1}^\infty\, \frac{1}{n!} \, \psi_n(\phi^{\circ\,\otimes n}) \ \in \  \Omega^0(\FR^{1,d-1},\frg)[-1] \ .
\end{equation}

We now consider the decomposition in terms of on-shell multiparticle solutions. This differs from the formula \eqref{eq:MCadjoint} that is used to reconstruct a Maurer--Cartan element in the full interacting theory from identical free fields. Denote by $\phi(i) \in \sfH^1(\AS)$ a field in the minimal model corresponding to the `$i$-th' adjoint scalar.
Start by decomposing the quasi-isomorphism maps in terms of plane waves with coefficients $J(1\cdots n)$, called the {Berends--Giele currents}:
\begin{equation}
    \psi_n\big(\phi(1), \ldots, \phi(n)\big) = J(1 \cdots n) \ \e^{\,\ii\, p_{1\cdots n}\cdot x} \ \in \ \Omega^0(\FR^{1,d-1},\frg)[-1] \ ,
\end{equation}
where $p_{1\cdots n}:=p_1+\cdots+ p_n$.
These currents are sometimes written as $\phi_{1\cdots n}$; they are simply the Fourier coefficients of the quasi-isomorphism applied to $n$ different external states.

The recursion relations for the currents are extracted from the quasi-isomorphisms $\psi_n$, and one finds for $n >1$
\begin{equation}
\begin{split}\label{eq:recursion for adjoint}
    J(1 \cdots n)
    &= \frac{g}{2} \,
    \sum_{i=1}^{n-1} \ 
    \sum_{\sigma \in \mathrm{Sh}(i;n)} \,
    G_{\rm F} \circ \llbracket   J(\sigma(1), \ldots, \sigma(i)) , J(\sigma(i+1), \ldots, \sigma(n))\rrbracket_{\textrm{\tiny AS}} \ ,
\end{split}
\end{equation}
where we understand $G_{\rm F}$ as acting on the plane wave basis representing the double bracket. 
Writing $\phi(i) = \phi_i \, \e^{\,\ii\,p_i\cdot x}$ for $\phi_i \in \Omega^0(\mathbbm{R}^{1,d-1}, \mathfrak{g})$, the first few currents are given by
\begin{equation}
\begin{split}
    J(1) &= \phi_1 \ , \\[4pt]
    J(12)& =  -\ii\,g \, \frac{\llbracket \phi_1, \phi_2\rrbracket_{\textrm{\tiny AS}}}{s_{12}} \ , \\[4pt]
    J(123)& = -(\,\ii\,g)^2 \, \bigg(
    \frac{\llbracket \phi_1,  \llbracket \phi_2, \phi_3\rrbracket_{\textrm{\tiny AS}} \rrbracket_{\textrm{\tiny AS}}}{s_{23}\,s_{123}} + 
    \frac{\llbracket \phi_2, \llbracket \phi_3, \phi_1\rrbracket_{\textrm{\tiny AS}}\rrbracket_{\textrm{\tiny AS}}}{s_{31}\,s_{123}} +
    \frac{\llbracket\phi_3,\llbracket \phi_1, \phi_2\rrbracket_{\textrm{\tiny AS}} \rrbracket_{\textrm{\tiny AS}}}{s_{12}\,s_{123}}
    \bigg) \ ,
\end{split}
\end{equation}
and so on, where $s_{ij\cdots}=(p_i+p_j+\cdots)^2$ are Mandelstam invariants.
This verifies that the currents are cyclic symmetric.

The brackets on the minimal model, for the projection $\sfp: L^1 \longrightarrow \sfH^1(\AS)$, are given by \cref{eq:min model brackets}:
\begin{align}\label{eq:recursion for adjoint2}
\begin{split}
    & \mu_n^{\circ{\textrm{\tiny AS}}}\big(\phi(1), \ldots, \phi(n)\big) \\[4pt]
    & \hspace{2cm} = -\frac{g}{2} \,
    \sum_{i=1}^{n-1} \ 
    \sum_{\sigma \in \mathrm{Sh}(i;n)}\,
    \sfp \circ
    \llbracket 
    J(\sigma(1), \ldots, \sigma(i)) , J( \sigma(i + 1), \ldots, \sigma(n))
    \rrbracket_{\textrm{\tiny AS}} \
    \e^{\,\ii\, p_{1 \cdots n}\cdot x} \ .
\end{split}
\end{align}
For the first few brackets this gives
\begin{align}
\begin{split}
    \mu^{\circ\textrm{\tiny AS}}_2(\phi_1,\phi_2) &= -g\,\llbracket \phi_1, \phi_2\rrbracket_{\textrm{\tiny AS}} \ \e^{\,\ii\,p_{12}\cdot x} \ , \\[4pt]
    \mu^{\circ\textrm{\tiny AS}}_3(\phi_1,\phi_2,\phi_3)&= \ii\,g^2 \, \bigg(
    \frac{ \llbracket \phi_1,  \llbracket \phi_2, \phi_3\rrbracket_{\textrm{\tiny AS}}\rrbracket_{\textrm{\tiny AS}}}{s_{23}} +
    \frac{\llbracket \phi_2, \llbracket \phi_3, \phi_1\rrbracket_{\textrm{\tiny AS}}\rrbracket_{\textrm{\tiny AS}}}{s_{31}} \\
    & \hspace{7cm} +
    \frac{\llbracket \phi_3, \llbracket \phi_1, \phi_2\rrbracket_{\textrm{\tiny AS}}\rrbracket_{\textrm{\tiny AS}}}{s_{12}}
    \bigg) \ \e^{\,\ii\,p_{123}\cdot x} \ , 
\end{split}
\end{align}
and so on.

Ordered tree-level amplitudes are given by the cyclic structure of the minimal model, taken on distinct perturbiner elements as in \cref{eq:treelevelscattering}:
\begin{align}\label{eq:ordered tree amplitude AS}
\CM_n^{\textrm {\tiny AS}}(1,\dots,n) = \langle\phi_1,\mu_{n-1}^{\circ\textrm{\tiny AS}}(\phi_2,\dots,\phi_n)\rangle_{\circ\textrm{\tiny AS}} \ .
\end{align}
The full amplitude $\CCA_n^{\textrm {\tiny AS}}(p,a)$ is given by the sum over all planar ordered permutations (cf.~\cref{sub:biadjoint}). 

For example, the full tree-level off-shell four-point amplitude is given by a sum of three terms
\begin{align}\label{eq:CAadstar}
\CCA^{\textrm {\tiny AS}}_4(p,a) =    \sum_{\sigma \in \mathrm{Sh}(1;3)} \, \CM^{\textrm {\tiny AS}}_4\big(1, \sigma(2), \sigma(3), \sigma(4)\big) =  \frac{c_s\,n_s}s + \frac{c_t\,n_t}t + \frac{c_u\,n_u}u \ ,
\end{align}
where the kinematical numerators are
\begin{align}
\begin{split}
    n_s &= g \,\int\,\dd^d q \ \bar F(p_1,p_2,q) \, \bar F(q,p_3,p_4) 
    = -4\,g\, \sin\big(\tfrac t2\,p_1\cdot\theta\, p_2\big) \sin\big(\tfrac t2\,p_3\cdot \theta\,p_4\big) \ , \\[4pt]
    n_t &= g\,\int\,\dd^d q \ \bar F(p_3,p_1,q) \, \bar F(q,p_2,p_4) 
    = -4\,g\, \sin\big(\tfrac t2\,p_3\cdot\theta\, p_1\big) \sin\big(\tfrac t2\,p_2\cdot \theta\,p_4\big) \ , \\[4pt]
    n_u &= g\,\int\,\dd^d q \ \bar F(p_2,p_3,q) \, \bar F(q,p_1,p_4) 
    = -4\,g\, \sin\big(\tfrac t2\,p_2\cdot\theta\, p_3\big) \sin\big(\tfrac t2\,p_1\cdot \theta\,p_4\big) \ ,
\end{split}
\end{align}
and we suppressed the overall delta-functions enforcing momentum conservation.
These satisfy the off-shell kinematic Jacobi identity $n_s+n_t+n_u=0$ for any deformation parameter $t$,  as a consequence of the Jacobi identity for the infinite-dimensional Lie algebra $\fru_\star(1)$. This mirrors the Jacobi identity $c_s+c_u+c_t=0$ satisfied by the colour numerators from \cref{eq:colournum}.
\paragraph{Colour-kinematics duality.} 

Using these perturbative computations we show colour-kinematics duality directly from the $L_\infty$-recursion relations.
We use the evaluation map $\mathrm{ev} : \mathrm{Sh}(i, n-1) \longrightarrow \CCW _i \otimes \CCW _{n-i}$ from \cref{eq:evaluation map}, into the tensor product of ordered words, to translate the recursion relations into the language of binary trees \cite{Bridges:2019siz}.

For the adjoint scalar theory, the Berends--Giele currents \eqref{eq:recursion for adjoint} and minimal model brackets \eqref{eq:recursion for adjoint2} can be written as a sum over deconcatenations of the ordered word $w=1\cdots n$ into non-empty ordered words $w_1$ and $w_2$:
\begin{equation}
\begin{split}
    J(w)
    &= \frac{g}{2} \,
    \sum_{w = w_1\sqcup w_2} \,
    G_{\rm F} \circ \llbracket J(w_1) , J(w_2)\rrbracket_{\textrm{\tiny AS}} \ , \\[4pt]
    \mu_n^{\circ\textrm{\tiny AS}}\big(\phi(1), \ldots, \phi(n)\big) 
    &= -\frac{g}{2} \, 
    \sum_{w =w_1\sqcup w_2}\,
    \sfp \circ
    \llbracket 
    J(w_1) , J( w_2)
    \rrbracket_{\textrm{\tiny AS}} \ 
    \e^{\,\ii\, p_{1 \cdots n}\cdot x}  \ .
\end{split}
\end{equation}
We have observed in \cref{eq:adell2fact} that the bracket $\llbracket-,-\rrbracket_{\textrm{\tiny AS}}$ of the adjoint scalar theory has a tensor product factorisation into the Lie bracket on the colour Lie algebra $\mathfrak{g}$ and the Lie bracket bracket on the kinematic Lie algebra $\mathfrak{u}_\star(1)$, as illustrated by the coefficient of the three-point vertex \eqref{eq:3pointvertexadjoint}.

Given a word $w=k_1\cdots k_n \in \CCW _n$ labelling $n$ external particles, we denote its length by $|w|=n$ and factorise the symmetric bracket into the tensor product of maps,
\begin{equation}
\begin{split}
   C: \CCW_n  \xrightarrow{\ \ell_{\rm c} \ } (\CCL _n, [-,-]_{\mathfrak{g}}) \xrightarrow{\ {\rm col}\ } \frg \quad , \quad
   K: \CCW_n  \xrightarrow{\ \ell_{\rm k} \ } (\CCL _n, [-\stackrel{\star}{,}-]_{\mathfrak{u}(1)}) \xrightarrow{\ {\rm kin}\ } \fru_\star(1) \ ,
\end{split}
\end{equation} 
as $\llbracket -,-\rrbracket_{\textrm{\tiny AS}}(w) = \big({\rm col}\circ\ell_{\rm c}(w)\big)\otimes \big({\rm kin}\circ\ell_{\rm k}(w)\big)$.
We have made explicit the left bracketing maps $\ell_{\rm c}$ and $\ell_{\rm k}$ from \cref{eq:left bracketing map}  for colour and kinematical numerators into the multilinear Lie polynomials with colour and kinematical Lie brackets, respectively.
In components we find
\begin{equation}
\begin{split}
    C(w)& = [ \cdots [[T^{a_{k_1}},T^{a_{k_2}}]_{\mathfrak{g}}, T^{a_{k_3}}]_{\mathfrak{g}},\ldots, T^{a_{k_n}}]_{\mathfrak{g}} \ , \\[4pt]
    K(w) &= [ \cdots [[\phi^{a_{k_1}}_{k_1} \stackrel{\star}{,} \phi^{a_{k_2}}_{k_2}]_{\mathfrak{u}(1)} \stackrel{\star}{,}  \phi^{a_{k_3}}_{k_3}]_{\mathfrak{u}(1)}\ds\ldots \stackrel{\star}{,}  \phi^{a_{k_n}}_{k_n}]_{\mathfrak{u}(1)} \ . 
\end{split}
\end{equation}

Finally, one arrives at 
\begin{equation}\label{eq:Jwrecursion}
    J(w) =  -\frac{1}{2}\,\sum_{w=w_1\sqcup w_2}\, \frac{(\,\ii\,g)^{|w|-1}}{s_w} \, C(w) \otimes K(w)\ ,
\end{equation}
with recursion relations for the colour and kinematic numerators given by
\begin{equation}
\begin{split}
    C(w) &=  {\rm col} \circ \sum_{w=w_1\sqcup w_2}\, [\ell_{\rm c}(w_1) , \ell_{\rm c}(w_2)]_{\mathfrak{g}} \ , \\[4pt]
    K(w) &=  {\rm kin} \circ \sum_{w=w_1\sqcup w_2}\, [\ell_{\rm k}(w_1) \ds \ell_{\rm k}(w_2)]_{\mathfrak{u}(1)} \ .
\end{split}
\end{equation}
One can replace the recursion \eqref{eq:Jwrecursion} in terms of the binary tree map from \cref{eq:binary tree map} by defining $[C\otimes K]\circ w := C(w)\otimes K(w)$ for any word $w$, which yields
\begin{align}\label{eq:BGadjointfact}
J(w) = -\tfrac12 \, (\,\ii\,g)^{|w|-1} \, [C\otimes K]\circ b(w) \ .
\end{align}

One obtains the tree-level amplitude with partial ordering $1w \in \CCW  _n$ by simply cancelling the overall pole in the current $J(w)$. 
Indeed, from \cref{eq:ordered tree amplitude AS}, cyclicity of the inner product on the minimal model implies that there is only one such component, namely
\begin{equation}
    \CM^{\textrm {\tiny AS}}_{n}(1,w) = - \mathrm{i} \,  s_{1w} \, J(w) \ (2 \pi)^d \, \delta(p_{1w}) \ ,
\end{equation}
where the overall momentum conservation comes from integrating over spacetime. 
Notice that the Moyal--Weyl product is trivialised by momentum conservation. 

The full $n$-point amplitude, obtained by summing over all ordered words, can then be written as
\begin{align}
\CCA^{\textrm {\tiny AS}}_{n}(p,a) = \sum_{\sfGamma\,\in\,\CCT_{3,n}}\,\frac{c_\sfGamma\,n_\sfGamma}{D_\sfGamma} \ ,
\end{align}
generalising \cref{eq:CAadstar}. Here $\CCT_{3,n}$ denotes the set of all trivalent trees with $n$ external edges; it has cardinality $(2n-5)!!$. Associated to any tree \smash{$\sfGamma\in\CCT_{3,n}$} there is a denominator $D_\sfGamma=\prod_{e\in \sfGamma}\, s_e$ with propagators $s_e$ assigned to each internal edge $e$ of $\sfGamma$. The colour factors $c_\sfGamma$ are contractions of the structure constants of the colour Lie algebra $\frg$ associated to $\sfGamma$, while $n_\sfGamma$ are the kinematic parts of the numerators involving analogous contractions of the structure constants of the kinematic Lie algebra $\fru_\star(1)$.

We conclude that, since the colour algebra based on $\frg$ naturally obeys the generalised Jacobi identities, or equivalently the STU-relations on any Jacobi subgraph, the kinematic numerators also obey them.
This is an instance of colour-kinematics duality: the infinite-dimensional kinematic Lie algebra $\fru_\star(1)$ which determines the kinematic numerators is dual to the colour Lie algebra $\frg$. Moreover, from the perspective of \cref{eq:BGadjointfact}, the effect of the zeroth copy construction is to replace the kinematic numerators $K(w)$ with a second set of colour numerators $\bar C(w)$. 

The colour-kinematics duality here is in fact implied by the factorisation of the three-point vertex \eqref{eq:3pointvertexadjoint} into $f^{abc}\,\bar F(k,p,q)$ and holds off-shell, as is evident from the homotopy algebraic perspective. At loop-level all integrands are computed using \cref{eq:3pointvertexadjoint}, and hence all loop-level kinematic Jacobi identities are automatically satisfied, even off-shell. Below we shall construct the action functional of the double copied theory using Maurer--Cartan theory. 

\subsection{Homotopy double copy}
\label{sub:DCad}
Our discussion of colour-kinematics duality in \S\ref{sub:adjointCKduality} justifies the use of the double copy prescription. The factorisation of the Berends--Giele currents in \cref{eq:BGadjointfact}, with respect to the (untwisted) tensor product $\frg\otimes\fru_\star(1)$ of Lie algebras, is a manifestation of the factorisation \eqref{eq:adfact} of the strict $L_\infty$-algebra $\AS$. This permits us to exploit the powerful techniques of $L_\infty$-algebras:  the homotopy double copy construction~\cite{Borsten:2021hua} replaces the colour factor $\frg$ in the factorization \eqref{eq:adfact} of $\AS$ with another copy of the ``twisted'' kinematic factor $\Kin_{\wedge^0}=\FR$, producing the cyclic strict $L_\infty$-algebra
\begin{align}\label{eq:DCdgLieadstar}
    \widehat{\AS}= \Kin_{\wedge^0}\otimes_{\tau^{\fru_{\bar\star}(1)}}(\Kin_{\wedge^0}\otimes_{\tau^{\fru_\star(1)}} \mathfrak{Scal})
\end{align}
as a twisted tensor product between the graded vector space $\Kin_{\wedge^0}$ (concentrated in degree~$0$) and the $C_\infty$-algebra \smash{$\frC_{\fru_\star(1)}=\Kin_{\wedge^0}\otimes_{\tau^{\fru_\star(1)}} \mathfrak{Scal}$}.

The underlying graded vector space is identified as $\FR\otimes(\FR\otimes L)\simeq L$, and the twisted differential is \smash{$\widehat{\mu}_1^{\textrm{\tiny AS}}=m_1^{\tau^{\fru_\star(1)}}=\square$}, so that again the cochain complex \smash{$\Ch(\widehat{\AS}$)} is the cochain complex $\Ch_{\rm Scal}$ from \cref{eq:scalarcochaincomplex}. The bracket of the double copy is given by the doubly twisted bracket
\begin{align}
\begin{split}
\widehat{\mu}_2^{\textrm{\tiny AS}}(\phi_1,\phi_2) &= \mu_2^{\rm Scal}\circ 2\,\ii\,\sin\big(\tfrac {\bar t}2\,\bar\theta^{\mu\nu} \, \partial_\mu \otimes \partial_\nu\big) \circ 2\,\ii\,\sin\big(\tfrac t2\,\theta^{\lambda\rho} \, \partial_\lambda \otimes \partial_\rho\big) (\phi_1\otimes\phi_2) \ ,
\end{split}
\end{align}
where we use two independent expansion parameters $t,\bar t\in\FR$ to keep track of the double Moyal--Weyl deformation, determined by star-products $\star,\bar\star$ which quantize two generally independent constant Poisson bivectors $\theta,\bar\theta$ on $\FR^{1,d-1}$.
Note that this is symmetric in $\phi_1$ and $\phi_2$, as required for a strict $L_\infty$-algebra bracket.
It may be written explicitly as a formal power series
\begin{align}
\begin{split}
    &\widehat{\mu}^{\textrm{\tiny AS}}_2(\phi_1,\phi_2) \\[4pt]
    & \hspace{1cm} = \ii\, \kappa\,\sum_{n=0}^\infty\, \frac{(-1)^n\,\bar t^{\,2n+1}}{(2n+1)!} \, \bar\theta^{\mu_1\nu_1}\cdots \bar\theta^{\mu_{2n+1}\nu_{2n+1}} \, [\partial_{\mu_1}\cdots\partial_{\mu_{2n+1}}\phi_1 \ds \partial_{\nu_1}\cdots\partial_{\nu_{2n+1}}\phi_2]_{\fru(1)} \ ,
\end{split}
\end{align}
where we denote the coupling as $g=\frac\kappa2$ in the double copied theory, as it acquires different engineering dimension.

The Maurer--Cartan equation associated to $\widehat{\AS}$ yields the field equation which governs the dynamics of the double copy as
\begin{align} \label{eq:DCMCeqn}
\square\,\phi + \frac{\ii \, \kappa}2\,\sum_{n=0}^\infty\, \frac{(-1)^n\,\bar t^{\,2n+1}}{(2n+1)!} \, \bar\theta^{\mu_1\nu_1}\cdots\bar\theta^{\mu_{2n+1}\nu_{2n+1}} \, [\partial_{\mu_1}\cdots\partial_{\mu_{2n+1}}\phi \ds \partial_{\nu_1}\cdots\partial_{\nu_{2n+1}}\phi]_{\fru(1)} = 0 \ .
\end{align}
It can be derived as the stationary locus of the Maurer--Cartan functional, which using the cyclic inner product of $\mathfrak{Scal}$ from \cref{eq:ncscalarpairing} reads
\begin{align}\label{eq:DCMCaction}
\begin{split}
    \widehat{S}_{\textrm{\tiny AS}}[\phi] 
    = \int\,\dd^dx \ &  \frac12\,\phi\star \square\,\phi - \frac{\ii\,\kappa} 6\, \sum_{n=0}^\infty\, \frac{(-1)^n\,\bar t^{\,2n+1}}{(2n+1)!} \, \bar\theta^{\mu_1\nu_1} \cdots \bar\theta^{\mu_{2n+1}\nu_{2n+1}} \\
& \hspace{5cm} \times  \phi\, [\partial_{\mu_1}\cdots\partial_{\mu_{2n+1}}\phi \ds \partial_{\nu_1}\cdots\partial_{\nu_{2n+1}}\phi]_{\fru(1)} \ .
\end{split}
\end{align}
The three-point Feynman vertex following from \eqref{eq:DCMCaction} is
\begin{equation} \label{eq:ASdoublevertex}
\begin{split}
\begin{tikzpicture}[scale=0.7]
{\small
    \draw[fermionnoarrow] (-140:1.8)--(0,0);
    \draw[fermionnoarrow] (140:1.8)--(0,0);
    \draw[fermionnoarrow] (0:1.8) --(0,0);
    \draw[->, shift={(-0.2,0.45)}] (140:.5) node[above]{$p_1$} -- (0,0);
    \draw[->, shift={(-0.2,-0.45)}] (-140:.5) node[below]{$p_2$} -- (0,0);
    \draw[->, shift={(0.2,0.2)}] (0:.5) node[above]{$p_3$} -- (0,0);
    }\normalsize
    \node at (6.3, 0) {$\displaystyle = \ -\frac{\ii\,\kappa}2\,F(p_1,p_2,p_3) \, \bar F(p_1,p_2,p_3) \ ,$};
\end{tikzpicture}
\end{split}
\end{equation}
where the kinematic structure constants are given by \cref{eq:starcommrels}.

\remark{
This scalar theory is a double noncommutative deformation of the cubic scalar field theory of \S\ref{sub:nc_strict_scalar_field_theory}. It can be obtained from the adjoint scalar theory by performing substitutions analogous to \cref{eq:adprescr} for the remaining colour factor, or equivalently by the substitutions
\begin{align}\label{eq:DCprescr}
    \phi^{a\bar a}\longrightarrow \phi \ , \quad f^{abc}\, \bar f^{\bar a\bar b\bar c} \longrightarrow F(k,p,q) \, \bar F(k,p,q)  \qquad \mbox{and} \qquad \Tr_{\frg\otimes\bar\frg} \longrightarrow \textrm{\small$\int$}
\end{align} 
directly in the Lagrangian of the commutative biadjoint scalar theory of \S\ref{sub:biadjoint}. 
Hence the double copy of the adjoint scalar theory can in this sense be regarded as a biadjoint scalar theory based on the twisted tensor product $\fru(1)\otimes_{\tau^{\fru_\star(1)}}\fru_{\bar\star}(1)$ of  Lie algebras. 
}

\paragraph{Perturbative calculations.} 
\label{par:perturbative_calculations}
To get the Berends--Giele currents and $n$-point amplitudes of the double copy, one simply needs to perform calculations analogous to those presented for the adjoint scalar in \S\ref{sub:adjointCKduality}.
The double copied homotopy algebra is a strict $L_\infty$-algebra with bracket $\widehat{\mu}^{\textrm{\tiny AS}}_2$ and contracting homotopy $\widehat{\sfh} = \widehat{G}_{\rm F}$ satisfying
\begin{equation}
    (\widehat{\mu}^{\textrm{\tiny AS}}_1 \circ \widehat{G}_{\rm F}) (x,y) = \ii\, \delta(x-y) \qquad \mbox{with} \quad
    \widehat{G}_{\rm F}( \e^{\,\ii\,p\cdot x}) = -\frac{\ii}{p^2} \, \e^{\,\ii\,p\cdot x}\ .
\end{equation}
This is just the scalar contracting homotopy acting on the double copied fields, which are scalars.

For example, the tree-level off-shell four-point amplitude is found in this way to be
\begin{align}
\widehat{\CCA}^{ \, \textrm {\tiny AS}}_4(p) = \sum_{\sigma \in \mathrm{Sh}(1;3)} \, \widehat{\CM}^{\textrm {\tiny AS}}_4\big(1, \sigma(2), \sigma(3), \sigma(4)\big) = \frac{n_s \, \bar n_s}s + \frac{n_t \, \bar n_t}t + \frac{n_u \, \bar n_u}u \ ,
\end{align}
as expected from the replacement rule \eqref{eq:DCprescr}. More generally, the $n$-point amplitudes follow by replacing the colour factors $C(w)$ with a second set of kinematic numerators $\bar K(w)$ in \cref{eq:BGadjointfact} and can be written in the form
\begin{align}
\widehat{\CCA}^{ \, \textrm {\tiny AS}}_n(p) = \sum_{\sfGamma\,\in\,\CCT_{3,n}} \, \frac{n_\sfGamma\,\bar n_\sfGamma}{D_\sfGamma} \ .
\end{align}

\subsection{Binoncommutative biadjoint scalar theory} 
\label{sub:the_other_single_copy}

The homotopy double copy construction of \S\ref{sub:DCad} inspires a more general class of noncommutative deformations of the biadjoint scalar theory from \S\ref{sub:biadjoint}. They are obtained  from the noncommutative biadjoint scalar theory of \cref{sub:NCbiadjoint} by replacing the tensor product $\frg\otimes\fru_{\bar\star}(\bar N)$ with the twisted tensor product \smash{$\fru(N)\otimes_{\tau^{\fru_\star(N)}} \fru_{\bar\star}(\bar N)$}. The double copy theory of \S\ref{sub:DCad} is then recovered in the rank one limits $N=\bar N=1$. This `binoncommutative' biadjoint scalar theory will turn out to be the zeroth copy theory for noncommutative gauge theories. In particular, it will play an important role when we study the KLT relations in association with noncommutative Yang--Mills theory in \S\ref{sec:klt_with_b_field}.

The binoncommutative biadjoint scalar theory is organised by a cyclic strict $L_\infty$-algebra $\BAS_{\star\bar\star}$ which is given by the twisted tensor product construction of  \cref{sub:factNCbiadjoint}, now applied to both Lie algebra factors:
\begin{equation} \label{eq:BASdoublestar}
    \mathfrak{BAS}_{\star\bar\star} := \mathfrak{u}(N) \otimes_{\tau^{\mathfrak{u}_\star(N)}} \big( \mathfrak{u}( \bar N) \otimes_{\tau^{\mathfrak{u}_{\bar\star}(\bar N)}} \mathfrak{Scal} \big) \ .
\end{equation}
The underlying cochain complex $\Ch(\mathfrak{BAS}_{\star\bar\star})$ is given by \cref{eq:nc biadjoint cochain complex} with $\frg=\fru(N)$ and $\bar\frg=\fru(\bar N)$, which as usual identifies the differential as $\mu_1^{\star\bar\star\textrm{\tiny BAS}} =  \mu_1^{\textrm{\tiny BAS}} = \square$. The higher bracket $\mu_2^{\star\bar\star\textrm{\tiny BAS}}$ is the binoncommutative product
\begin{equation}\label{eq:2bracketBASdouble}
\begin{split}
\mu_2^{\star\bar\star\textrm{\tiny BAS}}(\phi_1,\phi_2) &=
    T^{c} \otimes \bar T^{\bar c} \otimes
    \mu_2^{\rm Scal}\circ \Big( f^{ a b c} \cos\big(\tfrac {t}2\,\theta^{\mu\nu} \, \partial_\mu \otimes \partial_\nu\big) -
    d^{ a  bc}\sin\big(\tfrac {t}2\,\theta^{\mu\nu} \, \partial_\mu \otimes \partial_\nu\big) 
    \Big)\\
    & \hspace{2cm} \circ \Big(  \bar f^{ \bar a \bar b \bar c} \cos\big(\tfrac {\bar t}2\,\bar\theta^{\lambda\rho} \, \partial_\lambda \otimes \partial_\rho\big)
    - \, \bar d^{ \bar a \bar b \bar c}\sin\big(\tfrac {\bar t}2\,\bar\theta^{\lambda\rho} \, \partial_\lambda \otimes \partial_\rho\big)
    \Big)
    \big(\phi_1^{a \bar a}\otimes\phi_2^{b \bar b}\big) 
\end{split}
\end{equation}
of elements $\phi_1=
T^{ a} \otimes \bar T^{\bar a} \otimes\phi_1^{ a\bar a}$ and $\phi_2=T^{b} \otimes \bar T^{\bar b} \otimes \phi_2^{b\bar  b}$ in $\Omega^0\big(\FR^{1,d-1},\fru(N)\otimes\fru(\bar N)\big)[-1]$. The cyclic inner product is again identified as the pairing \eqref{eq:bipairing} of fields $\phi \in \Omega^0\big(\mathbbm{R}^{1,d-1}, \mathfrak{u}(N) \otimes \mathfrak{u}(\bar N)\big)[-1]$ with antifields $\phi^+ \in \Omega^0\big(\mathbbm{R}^{1,d-1}, \mathfrak{u}( N) \otimes \mathfrak{u}(\bar N)\big)[-2]$.

One can expand the bracket \eqref{eq:2bracketBASdouble} analogously to what we did in \S\ref{sub:DCad}, and formally develop the Maurer--Cartan theory underlying the cyclic $L_\infty$-algebra $\mathfrak{BAS}_{\star\bar\star}$. Similarly to the noncommutative biadjoint scalar theory of \S\ref{sub:NCbiadjoint}, this theory has a rigid symmetry under the adjoint action of $\fru(N)\oplus\fru(\bar N)$ and can be extended to a four-parameter family of binoncommutative biadjoint scalar theories with structure constants $\lambda^{abc}_{u,v}$ and $\bar\lambda^{\bar a\bar b\bar c}_{\bar u,\bar v}$ (cf. Remark~\ref{rem:NCbiadjointfamily}).
The resulting formulas are even more complicated than those of \S\ref{sub:DCad}, and they will not be needed explicitly in this paper. 

In momentum space, this theory has a remarkably much simpler representation, giving the three-point vertex in its Feynman diagram expansion in terms of two copies of the structure constants from \cref{eq:u(n) star structure constant} as
\begin{equation}\label{eq:vertex double deformed}
\begin{split}
\begin{tikzpicture}[scale=0.8]
{\small
    \draw[fermionnoarrow] (140:1.8)node[above left] {$a, \bar a$}--(0,0);
    \draw[fermionnoarrow] (-140:1.8)node[below left] {$b, \bar b$}--(0,0);
    \draw[fermionnoarrow] (0:1.8)node[below left] {$c, \bar c$} --(0,0);
    \draw[->, shift={(-0.2,0.45)}] (140:.5) node[above]{$p_1$} -- (0,0);
    \draw[->, shift={(-0.2,-0.45)}] (-140:.5) node[below]{$p_2$} -- (0,0);
    \draw[->, shift={(0.2,0.2)}] (0:.5) node[above]{$p_3$} -- (0,0);
    }\normalsize
    \node at (6.3, 0) {$ = \ -\ii\,g\, F^{abc}(p_1,p_2,p_3) \, \bar F^{\bar a \bar b \bar c}(p_1,p_2,p_3)\ . $};
\end{tikzpicture}
\end{split}
\end{equation}
Note that the momentum-dependent structure constants $F^{abc}$ and $\bar F^{\bar a\bar b\bar c}$ obey the Jacobi identities of the infinite-dimensional Lie algebras $\fru_\star(N)$ and $\fru_{\bar\star}(\bar N)$, respectively, which from this perspective can therefore also be regarded as kinematic Lie algebras. Looking at the twisted homotopy factorization \eqref{eq:NCbiadfact}, this means that the binoncommutative biadjoint scalar theory can be regarded as a double copy between the noncommutative biadjoint scalar theory and itself, with the replacement of colour by kinematics involving twisted tensor products.

\paragraph{Colour ordering and decomposition.} 

The tree-level amplitudes of the binoncommutative biadjoint scalar theory admit a colour decomposition on both copies of the colour algebra.
By writing the structure constants of $\fru(N)$ as $f^{abc} = 2 \,\mathrm{Tr}_{\fru(N)}([T^a, T^b]_{\fru(N)}\, T^c)$ and $\ii\, d^{abc} =  2\, \mathrm{Tr}_{\fru(N)}(\{ T^a, T^b \}_{\fru(N)}\, T^c)$, the structure constants of $\fru_\star(N)$ given in \cref{eq:u(n) star structure constant} can be written as
\begin{equation}
\begin{split}\label{eq:colour ordering structure}
    F^{abc}(p_1, p_2, p_3) &= 2 \,\big(\mathrm{Tr}_{\fru(N)}(T^a\, T^b\, T^c) \ \e^{-\frac{\ii\, t}{2}\, p_1 \cdot \theta\, p_2}
    -  \mathrm{Tr}_{\fru(N)}(T^b\, T^a\, T^c) \ \e^{- \frac{\ii\, t}{2}\,p_2 \cdot \theta\, p_1}\big) \\
& \hspace{9cm} \times \delta(p_1+p_2+p_3) \ .
\end{split}
\end{equation}
This is an explicit factorisation for a given colour ordering, a common feature of noncommutative field theories with Moyal--Weyl deformation. In the commutative limit $t\to0$, this returns $f^{abc}$ as expected.

In the binoncommutative biadjoint scalar theory, the same factorisation appears as a decomposition in terms of two sets of orderings of colour indices as
\begin{equation}
\begin{split}\label{eq:double decomp}
  &   F^{abc}(p_1, p_2, p_3)\,\bar F^{\bar a \bar b \bar c}(p_1, p_2, p_3) \\[4pt]
  & \hspace{1cm} = 4\,\big(
     \mathrm{Tr}_{\fru(N)}(T^a\, T^b\, T^c) \ \e^{- \frac{\ii\, t}{2}\, p_1 \cdot \theta\, p_2} - 
     \mathrm{Tr}_{\fru(N)}(T^b\, T^a\, T^c) \ \e^{- \frac{\ii\, t}{2}\, p_2 \cdot \theta\, p_1}
    \big) \\
 &\hspace{3cm} \times  \big(
   \mathrm{Tr}_{\fru(\bar N)}(\bar T^{\bar a}\, \bar T^{\bar b}\,\bar T^{\bar c}) \ \e^{- \frac{\ii\, \bar t}{2}\, p_1 \cdot \bar\theta\, p_2}  - 
     \mathrm{Tr}_{\fru(\bar N)}(\bar T^{\bar b}\,\bar T^{\bar a}\,\bar T^{\bar c}) \ \e^{- \frac{\ii\, \bar t}{2}\, p_2 \cdot \bar\theta\, p_1} \big) \\
 & \hspace{11cm} \times \delta(p_1+p_2+p_3) \ .
\end{split}
\end{equation}
The commutative limit $t, \bar t \rightarrow 0$ similarly reduces to $f^{abc}\,\bar f^{\bar a \bar b \bar c} $.
The rank one theory studied in \cref{sub:DCad} also follows from this formula, which for $N=\bar N=1$ recovers only the sinus part of \cref{eq:double decomp}.

The binoncommutative biadjoint scalar theory studied here differs from the \textit{commutative} biadjoint scalar theory of \S\ref{sub:biadjoint} only in the three-point vertex. Crucially, this vertex admits a colour factorisation with only phase factors that depend solely on the orderings of momenta.
We conclude that the $n$-point partial amplitudes of the binoncommutative biadjoint scalar theory, defined as the summands in the expansion of the full amplitudes
\begin{equation}\label{eq:partial amplitude def deformed biad}
\CCA^{\star\bar\star\textrm{\tiny BAS}}_n(p,a,\bar a) = \sum_{\sigma,\sigma'\in S_n/\RZ_n\rtimes\RZ_2} \,  C(\sigma) \, \bar C(\sigma') \ \CA^{\star\bar\star\textrm{\tiny BAS}}_n\big( \sigma(1), \dots, \sigma(n)\big|\sigma'(1), \dots, \sigma'(n)\big) \ ,
\end{equation}
factor through these phases.

For example, the first term in the three-point partial amplitude given by the diagram \eqref{eq:vertex double deformed} and the decomposition \eqref{eq:double decomp} is
\begin{equation}
\begin{split}
    \CA_3^{\star\bar\star\textrm{\tiny BAS}}(1,2,3|1,2,3) &= - 4\,\ii\, g \ \e^{- \frac{\ii\, t}{2}\,p_1 \cdot \theta\, p_2 -\frac{\ii\, \bar t}{2}\, p_1 \cdot \bar\theta\, p_2} \ \delta(p_1 + p_2 + p_3)\\[4pt]
    &= \e^{- \frac{\ii\, t}{2}\,p_1 \cdot \theta\, p_2 -\frac{\ii\, \bar t}{2}\, p_1 \cdot \bar\theta \,p_2} \ \CA^{\textrm{\tiny BAS}}_3(1,2,3|1,2,3) \ ,
\end{split}
\end{equation}
where $\CA^{\textrm{\tiny BAS}}_n$ is the commutative $n$-point biadjoint scalar partial amplitude, defined in \cref{sub:biadjoint}.

We interpret this factorisation as a slight generalisation of the well-known fact that noncommutative scalar field theory differs from its commutative counterpart  only by a phase in planar graphs~\cite{Filk:1996dm,Szabo:2001kg}. The upshot is that one can dress commutative colour-stripped biadjoint scalar amplitudes by the corresponding phase factors to get tree-level binoncommutative biadjoint scalar amplitudes.
This fact will be of use in~\S\ref{sec:klt_with_b_field}.

\subsection{Applications: Special galileons and self-dual gravity}
\label{sub:topgravity}

As some concrete applications of the formalism we have developed thus far, as well as a glimpse towards some of our later double copy constructions, let us now look at two special ``topological'' realisations of the generic double copy map from the adjoint scalar theory of \cref{sub:adjoint_scalar_theory_as_a_strict_linf} to the theory of \cref{sub:DCad}. Here we work mostly at the level of equations of motion for brevity, but analogous statements also hold off-shell at the level of Maurer--Cartan functionals.

\paragraph{Zakharov--Mikhailov theory and special galileons.}

In $d=2$ dimensions with the Moyal--Weyl star-product \eqref{eq:starproduct}, the  adjoint scalar theory was considered in~\cite{Cheung:2022mix} as a `stringy deformation' of Zakharov--Mikhailov theory~\cite{ZM1978}, which is classically equivalent to the principal chiral model. We set $t=\bar t=\alpha'$, regarded as a string tension parametrizing an infinite tower of higher dimension operators, and take $\theta^{\mu\nu}=\bar\theta^{\mu\nu}=\epsilon^{\mu\nu}$ to be the Levi--Civita symbol in two dimensions with $\epsilon^{01}=1$. Let $g=\frac{\bar g}{\alpha'}$, and take the semi-classical limit $\alpha'\to0$ with $\bar g$ finite. 

Recall from \cref{sub:MWdeformation} that in the semi-classical limit the star-commutator $\ii\,g\,[\phi_1\ds\phi_2]_{\fru(1)}$ is replaced by the Poisson bracket $\bar g\,\{\phi_1,\phi_2\}$, where
    \begin{align}
        \{\phi_1,\phi_2\} = \epsilon^{\mu\nu} \, \partial_\mu\phi_1 \, \partial_\nu\phi_2 \ .
    \end{align}
The kinematic algebra $\fru_\star(1)$ reduces to the Poisson--Lie algebra $\mathfrak{sdiff}(\FR^{1,1})$ of area-preserving diffeomorphisms of $\FR^{1,1}$, and the kinematical numerators $n_s$ become $-4\,g\,(p_1\times p_2)\,(p_3\times p_4)$, and so on, where $p_i\times p_j := \epsilon^{\mu\nu}\,p_i{}_\mu\,p_j{}_\nu$.

The equations of motion \eqref{eq:adMCeq} of the adjoint scalar theory reduce to
\begin{align}\label{eq:ZMeom}
\square \, \phi-\tfrac{\bar g}2\,\epsilon^{\mu\nu}\,[\partial_\mu\phi,\partial_\nu\phi]_\frg = 0 \ .
\end{align}
These are just the equations of motion for the non-linear sigma-model in two dimensions, with fields $h\in\Omega^0(\FR^{1,1},\sG)$, which read as $\partial^\mu j_\mu=0$ for the left-invariant currents $j= h^{-1}\,\dd h\in\Omega^1(\FR^{1,1},\frg)$. These imply that there exists an adjoint scalar $\phi\in\Omega^0(\FR^{1,1},\frg)$ such that $j_\mu = \bar g\, \epsilon_{\mu\nu}\,\partial^\nu\phi$, and the Maurer--Cartan equation $\dd j+\frac12\,[j,j]_\frg=0$ coincides with the equations of motion \eqref{eq:ZMeom}. Thus in this limit the adjoint scalar field theory for $d=2$ reduces to the usual Zakharov--Mikhailov theory~\cite{ZM1978}.

On the other hand, the equation of motion \eqref{eq:DCMCeqn} of the homotopy double copy theory in the limit $\alpha'\to0$ becomes
    \begin{align}\label{eq:SGeom}
    \square\,\phi - \tfrac{\bar\kappa}4\,\epsilon^{\mu\nu}\,\{\partial_\mu\phi , \partial_\nu\phi\} = 0 \ ,
    \end{align}
which is the equation of motion for the special galileon theory in two dimensions~\cite{CKNT2014,CKNST2015,Hinterbichler2015}.
The special galileon theory is invariant under the Galilean-type transformations $\phi(x)\longmapsto\phi(x)+b\cdot x+c$ of the scalar field $\phi\in\Omega^0(\FR^{1,1})$ for $b\in\FR^{1,1}$ and $c\in\FR$, and it is quasi-isomorphic to a two-dimensional free theory.
Hence the double copy of the adjoint scalar theory can be regarded as a `stringy deformation' of the special galileon theory in two dimensions. 

One application of the semi-classical double copy construction is to integrability. Since the Zakharov--Mikhailov theory is classically equivalent to the principal chiral model, it is likewise integrable. Applying the colour-kinematics duality map, the special galileon theory is also integrable; see~\cite{Cheung:2022mix} for the explicit map of the Lax connection which furnishes an infinite tower of conserved currents. However, integrability does not seem to persist generally beyond the semi-classical limits in the full noncommutative theories, as we discuss below.

\paragraph{Self-dual Yang--Mills theory and gravity.}

Self-dual Yang--Mills theory and gravity provide a four-dimensional realization of our homotopy double copy construction: for $d=2$ the adjoint scalar theory was considered in~\cite{Chacon:2020fmr} as a noncommutative deformation of self-dual Yang--Mills theory, while \cref{eq:DCMCeqn} is the equation of motion for doubly deformed self-dual gravity considered in~\cite{Chacon:2020fmr}. Below we discuss this realisation in a bit more detail. In the semi-classical limit, the explicit relationship of these $d=4$ theories and their double copy duality to the $d=2$ theories discussed above is elucidated by~\cite{Armstrong-Williams:2022apo}.

Let $A\in\Omega^1(\FR^{1,3},\frg)$ be a gauge field with curvature $F=\dd A+\frac{\bar g}2\,[A,A]_\frg$. We decompose four-dimensional Minkowski space as $\FR^{1,3}\simeq\FR^{1,1}\times\FR^2$, with light-cone coordinates $(x^+,x^-)$ on $\FR^{1,1}$ and complex coordinates $(z,\bar z)$ on $\FR^2\simeq\FC$. Let $\ast_{\textrm{\tiny H}}$ be the Hodge duality operator on $\FR^{1,3}$, which acts on two-forms with $\ast_{\textrm{\tiny H}}^2=-\ident$. Then the component form of the self-duality equation $F = \ii\ast_{\textrm{\tiny H}} F$ can be written as
\begin{align}
F_{-z} = F_{+\bar z} = 0 \qquad \mbox{and} \qquad F_{-+} = F_{z\bar z} \ .
\end{align}

From $F_{-z}=0$ we may choose the light-cone gauge $A_-=A_z=0$. From $F_{-+} = F_{z\bar z}$ it follows that there exists an adjoint scalar $\phi\in\Omega^0(\FR^{1,3},\frg)$ such that $A_+ = -\partial_z\phi$ and $A_{\bar z}=-\partial_-\phi$; the scalar field $\phi$ represents the single polarization state remaining in a gluon after projection to the self-dual sector. The remaining equation $F_{+\bar z}=0$ then yields
\begin{align}
\square \, \phi -\bar g \, [\partial_z\phi,\partial_-\phi]_{\frg} = 0 \ ,
\end{align}
where $\square = -\partial_+\,\partial_- + \partial_z\,\partial_{\bar z}$, which is just the equation of motion \eqref{eq:ZMeom} with $\mu,\nu\in\{z,-\}$. In this case the semi-classical kinematic algebra generates area-preserving diffeomorphisms of the null $(z,x^-)$-plane, which is the known kinematic Lie algebra of self-dual Yang--Mills theory~\cite{Monteiro2011}. The corresponding semi-classical limit of the Maurer--Cartan functional \eqref{eq:adMCaction} for the adjoint scalar theory is the cubic action functional for self-dual Yang--Mills theory in the Leznov gauge~\cite{Leznov:1986mx,Park:1989vq,Parkes1992}.

The analogue in asymptotically flat gravity is the self-duality equation for the Riemann curvature tensor, which encodes both the equations of motion and the algebraic Bianchi identity. With the light-cone gauge choice for the metric, the semi-classical double copy equation of motion \eqref{eq:SGeom} coincides with Pleba\'nski's second heavenly equation for self-dual gravity~\cite{Plebanski:1975wn}:
\begin{align}
\square\,\phi - \tfrac{\bar\kappa}2\,\{\partial_z\phi,\partial_-\phi\} = 0 \ ,
\end{align}
where here the scalar field $\phi\in\Omega^0(\FR^{1,3})$ represents the positive helicity state of the graviton, and the constant $\bar\kappa$ controls the deformation away from flat space. This involves two copies of the area-preserving diffeomorphism algebra. The corresponding semi-classical limit of the Maurer--Cartan functional \eqref{eq:DCMCaction} is the action functional for the string field theory of $\CN=2$ strings~\cite{Ooguri:1991fp}.

Like the Zakharov--Mikhailov and special galileon theories, self-dual Yang--Mills theory and gravity are integrable theories, each admitting an infinite tower of conserved charges: On the gauge theory side there is a Lax pair $(\CL,\CM)$, which with our gauge choice is given by
\begin{align}
\CL=\partial_+-\lambda\,(\partial_{\bar z}-\partial_+\phi) \qquad \mbox{and} \qquad \CM=\partial_z-\lambda\,(\partial_--\partial_z\phi) 
\end{align}
for a spectral parameter $\lambda\in\FC\mathbbm{P}^1$, and analogously on the gravity side where integrability is linked to its infinite-dimensional $w_{1+\infty}$ symmetry. 
This is consistent with Ward's conjecture: all integrable theories are related to self-dual Yang--Mills theory by replacing its structure constants $f^{abc}$ by other structure constants, which can be thought of as a ``symmetry reduction''. Here we observe the natural sequence of symmetry reductions similarly to \cref{eq:symred}, starting from the biadjoint scalar theory of \cref{sub:biadjoint}, to self-dual Yang--Mills theory, and finally to self-dual gravity. 

Let us now look at the generic case of finite deformation parameters $t\neq\bar t$, with the understanding as above that spacetime indices are always restricted to directions along the null $(z,x^-)$-plane. Then the Maurer--Cartan equation \eqref{eq:DCMCeqn} for the homotopy double copy theory coincides with the ``doubly-deformed'' Pleba\'nski equation for self-dual gravity considered in~\cite{Chacon:2020fmr}. This theory of noncommutative gravity is not integrable, due to the breakdown of the Jacobi identities. In the semi-classical limit $\bar t\to0$ it reduces to the deformed Pleba\'nski equation of noncommutative gravity~\cite{Strachan:1992em,Takasaki:1992jf,Plebanski:1995gk}
\begin{align} \label{eq:NCgrav}
\square\,\phi - \tfrac{\ii\,\bar\kappa}2 \, [\partial_z\phi\ds\partial_-\phi]_{\fru(1)} = 0 \ .
\end{align}
This is now integrable with an infinite tower of conserved currents; integrability here relies crucially on the semi-classical form of the kinematic structure constants $\bar F_0(p_1,p_2,p_3)$ of the symplecton $W_{1+\infty}$ algebra~\cite{Chacon:2020fmr}. The Moyal--Weyl deformation of self-dual gravity is one-loop exact and has also appeared in recent parallel discussions of celestial holography~\cite{Monteiro:2022lwm,Bu:2022iak,Guevara:2022qnm, Monteiro:2022xwq,Bittleston:2023bzp}; in~\cite{Shyam:2022iwd} it is interpreted as the $T\overline{T}$-deformation of self-dual gravity.

Consider now what happens beyond the semi-classical limit of the adjoint scalar theory. The self-dual Yang--Mills equations are then deformed to
\begin{align} \label{eq:deformedSDYMeqs}
\square\,\phi+\tfrac{g}2\,\llbracket\phi,\phi\rrbracket_{\textrm{\tiny AS}} = 0 \ ,
\end{align}
where the bracket $\llbracket-,-\rrbracket_{\textrm{\tiny AS}} $ is defined in \cref{sub:adjoint_scalar_theory_as_a_strict_linf}.
This is obtained as one of two single copies of deformed self-dual gravity in the semi-classical limit, as $\bar t\to0$ alone, of the $L_\infty$-algebra \eqref{eq:DCdgLieadstar}: replacing the inner kinematical vector space $\Kin_{\wedge^0}$ with the colour algebra $\frg$ gives the standard self-dual Yang--Mills theory above, which is integrable, whereas replacing the outer factor of $\Kin_{\wedge^0}$ with $\frg$ leads to the theory with deformed equations of motion \eqref{eq:deformedSDYMeqs}, which is not integrable. The main technical issue is that the bracket operation $\llbracket-,-\rrbracket_{\textrm{\tiny AS}} $ is not a Lie bracket, as it violates the (ungraded) Jacobi identities. 

As we have shown in \cref{sub:adjointCKduality}, \cref{eq:deformedSDYMeqs} are the natural gauge field equations that arise in the homotopy double copy prescription and which are consistent with colour-kinematics duality. However they differ from the field equations of the usual noncommutative self-dual Yang--Mills theory, which has a non-trivial commutative limit to ordinary self-dual Yang--Mills theory and is defined using the star-commutator bracket:
\begin{align} \label{eq:NCinst}
\square\,\phi-{\ii\,g}\,[\partial_z\phi\ds\partial_-\phi]_{\fru(N)} = 0 \ .
\end{align}
These are the Euler--Lagrange equations which follow from varying the noncommutative cubic action functional~\cite{Lechtenfeld:2000nm}
\begin{align} \label{eq:SDYMNC}
S_{\text{\tiny YM+}}^\star[\phi] = \int\,\dd^4x \ \Tr_{\fru(N)}\Big(\frac12\,\phi\star\square\,\phi -\frac{\ii\,g}{3!} \, \epsilon^{\mu\nu} \, \phi\star[\partial_\mu\phi \ds \partial_\nu\phi]_{\fru(N)} \Big)
\end{align}
for the Leznov prepotential $\phi\in\Omega^0(\FR^{1,3},\fru(N))$. 

Noncommutative $\sU(N)$ self-dual Yang--Mills theory with action functional~\eqref{eq:SDYMNC} is the semi-classical limit $\bar t\to0$ of the rank $\bar N=1$ binoncommutative biadjoint scalar theory of \cref{sub:the_other_single_copy} in $d=4$ dimensions.
As we discussed in \cref{sub:MWdeformation}, the star-commutator bracket is not compatible with standard colour-kinematics duality, due to the appearance of anticommutators of Lie algebra generators in \cref{eq:factlack} which obstructs an immediate (untwisted) homotopy factorisation of the $L_\infty$-structure. On the other hand, this theory is integrable~\cite{Takasaki:2000vs,Hamanaka:2006re}.

The noncommutative instanton equation \eqref{eq:NCinst} in the rank one limit $N=1$ coincides with the noncommutative Pleba\'nski equation \eqref{eq:NCgrav}, while the action functional \eqref{eq:SDYMNC} for $N=1$ is the semi-classical limit of the Maurer--Cartan functional \eqref{eq:DCMCaction} as $\bar t\to0$. 
It follows that self-dual $\fru_\star(1)$ Yang--Mills theory is the double copy of the adjoint scalar theory with ordinary self-dual Yang--Mills theory, for any gauge algebra $\frg$. This double copy theory is precisely noncommutative self-dual gravity. In particular, the integrability of gravity is ``inherited'' from the corresponding gauge theories via the double copy and is guaranteed by integrability of at least one of the two gauge theory single copies~\cite{Chacon:2020fmr,Monteiro:2022lwm}. 
In what follows we shall assert that such double copy interpretations of rank one noncommutative gauge theories hold in general.

\section{Noncommutative Chern--Simons theory} 
\label{sec:cs}

\subsection{Moyal--Weyl deformation of differential forms}
\label{sub:MWdeformationforms}

To study general noncommutative gauge theories, we first discuss the extension of the Moyal--Weyl deformation of scalar fields from \S\ref{sub:MWdeformation} to include differential forms of arbitrary degree. Let $\Omega^\bullet(\FR^{1,d-1},\fru(N))$ be the exterior algebra of differential forms on $\FR^{1,d-1}$ valued in the matrix Lie algebra $\frg=\fru(N)$. This is a strict $A_\infty$-algebra with the de~Rham differential $\dd$ and the composition of matrix multiplication with the exterior product of forms, which as usual we make into a strict $L_\infty$-algebra under the commutator bracket:
\begin{align}\label{eq:commutatorforms}
[\alpha,\beta]_{\fru(N)} = \alpha\wedge\beta - (-1)^{|\alpha|\,|\beta|} \, \beta\wedge\alpha \ ,
\end{align}
where $|\alpha|$ denotes the degree of a homogeneous form $\alpha\in\Omega^\bullet(\FR^{1,d-1},\fru(N))$.

As a vector space
\begin{align}\label{eq:vectorspaceiso}
\Omega^\bullet(\FR^{1,d-1},\fru(N)) \simeq \fru(N)\otimes\Omega^\bullet(\FR^{1,d-1}) \ ,
\end{align}
and this is also true at the level of homotopy algebras: the strict $L_\infty$-algebra $\Omega^\bullet(\FR^{1,d-1},\fru(N))$ factorizes into the tensor product of the Lie algebra $\fru(N)$ with the strict $C_\infty$-algebra $\Omega^\bullet(\FR^{1,d-1})$:
\begin{align}
\begin{split}
\dd\alpha&=\dd(T^a\otimes\alpha_a)=T^a\otimes \dd\alpha_a \ , \\[4pt]
[\alpha,\beta]_{\fru(N)}&=[T^a\otimes \alpha_a,T^b\otimes\beta_b]_{\fru(N)} = [T^a,T^b]_{\fru(N)} \otimes (\alpha_a\wedge\beta_b) \ ,
\end{split}
\end{align}
where $\alpha_a,\beta_b\in\Omega^\bullet(\FR^{1,d-1})$ and we used graded commutativity of the exterior product of forms: $\alpha_a\wedge\beta_b = (-1)^{|\alpha_a|\,|\beta_b|} \, \beta_b\wedge\alpha_a$. This is a special instance of a more general statement that is relevant for us: The tensor product of any $L_\infty$-algebra $\frL$ with $\Omega^\bullet(\FR^{1,d-1})$ is an $L_\infty$-algebra.

The Moyal--Weyl deformation of $\Omega^\bullet(\FR^{1,d-1},\fru(N))$ is defined by deforming the exterior product to the star-product
\begin{align}
\alpha\wedge_\star\beta :=  \alpha \wedge \, \exp\Big(-\frac{\ii\,t}2 \ \lvlie_\mu \, \theta^{\mu\nu}\, \rvlie_\nu \Big) \, \beta \ ,
\end{align}
where $\pounds_\mu:=\pounds_{\partial_\mu}$ are Lie derivatives along the holonomic frame of vector fields on $\FR^{1,d-1}$. The conventional noncommutative gauge theories are realised by deforming the commutator \eqref{eq:commutatorforms} to the star-commutator on $\Omega^\bullet(\FR^{1,d-1},\fru(N))$:
\begin{align}\label{eq:commutatorformsstar}
[\alpha\ds\beta]_{\fru(N)} = \alpha\wedge_\star\beta - (-1)^{|\alpha|\,|\beta|} \, \beta\wedge_\star\alpha \ .
\end{align}
This bracket again makes $\Omega^\bullet(\FR^{1,d-1},\fru(N))$ into a strict $L_\infty$-algebra. 

As before, while the vector space isomorphism \eqref{eq:vectorspaceiso} still holds, it is no longer an isomorphism of strict $L_\infty$-algebras, as
\begin{align}
\begin{split}\label{eq:brkt_col_first}
[\alpha\ds\beta]_{\fru(N)}=[T^a\otimes \alpha_a\ds T^b\otimes\beta_b]_{\fru(N)} &= \tfrac12\,[T^a,T^b]_{\fru(N)}\otimes\{\alpha_a\ds\beta_b\}_{\fru(1)}  \\
& \quad \, +\tfrac12\, \{T^a,T^b\}_{\fru(N)} \otimes [\alpha_a\ds\beta_b]_{\fru(1)} \ ,
\end{split}
\end{align}
where $\{\alpha_a\ds\beta_b\}_{\fru(1)} := \alpha_a\wedge_\star\beta_b + (-1)^{|\alpha|\,|\beta|} \, \beta_b\wedge_\star\alpha_a$. As previously, in what follows we shall understand colour-stripping in noncommutative gauge theories in a twisted sense, which is compatible with colour-kinematics duality in this setting.

\subsection{The $L_\infty$-structure of noncommutative Chern--Simons theory} 
\label{sub:noncommutative_chern_simons}

Consider standard noncommutative $\sU(N)$ Chern--Simons gauge theory on $\FR^{1,2}$, which is defined by the action functional
\begin{align}\label{eq:NCCSaction}
    S_\CS^\star[A] = \int\, \mathrm{Tr}_{\mathfrak{u}(N)} \Big(\frac12\,A\wedge_\star\dd A + \frac g{3!}\, A\wedge_\star [A \stackrel{\star}{,} A]_{\fru(N)} \Big) \ ,
\end{align}
where $A\in\Omega^1(\FR^{1,2}, \mathfrak{u}(N))$ is the gauge field and $g=\sqrt{2\pi/k}$ is the gauge coupling constant with $k$ the Chern--Simons level. 
Solutions of the corresponding field equation are flat noncommutative connections, $F_A^\star=0$, where 
\begin{align}\label{eq:fieldstrength}
F_A^\star = \dd A + \tfrac g2\,[A \stackrel{\star}{,}A]_{\fru(N)}
\end{align}
is the field strength in $\Omega^2(\FR^{1,2},\fru(N))$.
The noncommutative Chern--Simons functional is invariant under infinitesimal star-gauge transformations $\delta^\star_c A=\dd c+g\,[c \stackrel{\star}{,}A]_{\fru(N)}$ with $c\in\Omega^0(\FR^{1,2},\mathfrak{u}(N))$. 
The Bianchi identity $\nabla_A^\star F^\star_A=0$ in $\Omega^3(\FR^{1,2},\fru(N))$ is the Noether identity corresponding to this gauge symmetry, where $\nabla_A^\star  : \Omega^p(\FR^{1,2},\fru(N))\longrightarrow \Omega^{p+1}(\FR^{1,2},\fru(N))$ is the covariant derivative
\begin{align}\label{eq:covariantderivative}
\nabla_A^\star = \mathrm{d} + g\, [A \stackrel{\star}{,} -]_{\mathfrak{u}(N)} \ .
\end{align}

This noncommutative gauge theory is organised by the cyclic strict $L_\infty$-algebra
\begin{align}\label{eq:Linf of unCS}
    \ChS_\star = \Big(\big(\Omega^\bullet(\FR^{1,2}, \mathfrak{u}(N)),\mu_1^{\star \CS}=\dd\big) \,,\, \mu_2^{\star \CS}= g\,[-\ds-]_{\fru(N)} \,,\,\langle-,-\rangle_\CS^\star\Big) \ ,
\end{align}
whose underlying cochain complex $\Ch(\ChS_\star) = \big(\Omega^\bullet(\FR^{1,2}, \mathfrak{u}(N)),\dd\big)$ is the de~Rham complex of differential forms on $\FR^{1,2}$  valued in the matrix Lie algebra $\fru(N)$. The cyclic structure of degree~$-3$ is given by the pairing of $\fru(N)$-valued differential forms in complementary degrees:
\begin{align}
\langle\alpha,\alpha^+\rangle^\star_\CS = \int\,\Tr_{\fru(N)}(\alpha\wedge_\star\alpha^+) = \int\,\Tr_{\fru(N)}(\alpha\wedge\alpha^+) \ ,
\end{align}
for $\alpha\in\{c,A\}$, with cyclicity ensured by the $\mathrm{ad}(\mathfrak{u}(N))$-invariance of the trace pairing $\mathrm{Tr}_{\fru(N)}: \mathfrak{u}(N) \otimes \mathfrak{u}(N) \longrightarrow \mathbbm{R}$. 

For this strict $L_\infty$-algebra, the Maurer--Cartan theory from \S\ref{sub:preliminaries} reproduces the standard curvature and Bianchi identity given in degree~$2$ and~$3$ respectively by
\begin{equation}
\begin{split}\label{eq:NCCS eom and noether}
    f_A^{\star\CS} &= \mu_1^{\star\CS}(A) + \tfrac{1}{2}\,\mu_2^{\star\CS}(A,A) = F_A^\star \ , \\[4pt]
    \dd_A^{\star\CS} f_A^{\star\CS} &= \mu_1^{\star\CS}(f_A^{\star\CS}) + \mu_2^{\star\CS}(A,f_A^{\star\CS}) = \nabla_A^\star F_A^\star = 0 \ , 
\end{split}
\end{equation}
and also the noncommutative Chern--Simons functional
\begin{align}
S_{\CS}^\star[A] = \tfrac12\,\langle A,\mu_1^{\star\CS}(A)\rangle_\CS^\star + \tfrac1{3!} \, \langle A,\mu_2^{\star\CS}(A,A)\rangle_\CS^\star \ ,
\end{align}
invariant under the star-gauge transformations $\delta_c^{\star}A = \mu_1^{\star\CS}(c) + \mu_2^{\star\CS}(c,A)$. 

\paragraph{Batalin--Vilkovisky formalism.}

The Batalin--Vilkovisky (BV) complex corresponding to the $L_\infty$-algebra $\ChS_\star$ consists of BRST ghosts $c\in\Omega^0(\FR^{1,2},\fru(N))$, gauge fields $A\in\Omega^1(\FR^{1,2},\fru(N))$ along with their antifields $c^+\in\Omega^3(\FR^{1,2},\fru(N))$ and $A^+\in\Omega^2(\FR^{1,2},\fru(N))$. The non-zero differentials $\mu_1^{\star\CS}:\Omega^p(\FR^{1,2},\fru(N))\longrightarrow\Omega^{p+1}(\FR^{1,2},\fru(N))$ are
\begin{subequations}
\begin{equation}
\begin{split}
\mu_1^{\star\CS}(c) =  \mathrm{d} c \quad , \quad 
\mu_1^{\star\CS}(A) = \mathrm{d} A \quad \mbox{,} \quad
\mu_1^{\star\CS}( A^+) =  \mathrm{d} A^+ \ ,
\end{split}
\end{equation}
while the non-zero $2$-brackets $\mu_2^{\star\CS}:\Omega^p(\FR^{1,2},\fru(N))\,\otimes\, \Omega^q(\FR^{1,2},\fru(N))\longrightarrow\Omega^{p+q}(\FR^{1,2},\fru(N))$ are given by
\begin{equation}
\begin{split}    
\mu_2^{\star\CS}(c_1, c_2) =  g\,[c_1 \stackrel{\star}{,} c_2]_{\fru(N)} \quad , & \quad
\mu_2^{\star\CS}(c,A) =  g\,[c \stackrel{\star}{,} A]_{\fru(N)} \ , \\[4pt]
\mu_2^{\star\CS}(c, A^+)  = g\,[c \stackrel{\star}{,} A^+]_{\fru(N)} \quad , & \quad
\mu_2^{\star\CS}(A_1, A_2) = g\, [A_1 \stackrel{\star}{,} A_2]_{\fru(N)} \ , \\[4pt]
\mu_2^{\star\CS}(A, A^+) = g\,[A \stackrel{\star}{,} A^+]_{\fru(N)} \quad , & \quad
\mu_2^{\star\CS}(c, c^+) = g\,[c \stackrel{\star}{,} c^+]_{\fru(N)} \ .
\end{split}
\end{equation}
\end{subequations}

Consider now a superfield element $\mbf A\in \mathsf{Fun}\big(\Omega^\bullet(\FR^{1,2},\fru(N))[1]\big)\otimes \Omega^\bullet(\FR^{1,2},\fru(N))$. With the local coordinate functions $\vartheta^{\mu_1 \cdots \mu_p}$ on $\Omega^\bullet(\FR^{1,2},\fru(N))[1]$ of degree $|\vartheta^{\mu_1 \cdots \mu_p}| = 1-p$, we can express the superfield as
\begin{equation}\label{eq:CS superfield}
    \mbf{A} = \vartheta \otimes c + \vartheta^\mu \otimes A_\mu + \vartheta^{\mu\nu} \otimes A^{+}_{\mu\nu} + \vartheta^{\mu\nu\rho} \otimes c^+_{\mu\nu\rho} \ .
\end{equation}
Since $\mbf A$ is a degree~$1$ element by construction, we  retrieve the full BV action functional from the Maurer--Cartan functional of $\mbf A$ as (see \S\ref{sub:preliminaries})
\begin{align}\label{eq:CS action with BV}
\begin{split}
    S_{\textrm{\tiny BV}}^\star[\mbf{ A}]
    &= \int\, \mathrm{Tr}_{\mathfrak{u}(N)} \Big(
    \frac{1}{2}\, A \wedge_\star \mathrm{d}A +\frac{g}{3!}\, A \wedge_\star [A \stackrel{\star}{,} A]  _{\fru(N)} + A^+\wedge_\star \nabla_A^\star\, c
    + \frac{g}{2}\, c^+ \wedge_\star [c \stackrel{\star}{,} c]_{\fru(N)}
    \Big) \ . 
\end{split}
\end{align}
In the commutative limit $t=0$, this reproduces the superfield Chern--Simons action functional of~\cite{Ben-Shahar:2021zww} entirely from the underlying $L_\infty$-structure.

The Seiberg--Witten map~\cite{Seiberg:1999vs} gives rise to a quasi-isomorphism between a noncommutative gauge theory and a commutative gauge theory.\footnote{See e.g.~\cite{Blumenhagen:2018shf,Kupriyanov:2023zfh} for recent explicit elucidations in the language of homotopy algebras.} The local BRST cohomology of standard Chern--Simons theory implies that any consistent deformation of commutative $\sU(N)$ Chern--Simons theory is trivial; in particular, the noncommutative deformation is trivial~\cite{Barnich:2002tz,Barnich:2003wq}. It follows that, under the Seiberg--Witten map, the $L_\infty$-algebra $\ChS_\star$ of noncommutative $\sU(N)$ Chern--Simons theory is quasi-isomorphic to the $L_\infty$-algebra $\ChS$ of ordinary $\sU(N)$ Chern--Simons theory. This will account for some classical features of the homotopy double copy construction that we explain below. However, it does \emph{not} imply that the corresponding quantum field theories are equivalent, and indeed we shall see that the amplitudes in the two theories are not the same.

\subsection{Twisted homotopy factorisation}
\label{sub:factNCCS}

Colour-stripping of noncommutative Chern--Simons theory can be done in two equivalent ways, each of which will prove useful for our subsequent analysis. Following what we did in \S\ref{sub:factNCbiadjoint}, we can first strip off the infinite-dimensional colour Lie algebra $\fru_\star(N)$ by extending the twist datum \smash{$\tau^{\fru_\star(N)} = \big(\tau_1^{\fru_\star(N)},\tau_2^{\fru_\star(N)}\big)$} to arbitrary degree differential forms, which by an abuse of notation we continue to denote with the same symbol. Thus the twist datum now
 consists of the map \smash{$\tau_1^{\fru_\star(N)}:\fru(N)\longrightarrow \fru(N)\otimes\sEnd\big(\Omega^\bullet(\FR^{1,2})\big)$} defined by
 \begin{subequations}\label{eq:twist with lie derivatives}
\begin{align}
\tau^{\fru_\star(N)}_1( T^{ a}) =  T^{ a}\otimes \mathbbm{1} \ ,
\end{align}
and the map \smash{$\tau_2^{\fru_\star(N)}:\fru(N)\otimes\fru(N)\longrightarrow \fru(N)\otimes\sEnd\big(\Omega^\bullet(\FR^{1,2})\big)\otimes \sEnd\big(\Omega^\bullet(\FR^{1,2})\big)$}
defined by
\begin{align}
\begin{split}
\tau_2^{\fru_\star(N)}( T^{ a},  T^{ b}) &= [ T^{ a}, T^{ b}]_{\fru(N)}\otimes\cos\big(\tfrac t2\,\theta^{\mu\nu} \, \pounds_\mu \otimes \pounds_\nu\big) \\
& \quad \, + \ii\,\{ T^{ a}, T^{ b}\}_{\fru(N)} \otimes \sin\big(\tfrac t2\,\theta^{\mu\nu} \, \pounds_\mu \otimes \pounds_\nu\big) \ .
\end{split}
\end{align}
\end{subequations}

By completely analogous calculations to those of \S\ref{sub:factNCbiadjoint}, it is now straightforward to show that the cyclic strict $L_\infty$-algebra $\ChS_\star$ admits a homotopy factorization
\begin{align}
\ChS_\star = \fru(N)\otimes_{\tau^{\fru_\star(N)}}\frC_{\Omega^\bullet}
\end{align}
into the twisted tensor product of the quadratic colour Lie algebra $\fru(N)$ with the kinematical strict $C_\infty$-algebra
\begin{align}
\frC_{\Omega^\bullet} = \Big(\big(\Omega^\bullet(\FR^{1,2}),m_1^{\Omega^\bullet}=\dd\big) \,,\, m_2^{\Omega^\bullet}=g\,\wedge \,,\, \langle-,-\rangle_{\Omega^\bullet}\Big) \ ,
\end{align}
with cyclic structure $ \langle-,-\rangle_{\Omega^\bullet}$ defined by integration of exterior products of forms in complementary degree. This is just the exterior algebra of differential forms on $\FR^{1,2}$, which represents the infinitesimal diffeomorphism invariance of classical Chern--Simons theory. 

Next we factorise $\frC_{\Omega^\bullet}$ into a finite-dimensional kinematic vector space $\Kin_{\wedge^\bullet}$ and a cyclic strict $L_\infty$-algebra encoding the trivalent interactions. As a graded vector space, we identify $\Omega^\bullet(\FR^{1,2})$ with the tensor product $\midwedge^\bullet(\FR^{1,2})^*\otimes\Omega^0(\FR^{1,2})$ by writing any one-form $\alpha=\alpha_\mu\,\dd x^\mu\in\Omega^1(\FR^{1,2})$ as $\alpha=e^\mu\otimes\alpha_{\mu
}$, where $\alpha_{\mu}\in\Omega^0(\FR^{1,2})$ and $e^\mu$ is the natural basis of covectors in $\FR^{1,2}$ relative to the rectangular coordinates $(x^\mu)$. 

At the level of homotopy algebras, the cyclic strict $C_\infty$-algebra $\frC_{\Omega^\bullet}$ factorizes as a twisted tensor product
\begin{align}\label{eq:CCSfact}
\frC_{\Omega^\bullet} = \Kin_{\wedge^\bullet} \otimes_{\tau^{\Omega^\bullet}} \Scal
\end{align}
of the $1$-shifted exterior algebra
\begin{align}
\Kin_{\wedge^\bullet} := \midwedge^\bullet(\FR^{1,2})^*[1]
\end{align}
with the cyclic strict $L_\infty$-algebra $\Scal$ from \S\ref{sub:nc_strict_scalar_field_theory} for scalar field theory with underlying graded vector space \eqref{eq:scalarvectorspace} on $\FR^{1,2}$. The twist datum $\tau^{\Omega^\bullet}=\big(\tau_1^{\Omega^\bullet},\tau_2^{\Omega^\bullet}\big)$ is defined as follows. The map $\tau_1^{\Omega^\bullet}:\Kin_{\wedge^\bullet} \longrightarrow \Kin_{\wedge^\bullet}\otimes\sEnd(L)$ is defined by
\begin{align}
\tau_1^{\Omega^\bullet}(v)= (e^\mu\wedge v)\otimes \frac{\partial_\mu}\square \ ,
\end{align}
while $\tau_2^{\Omega^\bullet}:\Kin_{\wedge^\bullet} \otimes \Kin_{\wedge^\bullet} \longrightarrow \Kin_{\wedge^\bullet}\otimes\sEnd(L) \otimes \sEnd(L)$ is given as
\begin{align}
\tau^{\Omega^\bullet}_2(v, w) = (v\wedge w) \otimes \mathbbm{1} \otimes \mathbbm{1} \ ,
\end{align}
for all $v,w\in \Kin_{\wedge^\bullet} = \midwedge^\bullet(\FR^{1,2})^*[1]$.

It is then straightforward to check that this twisting reproduces the differential: $m_1^{\tau^{\Omega^\bullet}}=m_1^{\Omega^\bullet}$; e.g. on degree~$1$ fields we find
\begin{align}
m_1^{\tau^{\Omega^\bullet}}(e^\mu\otimes\alpha_\mu) = (e^\nu\wedge e^\mu)\otimes\mu_1^{\rm Scal}\big(\tfrac{\partial_\nu}\square\,\alpha_\mu\big) = (e^\nu\wedge e^\mu)\otimes\partial_\nu\alpha_\mu = \dd \alpha \ .
\end{align}
Similarly we reproduce the product: $m_2^{\tau^{\Omega^\bullet}}=m_2^{\Omega^\bullet}$; e.g. on degree~$1$ fields we find
\begin{align}
m_2^{\tau^{\Omega^\bullet}}(e^\mu\otimes\alpha_\mu,e^\nu\otimes\beta_\nu) = (e^\mu\wedge e^\nu)\otimes \mu_2^{\rm Scal}(\alpha_\mu,\beta_\nu) = (e^\mu\wedge e^\nu)\otimes g\,\alpha_\mu\,\beta_\nu = g\,\alpha\wedge\beta \ .
\end{align}
Note, however, that due to the tensor product of the underlying graded vector spaces $\Kin_{\wedge^\bullet} = \midwedge^\bullet(\FR^{1,2})^*[1]$ and $L=\Omega^0(\FR^{1,2})[-1]\oplus\Omega^0(\FR^{1,2})[-2]$, the factorization \eqref{eq:CCSfact} introduces some redundancy. For example, the cochain complex underlying the twisted tensor product is
\begin{equation}\label{eq:twistedCScomplex}
\begin{tikzcd}[row sep=0ex,ampersand replacement=\&]
\Omega^0(\FR^{1,2}) \arrow[r,"\dd"] \& \Omega^1(\FR^{1,2})[-1] \arrow[r,"\dd"] \&  \Omega^2(\FR^{1,2})[-2] \arrow[r,"\dd"] \& \Omega^3(\FR^{1,2})[-3] \\
\& \oplus \& \oplus \&  \oplus \\
  \& \Omega^0(\FR^{1,2})[-1] \arrow[r,"0"] \& \Omega^1(\FR^{1,2})[-2] \arrow[r,"0"] \& \Omega^2(\FR^{1,2})[-3] 
\end{tikzcd}
\end{equation}

Finally, the cyclic structure is recovered by wedging forms in $ \midwedge^\bullet(\FR^{1,2})^*$ of complementary degrees and applying the Hodge duality operator $\Tr_{\wedge^\bullet}:\midwedge^3(\FR^{1,2})^*\longrightarrow \FR$ in three dimensions. Then $\Tr_{\wedge^\bullet}\otimes\langle-,-\rangle_{\rm Scal}$ reproduces the inner product $ \langle-,-\rangle_{\Omega^\bullet}$ on $\frC_{\Omega^\bullet}$.

Altogether, we have shown that the cyclic strict $L_\infty$-algebra of noncommutative Chern--Simons gauge theory admits the factorization
\begin{align}\label{eq:CSfact1}
\ChS_\star = \fru(N)\otimes_{\tau^{\fru_\star(N)}}\big(\Kin_{\wedge^\bullet} \otimes_{\tau^{\Omega^\bullet}} \Scal\big) \ .
\end{align}
In particular, replacing $\Kin_{\wedge^\bullet}$ with a copy of another colour Lie algebra $\fru(\bar N)$ and corresponding twisted tensor product yields the cyclic strict $L_\infty$-algebra \eqref{eq:BASdoublestar}. This identifies the $d=3$ binoncommutative biadjoint scalar theory of \S\ref{sub:the_other_single_copy} as the zeroth copy of noncommutative Chern--Simons theory. 

This last statement is perhaps more transparent if one notes that the factorization \eqref{eq:CSfact1} can be equivalently written as the twisted tensor product
\begin{align}\label{eq:CS as double copy}
\ChS_\star = \Kin_{\wedge^\bullet} \otimes_{\tau^{\Omega^\bullet}} \frC_{\fru_\star(N)} \ ,
\end{align}
where the cyclic strict $C_\infty$-algebra $\frC_{\fru_\star(N)}$ itself factorises as in \cref{eq:CCCbiadstarfact}. In \cref{eq:CS as double copy} the twist datum is defined in a completely analogous way to the twist datum above in terms of maps \smash{$\tau_1^{\Omega^\bullet}:\Kin_{\wedge^\bullet} \longrightarrow \Kin_{\wedge^\bullet}\otimes\sEnd\big(\fru(N)\otimes L\big)$} and \smash{$\tau_2^{\Omega^\bullet}:(\Kin_{\wedge^\bullet})^{\otimes 2}  \longrightarrow \Kin_{\wedge^\bullet}\otimes\sEnd\big(\fru(N)\otimes L\big)^{\otimes 2}$}. For example, on degree~$1$ fields we find
\begin{align}
\begin{split}
\mu_1^{\tau^{\Omega^\bullet}}(e^\mu\otimes\alpha_\mu) &= (e^\nu\wedge e^\mu)\otimes m_1^{\fru_\star(N)}\big(\tfrac{\partial_\nu}\square\,\alpha_\mu\big) = (e^\nu\wedge e^\mu)\otimes\partial_\nu\alpha_\mu = \dd \alpha \ , \\[4pt]
\mu_2^{\tau^{\Omega^\bullet}}(e^\mu\otimes\alpha_\mu,e^\nu\otimes\beta_\nu) &= (e^\mu\wedge e^\nu)\otimes \big(-m_2^{\fru_\star(N)}(\alpha_\mu,\beta_\nu)\big) \\[4pt]
&= (e^\mu\wedge e^\nu)\otimes g\,[\alpha_\mu\ds\beta_\nu]_{\fru(N)} = g\,[\alpha\ds\beta]_{\fru(N)} \ .
\end{split}
\end{align}
Compared to \cref{eq:CSfact1}, which is a factorisation into a colour Lie algebra and a kinematic $C_\infty$-algebra, in \cref{eq:CS as double copy} the factorisation is into a kinematic vector space and a colour $C_\infty$-algebra.

\subsection{Twisted colour-kinematics duality}
\label{sub:CKNCCS}

Perturbative computations in noncommutative Chern--Simons theory are facilitated by using the factorisation \eqref{eq:CS as double copy} to decompose degree~$1$ fields as $A=e^\mu\otimes A_\mu$ in terms of coordinate functions $A_\mu:\FR^{1,2}\longrightarrow\fru_(N)$, with bracket
\begin{align}
 [A_\mu \stackrel{\star}{,} B_\nu]_{\mathfrak{u}(N)} = A_\mu \star B_\nu - B_\nu \star A_\mu
\end{align}
for $A_\mu,B_\nu\in\Omega^0(\FR^{1,2},\fru(N))$. In this form, the three-gluon Feynman vertex can be computed directly by expanding the action functional \eqref{eq:NCCSaction} as
\begin{align}
S_\CS^\star[A] = \int\, \dd^3x \ \epsilon^{\mu\nu\rho} \ \Tr_{\fru(N)}\Big(A_\mu\star\partial_\nu A_\rho +\frac g{3!} \, A_\mu\star[A_\nu\ds A_\rho]_{\fru(N)} \Big) \ ,
\end{align}
where $\epsilon^{\mu\nu\rho}$ is the Levi-Civita symbol in three dimensions with $\epsilon^{012}=1$.
It is given in terms of the structure constants of the infinite-dimensional Lie algebra $\mathfrak{u}_\star(N) = \big(\Omega^0(\mathbbm{R}^{1,2}, \mathfrak{u}(N)), [ - \stackrel{\star}{,} -]_{\mathfrak{u}(N)}\big)$ as
\begin{equation}
\begin{split}
\begin{tikzpicture}[scale=0.8]
{\small
    \draw[gluon] (-140:1.8)node[below]{$\nu,b$}--(0,0);
    \draw[gluon] (140:1.8)node[above]{$\mu,a$}--(0,0);
    \draw[gluon] (0:1.8)node[right]{$\rho, c$} --(0,0);
    \draw[->, shift={(-0.25,0.65)}] (140:.5) node[above]{$p_1$} -- (0,0);
    \draw[->, shift={(-0.25,-0.65)}] (-140:.5) node[below]{$p_2$} -- (0,0);
    \draw[->, shift={(0.29,0.4)}] (0:.5) node[above]{$p_3$} -- (0,0);
    }\normalsize
    \node at (5.9, 0) {$= \ -\mathrm{i}\,g\, \epsilon_{\mu\nu\rho} \, F^{abc}(p_1,p_2,p_3) \ .$};
\end{tikzpicture}
\end{split}
\end{equation}

In the following we shall set up the perturbiner expansion for noncommutative Chern--Simons theory in the $L_\infty$-algebra formalism, and following the discussion from \S\ref{sub:adjointCKduality} we establish a twisted form of colour-kinematics duality compatible with the twisted colour-stripping. 

\paragraph{Perturbiner solutions.}

As Chern--Simons fields have no propagating degrees of freedom, all currents vanish on-shell and there are no interactions; in particular, the pure gauge theory has no non-trivial S-matrix elements. From the homotopy algebraic perspective, this is an easy consequence of the fact that the de~Rham cohomology of $\FR^{1,2}$ is trivial:
\begin{align}
\begin{split}
\sfH^0(\ChS_\star)=\ker(\dd)\otimes\fru(N) \quad , \quad \sfH^3(\ChS_\star) = \ker(\delta)\otimes\fru(N)\quad , \quad
 \sfH^1(\ChS_\star)&=\sfH^2(\ChS_\star)=0 \ ,
\end{split}
\end{align}
where $\delta=\ast_{\textrm {\tiny H}}\,\dd\,\ast_{\textrm {\tiny H}}$ is the codifferential corresponding to the exterior differential $\dd$ and the Hodge duality operator $\ast_{\textrm {\tiny H}}$ induced by the Minkowski metric. Correlation functions with external states lying in $\sfH^1(\ChS_\star)$ are thus always trivially zero.

As we did in \S\ref{sub:DCad}, below we will construct the off-shell action functional of the double copied theory using Maurer--Cartan theory, so we should first establish that colour-kinematics duality holds off-shell from the homotopy algebraic perspective.
Off-shell correlation functions are obtained by projecting onto ``harmonic'' states $\mathcal{H}^\bullet(\ChS_\star) = \ker(\square)$ rather than the minimal model~\cite{Axelrod:1991vq,Jurco:2018sby,Borsten:2022vtg}, where $\square = \mathrm{d}\, \delta + \delta\, \mathrm{d}:\Omega^\bullet(\mathbbm{R}^{1,2}, \mathfrak{u}(N)) \longrightarrow \Omega^\bullet(\mathbbm{R}^{1,2}, \mathfrak{u}(N))$ is the Hodge--d'Alembertian operator acting on forms; in Minkowski signature this is a bigger space than $\sfH^\bullet(\ChS_\star)$. Correlation functions on this space will be non-zero as, unlike the projection to the cohomology, the space of harmonic forms contains configurations which are not pure gauge and hence can propagate. 

Chern--Simons perturbation theory requires a ``Hodge inverse'' of the differential $\mu_1^{\star \CS} = \mathrm{d}$, which is constructed using~$\square$ together with the Hodge--Kodaira decomposition \eqref{eq:abstract hodge kodaira}. 
We look for a contracting homotopy $\mathsf{h}$ satisfying a Hodge--Kodaira decomposition for projection $\sfp_{\ker(\square)}:\Omega^\bullet(\FR^{1,2},\fru(N)) \longrightarrow \mathcal{H}^\bullet(\ChS_\star)$ onto harmonic states:
\begin{equation}\label{eq:HK for harmonic}
    \mathbbm{1} = \sfp_{\ker(\square)} + \mu_1^{\star \CS} \circ \mathsf{h} + \mathsf{h} \circ \mu_1^{\star \CS} \ .
\end{equation}
The contracting homotopy is the partial inverse of the differential $\mu_1^{\star \CS}$ on the space of pure gauge configurations $\mathrm{im}(\mathrm{d}) \oplus \mathrm{im}(\delta)$, extended trivially to all of $\Omega^\bullet(\mathbbm{R}^{1,2}, \mathfrak{u}(N))$ by
\begin{equation}
\begin{split}
    \mathsf{h} : \mathrm{im}(\mathrm{d}) \oplus \mathrm{im}(\delta) \longrightarrow \Omega^\bullet(\mathbbm{R}^{1,2}, \mathfrak{u}(N)) \qquad \text{and} \qquad
    \ker(\mathsf{h}) = \ker(\square)  \ .
\end{split}
\end{equation}
It can be constructed from the partial inverse $\mathsf D$ of the d'Alembertian $\square$ on the subspace of pure gauge configurations by extending it trivially to $\ker(\square)$:
\begin{equation}
    \mathsf{D} \big\vert_{\mathrm{im}(\mathrm{d}) \oplus \mathrm{im}(\delta)} = \frac1\square \qquad \text{and} \qquad
    \ker(\mathsf{D}) =  \ker(\square) \ .
\end{equation}

A choice of contracting homotopy $\sfh:\Omega^\bullet(\FR^{1,2},\fru(N))\longrightarrow \Omega^\bullet(\FR^{1,2},\fru(N))$ of degree~$-1$ completing the Hodge--Kodaira decomposition  \eqref{eq:HK for harmonic} is then
\begin{equation}
    \mathsf{h} =  \delta \circ \mathsf{D} \ .
\end{equation}
Acting on gauge fields and ghost antifields, the maps $\sfh^{(1)}:\Omega^1(\FR^{1,2},\fru(N))\longrightarrow \Omega^0(\FR^{1,2},\fru(N))$ and $\sfh^{(3)}:\Omega^3(\FR^{1,2},\fru(N))\longrightarrow \Omega^2(\FR^{1,2},\fru(N))$ are respectively given by
\begin{align}
\sfh^{(1)}(A) = \frac{\partial_\mu}{\square}\,A^\mu \qquad \mbox{and} \qquad \sfh^{(3)}(c^+)_{\mu\nu} = \frac{\partial^\rho}{\square} \, c^+_{\mu\nu\rho} \ .
\end{align}
We set $\sfh^{(1)}(A)=0$, which imposes the Lorenz gauge $\delta A=\partial_\mu A^\mu=0$. Acting on antifields in degree~$2$, the contracting homotopy is then given by the Feynman propagator for Chern--Simons theory: $\sfh^{(2)}=C_{\rm F}:\Omega^2(\FR^{1,2},\fru(N))\longrightarrow\Omega^1(\FR^{1,2},\fru(N))$ where
\begin{equation}\label{eq:CS propa}
  C_{\rm F}(A^+)_\rho = \epsilon_{\rho \sigma \alpha} \, \epsilon^{\mu \nu \sigma} \, \frac{\partial^\alpha}{\square}\, A^+_{\mu \nu}   \ .
\end{equation}
Indeed, using these formulas, and the fact that the distribution $\frac{1}{\square}$ commutes with the operators $\mathrm{d}$, $\delta$  and $\ast_{\textrm {\tiny H}}$, we verify the projector onto harmonic forms fits into \cref{eq:HK for harmonic}.

Given a perturbiner element $A(i) \in \mathcal{H}^1(\ChS_\star)$ corresponding to the `$i$-th' gluon, we look for a quasi-isomorphism $\psi_n : \mathcal{H}^1(\ChS_\star)^{\otimes n} \longrightarrow \Omega^1(\FR^{1,2},\fru(N))$ which is again given by the $L_\infty$-recursion relations. 
We work in terms of word combinatorics, where an ordered word $w= k_1\cdots k_n \in \CCW _n $ represents $n$ external gluons. 
The pullback to ordered words on $n$ letters has the physical interpretation of performing a colour ordering expansion, as is often done in Yang--Mills theory.
The Berends--Giele currents $J_\mu: \CCW  _n \longrightarrow \mathfrak{u}(N)$ are the coefficients of the plane wave expansions
\begin{equation}
    \psi_n\big(A(1),\dots,A(n)\big) = e^\mu\otimes J_\mu(w) \, \e^{\,\mathrm{i}\, p_{1\cdots n}\cdot x} \  \in \  \Omega^1(\FR^{1,2},\fru(N)) \ ,
\end{equation}
where $p_{1 \cdots n} := p_1 + \cdots + p_n$. The currents $J(w):=e^\mu\otimes J_\mu(w)$ are coclosed in $\ker(\sfh^{(1)})$, $p_w\cdot J(w)=0$, from the way in which we have built the contracting homotopy $\sfh$. 

Since $\square$ is a second order differential operator, the star-commutator of two currents is not generally in $\ker(\square)$.
Thus upon using the recursion relations from \cref{eq:minimal model recursion relations,eq:min model brackets}, successive application of the $2$-bracket $\mu_2^{\star\CS}$ of the full theory move us out of the harmonic states. Similarly, the codifferential can be expressed as $\delta = \frac{\partial}{\partial e^\mu}\otimes \partial^\mu$ and so is a second order differential operator of degree $-1$. Hence the $2$-bracket $\mu_2^{\star\CS}$ of two currents is also not generally coclosed, but this is restored upon following with the Chern--Simons propagator $\mathsf{h}^{(2)} = C_{\rm F}$. 

The recursion relations for the quasi-isomorphism give
\begin{equation}
\begin{split}
    J(w) &= - \frac{g}{2} \,
    \sum_{w = w_1 \sqcup w_2} \,
     e^\rho\otimes\big((C_{\rm F})^{\mu \nu}_\rho \circ  \,
    [  J_\mu(w_1) \stackrel{\star}{,} J_\nu(w_2)]_{\mathfrak{u}(N)} \big)
    \ , \\[4pt]
    \mu_n^{\circ\star\CS}\big(A(1),\dots,A(n)\big)
    &= \frac{g}{2} \,
    \sum_{w = w_1 \sqcup w_2} \, \big(e^\mu\wedge e^\nu\big) \otimes \big(
    \sfp_{\ker(\square)} \circ
    [
    J_{\mu}(w_1) \stackrel{\star}{,} J_\nu(w_2)
    ]_{\mathfrak{u}(N)} \,
    \e^{\,\mathrm{i}\, p_{1 \cdots n}\cdot x}\big) \ .
\end{split}
\end{equation}
The images of the  $n$-brackets $\mu_n^{\circ\star\CS}$ are coclosed in $\ker(\sfh^{(2)})$, by our construction of the contracting homotopy $\sfh$.
The final projection to the harmonic states $\sfp_{\ker(\square)}: \Omega^\bullet(\FR^{1,2},\fru(N)) \longrightarrow \mathcal{H}^\bullet(\ChS_\star)$ also enforces momentum conservation $s_{1 \cdots n} = 0$, where $s_w$ are the Mandelstam variables.

The colour ordered partial amplitudes are given by substitution into the cyclic structure on the harmonic states. This gives
\begin{equation}
    \CM^{\star\CS}_n(1, \ldots, n) = \langle A(1), \mu^{\circ\star\CS}_{n-1}(A(2), \ldots , A(n))\, \rangle^\star_{\circ\CS}
\end{equation}
for the $n$-gluon amplitudes. 

\paragraph{Colour-kinematics duality.}

With the Feynman propagator from \cref{eq:CS propa}, and understanding star-commutators as living in $\mathfrak{u}_\star(N)$, the recursion relations can be written as
    \begin{equation}
    \begin{split}
        J_\rho(w) &= - \frac{g}{2} \, \sum_{w = w_1 \sqcup w_2} \, \delta_{[\alpha \rho]}^{\mu \nu} \, \frac{p_w^\alpha}{s_{w}} \, [J_\mu(w_1) \stackrel{\star}{,} J_\nu(w_2)]_{\mathfrak{u}(N)} \\[4pt]
        &= - \frac{g}{2s_w} \, \sum_{w = w_1 \sqcup w_2} \,
        \Big( \big(p_{w_2} \cdot J^a(p_{w_1})^\sharp\big)\, J^b_\rho(p_{w_2}) - J^a_\rho(p_{w_1})\, \big(p_{w_1} \cdot J^b(p_{w_2})^\sharp\big)\Big)
        \\
        & \hspace{10cm} \times  F^{abc}(p_{w_1}, p_{w_2}, p_{w}) \, T^c  \ ,
    \end{split}
    \end{equation}
where \smash{$\delta^{\mu\nu}_{[\alpha\rho]} := \delta^\mu_\alpha\,\delta^\nu_\rho - \delta^\mu_\rho\,\delta^\nu_\alpha$}, and we have decomposed the Berends--Giele currents as tensor products \smash{$J_\mu(w) = J_\mu^a(p_w) \otimes \, e^a_{p_w}$} using the momentum space basis of the infinite-dimensional Lie algebra $\fru_\star(N)$ introduced in \cref{eq:ustarnbasis}. The superscript \smash{${}^\sharp$} indicates the operation which sends a covector in $(\FR^{1,2})^*$ to its dual vector in $\FR^{1,2}$. 

The coefficients $J^a(p_w)$ can be thought of in terms of a basis of covectors $\varepsilon_{p_w} \in (\mathbbm{R}^{1,2})^*$, which are dual to gluon polarisation vectors \smash{$\varepsilon_{p_w}^\sharp \in \mathbbm{R}^{1,2}$}.  In Lorenz gauge, a given momentum $p_w \in \mathbb{R}^{1,2}$ is orthogonal to the corresponding polarisation vector: \smash{$p_w \cdot \varepsilon_{p_w}^\sharp = 0$}, corresponding to projection to states lying in $ \ker(\delta)$.
These covectors on $\mathbb{R}^{1,2}$ naturally form an infinite-dimensional Lie algebra under the bracket
    \begin{equation}\label{eq:diff 1-2 algebra}
     [\varepsilon_{p_{w_1}}, \varepsilon_{p_{w_2}}]_{\mathfrak{diff}(\mathbb{R}^{1,2})} 
        := \big(\varepsilon_{p_{w_1}}^\sharp \cdot p_{w_2}\big) \, \varepsilon_{p_{w_2}} 
        - \big(\varepsilon^\sharp_{p_{w_2}} \cdot p_{w_1}\big) \, \varepsilon_{p_{w_1}}    \ .
    \end{equation}
In particular, the bracket of two covectors is transverse to the total momentum they carry, that is, \smash{$p_{12} \cdot [\varepsilon_{p_1} , \varepsilon_{p_2}]^\sharp_{\mathfrak{diff}(\mathbb{R}^{1,2})} = 0$}. This is recognised as the bracket of the  Lie algebra $\mathfrak{diff}(\FR^{1,2})$ of infinitesimal diffeomorphisms of $\FR^{1,2}$. In Lorenz gauge, this restricts to the Lie subalgebra $\mathfrak{diff}_{\text{vol}}(\FR^{1,2})$ of volume-preserving diffeomorphisms, i.e. divergence-free vector fields on $\FR^{1,2}$.

We can therefore express the recursion relations for the Berends--Giele currents in terms of a twisted tensor product \smash{$\fru(N)\otimes_{\tau^{\fru_\star(N)}} \mathfrak{diff}_{\text{vol}}(\FR^{1,2})$} of two Lie algebras, which defines a symmetric bracket operation
\begin{equation}\label{eq:BGCSproduct}
\begin{split}
    \llbracket J (w_1), J(w_2) \rrbracket_{\CS}^\star :=& \ e^\rho\otimes\delta^{\mu \nu}_{[\alpha \rho]} \, p^\alpha_{w_1\sqcup w_2} \,
    [J_\mu(w_1) \stackrel{\star}{,} J_\nu (w_2)]_{\mathfrak{u}(N)} \\[4pt]
    =& \ [J^a(p_{w_1}), J^b(p_{w_2})]_{\mathfrak{diff}(\mathbb{R}^{1,2}) } \otimes [e^a_{p_{w_1}} \stackrel{\star}{,} e^b_{p_{w_2}}]_{\mathfrak{u}(N)} \ .
\end{split}
\end{equation}
The intertwining of momenta in \cref{eq:BGCSproduct} simply reflects the twisted factorization of colour degrees of freedom in \cref{eq:CSfact1}, and the related absence of full diffeomorphism invariance in the Moyal--Weyl deformation of Chern--Simons gauge theory. 

The recursion relations thus become
\begin{equation} \label{eq:Jrecursion}
\begin{split}
    J(w) &= - \frac{g}{2} \, \sum_{w = w_1 \sqcup w_2} \, \frac{\llbracket J(w_1), J(w_2)\rrbracket^\star_{\CS}}{s_w} \ ,
\end{split}
\end{equation}
with currents $J(w) \in \ker(\mathsf{h}^{(1)})$.
At the level of amplitudes, we scatter harmonic gluon states $A(1) = e^\mu \otimes e^a_{p_1} \otimes A^a_\mu(1)$ with $A^a_\mu(1) \in \Omega^0(\mathbbm{R}^{1,2})$ satisfying $\square A^a_\mu(1) = 0$. Then the $n$-point colour ordered partial amplitude is written as
\begin{equation}
\begin{split}
    \CM_n^{\star\CS} (1, w) &=\Big \langle A(1)\: ,\: 
    {\frac{g}{2} \,
    \sum\limits_{w = w_1 \sqcup w_2} \, \sfp_{\ker(\square)}\big(
    [
    J_{}(w_1) \stackrel{\star}{,} J(w_2)
    ]_{\mathfrak{u}(N)} \,
    \e^{\,\mathrm{i}\, p_{w}\cdot x}\big)} \Big \rangle^\star_{\circ \CS} \ .
\end{split}
\end{equation}
The projection operator enforces $p_w^2 = 0$, while $p_1^2=0$ already for the first state. 
Altogether one finds
\begin{equation}
\begin{split}
    \CM_n^{\star\CS}(1,w) &= \frac{g}{2} \,
    \sum_{w = w_1 \sqcup w_2}\,  \epsilon^{\mu \nu \rho}\, F^{abc}(p_{w_1}, p_{w_2}, p_1) \, J^a_\mu(w_1)\, J^b_\nu(w_2) \,
 A^c_\rho(1)     \ (2 \pi)^3 \, \delta(p_{1w}) \ .
\end{split}
\end{equation}
The full amplitude is obtained by summing over planar ordered permutations of $n-1$ gluons, and performing the recursion \eqref{eq:Jrecursion} for the currents $J(w)$.

By running through the same argument from \S\ref{sub:adjointCKduality} for the adjoint scalar theory, this demonstrates a twisted form of colour-kinematics duality for currents of noncommutative Chern--Simons theory, between the Lie algebras underlying volume-preserving diffeomorphisms of $\FR^{1,2}$ and noncommutative gauge transformations. In the commutative limit $t=0$, this identifies $\mathfrak{diff}_{\text{vol}}(\mathbbm{R}^{1,2})$ as the kinematic Lie algebra underlying true (off-shell) colour-kinematics duality in ordinary Chern--Simons theory~\cite{Ben-Shahar:2021zww}. Furthermore, from the perspective of \cref{eq:BGCSproduct}, the zeroth copy construction corresponds to replacing the kinematic factors $J(w)$ with a set of twisted colour factors $C_{\bar\star}(w)$ for the second rigid Lie algebra $\fru(\bar N)$ of the $d=3$ binoncommutative biadjoint scalar theory from \S\ref{sub:the_other_single_copy}.

\subsection{Homotopy double copy}
\label{sub:DCCS}

The twisted colour-kinematics duality discussed in \S\ref{sub:CKNCCS} makes it clear that, in noncommutative gauge theories of the type considered in this paper, the double copy involves a  replacement of the full \emph{twisted} colour-stripping by kinematical factors. 
From the factorization \eqref{eq:CSfact1}, the double copy prescription is clear: the homotopy double copy construction replaces the twisted tensor product with the colour factor $\fru(N)$ by the twisted tensor product with another copy of the kinematical factor \smash{$\Kin_{\wedge^\bullet} = \midwedge^\bullet(\FR^{1,2})^*[1]$}, which gives the cyclic strict $L_\infty$-algebra
\begin{align}\label{eq:CSdoublecopy}
\widehat{\ChS} = \Kin_{\wedge^\bullet} \otimes_{\tau^{\Omega^\bullet}}\big(\Kin_{\wedge^\bullet} \otimes_{\tau^{\Omega^\bullet}} \Scal\big) \ .
\end{align}
The second kinematical twist datum is defined by maps $\tau^{\Omega^\bullet}_1:\Kin_{\wedge^\bullet}\longrightarrow \Kin_{\wedge^\bullet}\otimes \sEnd\big(\Kin_{\wedge^\bullet}\otimes L\big)$ and \smash{$\tau^{\Omega^\bullet}_2:\Kin_{\wedge^\bullet}\otimes \Kin_{\wedge^\bullet} \longrightarrow \Kin_{\wedge^\bullet} \otimes\sEnd\big(\Kin_{\wedge^\bullet}\otimes L\big) \otimes \sEnd\big(\Kin_{\wedge^\bullet}\otimes L\big)$}. 

Most notably, this construction asserts the general statement that double copies of noncommutative gauge theories are not deformed and coincide with their commutative counterparts. In particular, the double copy theory organised by the $L_\infty$-algebra \eqref{eq:CSdoublecopy} coincides with that obtained in~\cite{Ben-Shahar:2021zww} from a more physical perspective. In what follows we will unravel some further details about this non-local higher-spin theory, which involves two copies of the volume-preserving diffeomorphism algebra $\mathfrak{diff}_{\text{vol}}(\FR^{1,2})$.

In the following we regard the vector space $\midwedge^\bullet(\FR^{1,2})^* \otimes \midwedge^\bullet(\FR^{1,2})^* \otimes \Omega^0(\FR^{1,2})$ as the tensor product $\Omega^{\bullet,\bullet}(\FR^{1,2}) := \Omega^\bullet(\FR^{1,2})\otimes_{\Omega^0(\FR^{1,2})}\Omega^\bullet(\FR^{1,2})$ of $\Omega^\bullet(\FR^{1,2})$ with itself, considered as a module over $\Omega^0(\FR^{1,2})$; we can think of its elements as differential forms on $\FR^{1,2}$ valued in the exterior algebra $\Omega^\bullet(\FR^{1,2})$. For example, an element $H\in\Omega^{1,1}(\FR^{1,2})$ can be decomposed as $H=e^{\bar \mu}\otimes H_{\bar \mu}$ relative to a basis $e^{\bar\mu}$ corresponding to a rectangular coordinate system $(x^{\bar\mu})$, with $H_{\bar\mu}\in\Omega^1(\FR^{1,2})$. We may then further expand $H_{\bar\mu}=e^\mu\otimes H_{\bar \mu\mu}$ relative to the given basis $e^\mu$, where $H_{\bar\mu\mu}\in\Omega^0(\FR^{1,2})$; we regard $x^\mu$ and $x^{\bar \mu}$ as independent variables, so that in particular the corresponding partial derivatives $\partial_\mu$ and $\partial_{\bar\mu}$ commute. This enables the definition of a tensor product differential $\dd\otimes\dd$, sending $\Omega^{1,1}(\FR^{1,2})$ to $\Omega^{2,2}(\FR^{1,2})$

The graded vector space $\Kin_{\wedge^\bullet}\otimes\Kin_{\wedge^\bullet}\otimes L$ underlying the $L_\infty$-algebra $\widehat{\ChS}$ has $16$ homogeneous components. Any zero differential appearing in the redundant part of \cref{eq:twistedCScomplex} yields a zero differential in the double copy, making the cochain complex of $\widehat{\ChS}$ ``diagonal'' in the sense that its non-trivial part is 
\begin{equation}\label{eq:diagcochaincomplex}
\begin{tikzcd}[row sep=0ex,ampersand replacement=\&]
    \Omega^{0,0}(\FR^{1,2}) \arrow[r," \frac{\dd\otimes\dd}\square"] \& \Omega^{1,1}(\FR^{1,2})[-1]
    \arrow[r," \frac{\dd\otimes\dd}\square"] \&  \Omega^{2,2}(\FR^{1,2})[-2] \arrow[r," \frac{\dd\otimes\dd}\square"] \& \Omega^{3,3}(\FR^{1,2})[-3]
\end{tikzcd}
    \ .
\end{equation}
For example, on fields of degree~$1$, the differential $\widehat{\mu}_1^{\CS}$ of $\widehat{\ChS}$ is given by
\begin{align}
\begin{split}
\widehat{\mu}_1^{\CS} (e^{\bar\mu}\otimes e^\mu\otimes H_{\bar\mu\mu}) &= (e^{\bar\nu}\wedge e^{\bar\mu})\otimes m_1^{\tau^{\Omega^\bullet}}\big(e^\mu\otimes\tfrac{\partial_{\bar\nu}}\square\,H_{\bar\mu\mu}\big) \\[4pt]
&=  (e^{\bar\nu}\wedge e^{\bar\mu})\otimes (e^\nu\wedge e^\mu)\otimes \frac1\square\,\partial_\nu\,\partial_{\bar\nu}H_{\bar\mu\mu} 
\end{split}
\end{align}
for $H_{\bar\mu\mu}\in\Omega^0(\FR^{1,2})$, because we treat the left and right factors of the kinematical vector space $\Kin_{\wedge^\bullet}$ as independent.

The $2$-bracket $\widehat{\mu}_2^\CS : \Omega^{p,q}(\mathbbm{R}^{1,2})[-i] \otimes \Omega^{r,s}(\mathbbm{R}^{1,2})[-j] \longrightarrow \Omega^{p+r, q+s}(\mathbbm{R}^{1,2})[-i-j]$ of $\widehat{\ChS}$ in addition acts non-trivially on ``off-diagonal'' fields.
It is easily identified as the map which scales the tensor product $\frac\kappa2\,\wedge\otimes\ \wedge$ of independent exterior products on the left and right kinematical vector spaces. For example, the bracket of two fields of degree~$1$ is given by
\begin{align}
\begin{split}
\widehat{\mu}_2^\CS(e^{\bar\mu}\otimes e^{\mu}\otimes H_{\bar\mu\mu},e^{\bar\nu}\otimes e^\nu\otimes H'_{\bar\nu\nu}) &= (e^{\bar\mu}\wedge e^{\bar\nu})\otimes m_2^{\tau^{\Omega^\bullet}}(e^\mu\otimes H_{\bar\mu\mu},e^\nu\otimes H'_{\bar\nu\nu}) \\[4pt]
&= (e^{\bar\mu}\wedge e^{\bar\nu})\otimes (e^{\mu}\wedge e^{\nu})\otimes \tfrac\kappa2\,H_{\bar\mu\mu}\,H'_{\bar\nu\nu} \ .
\end{split}
\end{align}

The cyclic structure is obtained by composing the tensor product of Hodge duality operators with wedge products in complementary degrees in each kinematic factor:
\begin{align}
 \langle -,- \rangle_{\widehat{\CS}} := \mathrm{Tr}_{\wedge^{\bullet}} \otimes \mathrm{Tr}_{\wedge^\bullet} \otimes \langle -, - \rangle_{\rm Scal} \ . 
\end{align} 
Altogether, this makes the twisted tensor product \eqref{eq:CSdoublecopy} into a cyclic strict $L_\infty$-algebra.

In Maurer--Cartan theory for the $L_\infty$-algebra $\widehat{\ChS}$, the underlying homogeneous spaces of the ``diagonal'' cochain complex \eqref{eq:diagcochaincomplex} contain all relevant data of the double copy field theory, because only that part of the complex is non-trivial. The Maurer--Cartan equation
\begin{align}\label{eq:curvCSDC}
\widehat{f}^{\,\CS}_{H} := \widehat{\mu}_1^\CS(H) + \tfrac{1}{2}\, \widehat{\mu}_2^\CS(H,H) = 0
\end{align}
for $H\in\Omega^{1,1}(\FR^{1,2})$ gives the equation of motion
\begin{align}
\epsilon^{\bar\mu\bar\nu\bar\rho}\,\epsilon^{\mu\nu\rho} \, \Big(\frac1{\square} \, \partial_{\bar\nu}\,\partial_\nu H_{\bar\rho\rho} + \frac \kappa4\,H_{\bar\nu\nu}\,H_{\bar\rho\rho}\Big) = 0 \ .
\end{align}
These are the variational equations for the non-local classical action functional which is extracted from the Maurer--Cartan functional
\begin{align}\label{eq:MCDCCS}
\begin{split}
    \widehat{S}_\CS[H] &= \tfrac12\,\langle H,\widehat{\mu}_1^\CS(H)\rangle_{\widehat{\CS}} + \tfrac1{3!}\,\langle H,\widehat{\mu}_2^\CS(H,H)\rangle_{\widehat{\CS}} \\[4pt]
    &=  \int\, \mathrm{d}^3x \ \epsilon^{\bar \mu \bar \nu \bar\rho} \, \epsilon^{\mu \nu \rho} \, \Big(
    H_{\bar \mu \mu} \,
    \frac{1}{\square} \, \partial_{\bar \nu} \, \partial_\nu H_{\bar \rho \rho} 
    + \frac{\kappa}{6} \, H_{\bar \mu \mu} \, H_{\bar \nu \nu} \, H_{\bar \rho \rho}
    \Big) \ .
    \end{split}
\end{align}
This theory is invariant under the variations $\widehat{\delta}_CH = \widehat{\mu}_1^\CS(C) + \widehat{\mu}_2^\CS(H,H)$ for a gauge parameter $C\in\Omega^{0,0}(\FR^{1,2})\simeq\Omega^0(\FR^{1,2})$, which reads explicitly as
\begin{align}
\widehat{\delta}_CH_{\bar\mu\mu} = \frac1\square\,\partial_{\bar\mu}\,\partial_\mu C + \frac\kappa2\,C\,H_{\bar\mu\mu} \ .
\end{align}
The Noether identity corresponding to this local gauge symmetry is the off-shell Bianchi identity for the curvature in \cref{eq:curvCSDC}: $\widehat{\mu}_1^\CS(\widehat{f}^{\,\CS}_{H}) + \widehat{\mu}_2^\CS(H,\widehat{f}^{\,\CS}_{H}) = 0$.

The three-particle Feynman vertex of this double copy field theory is
\begin{align}
\begin{split}
\begin{tikzpicture}[scale=0.8]
{\small
    \draw[vector] (-140:1.8)node[below]{$\nu,\bar\nu$}--(0,0);
    \draw[vector] (140:1.8)node[above]{$\mu,\bar\mu$}--(0,0);
    \draw[vector] (0:1.8)node[right]{$\rho, \bar\rho$} --(0,0);
    \draw[->, shift={(-0.2,0.45)}] (140:.5) node[above]{$p_1$} -- (0,0);
    \draw[->, shift={(-0.2,-0.45)}] (-140:.5) node[below]{$p_2$} -- (0,0);
    \draw[->, shift={(0.2,0.4)}] (0:.5) node[above]{$p_3$} -- (0,0);
    }\normalsize
    \node at (6.4, 0) {$\displaystyle = \ -\frac{\mathrm{i}\,\kappa}2 \, \epsilon_{\mu\nu\rho} \, \epsilon_{\bar\mu\bar\nu\bar\rho}  \ \delta(p_1 + p_2 +p_3) \ .$};
\end{tikzpicture}
\end{split}
\end{align}
The Maurer--Cartan functional \eqref{eq:MCDCCS} matches precisely with the non-local gauge-fixed action functional of~\cite{Ben-Shahar:2021zww}, which was obtained by identifying $H_{\bar\mu\mu} = A_{\bar\mu}\otimes A_\mu$ as a tensor product of fields in momentum space, and writing down its propagator and interaction vertex by double copying the Chern--Simons kinematic numerators in Lorenz gauge. In our approach, the result of the double copy procedure applied to Chern--Simons theory is immediate, with its local gauge symmetry automatically identified, and with the gauge-fixing prescription needed for the colour-kinematics duality of~\S\ref{sub:CKNCCS} manifest in the homotopy double copy construction without additional input. 

Our perspective also allows for a simple construction of the superspace formulation of the double copy in~\cite{Ben-Shahar:2021zww} that includes all bosonic fields and ghosts: Applying the Maurer--Cartan functional to a superfield $\mbf{H} \in \mathsf{Fun}\big(\Omega^{\bullet,\bullet}(\FR^{1,2})[1]\big) \otimes \Omega^{\bullet,\bullet}(\FR^{1,2})$ yields the BV action functional
\begin{equation}
    \widehat{S}_{\textrm{\tiny BV}}[\mbf{H}] = \tfrac{1}{2}\,\langle \mbf{H}, \widehat{\mu}_1^{\CS\,\textrm{ext}}(\mbf{H}) \rangle_{\widehat{\CS}\,\textrm{ext}} + \tfrac{1}{3!} \, \langle \mbf{H}, \widehat{\mu}_2^{\CS\,\textrm{ext}}(\mbf{H}, \mbf{H}) \rangle_{\widehat{\CS}\,\textrm{ext}} \ ,
\end{equation}
where $\mbf{H} = \overline{\mbf A}\otimes\mbf A$ is a double copy of two Chern--Simons superfields \eqref{eq:CS superfield} and the notation is defined in \S\ref{sub:preliminaries}.
Thus the homotopy double copy framework offers a more systematic and rigorous method of tackling the problem.

\subsection{$\fru_\star(1)$ Chern--Simons theory as a double copy}
\label{sub: u1CS as DC}
Similarly to the adjoint scalar theory from \S\ref{sub:adjoint_scalar_theory_as_a_strict_linf}, the rank one limit $N=1$ of noncommutative Chern--Simons theory is an interacting theory of photons with no non-trivial commutative counterpart. The noncommutative $\sU(1)$ Chern--Simons gauge theory on $\FR^{1,2}$ is defined by the action functional
\begin{align}
\widehat{S}_{\CS_\star(1)}[A] = \int\,\frac12\,A\wedge_\star\dd A + \frac \kappa{6}\, A\wedge_\star A\wedge_\star A \ ,
\end{align}
for $A\in\Omega^1(\FR^{1,2})$. This theory is organised by the cyclic strict $L_\infty$-algebra
\begin{align}
\widehat{\ChS_\star(1)} = \Big(\big(\Omega^\bullet(\FR^{1,2}),\widehat{\mu}_1^{\CS_\star(1)}=\dd\big) \,,\, \widehat{\mu}_2^{\CS_\star(1)}=\tfrac\kappa2\,[-\ds-]_{\fru(1)} \,,\, \langle-,-\rangle_{\Omega^\bullet} \Big) \ ,
\end{align}
whose underlying cochain complex is just the de~Rham complex of differential forms on $\FR^{1,2}$. 

As the notation suggests, we regard this noncommutative gauge theory without colour degrees of freedom itself as a double copy of a field theory with colour symmetry. This is analogous to the rank one limits $N=\bar N=1$ of the binoncommutative biadjoint scalar theory of \S\ref{sub:the_other_single_copy}, which results in the homotopy double copy theory of \S\ref{sub:DCad}. The two equivalent twisted factorizations \eqref{eq:CSfact1} and \eqref{eq:CS as double copy} offer two (equivalent) perspectives on the origins of this double copied theory, which corroborates the old suggestions, discussed in \cref{sec:introduction}, that noncommutative $\sU(1)$ gauge theories realise models of gravity in certain senses, despite the absence of propagating spin two states. 

The twisted factorisation \eqref{eq:CS as double copy} for $N=1$ is given by
\begin{align}
\widehat{\ChS_\star(1)} = \Kin_{\wedge^\bullet} \otimes_{\tau^{\Omega^\bullet}} \frC_{\rm \fru_\star(1)}  \ .
\end{align}
We use the factorization \eqref{eq:CCCbiadstarfact}, together with the interpretation of the Lie algebra $\fru_\star(1)$ as a kinematical factor with vector space $\Kin_{\wedge^0}=\FR$ from \S\ref{sub:adjoint_scalar_theory_as_a_strict_linf}, to write the factorisation of the cyclic strict $L_\infty$-algebra of $\sU(1)$ noncommutative Chern--Simons theory as
\begin{align}\label{eq:NCCSfact}
\widehat{\ChS_\star(1)} = \Kin_{\wedge^\bullet} \otimes_{\tau^{\Omega^\bullet}}\big(\Kin_{\wedge^0}\otimes_{\tau^{\fru_\star(1)}} \Scal\big) \ .
\end{align}
The underlying cochain complex in this factorization is again given by \cref{eq:twistedCScomplex}.
Comparing \cref{eq:adfact,eq:NCCSfact}, and recalling the colour-kinematics duality of \S\ref{sub:adjointCKduality}, we conclude that the noncommutative $\sU(1)$ Chern--Simons theory is a double copy of the $d=3$ adjoint scalar theory of \S\ref{sub:adjoint_scalar_theory_as_a_strict_linf} for any colour algebra $\frg$.

To better understand this statement, we note that the construction of \S\ref{sub:factNCCS} shows that \emph{ordinary} Chern--Simons gauge theory based on a quadratic Lie algebra $\frg$ is organised by a cyclic strict $L_\infty$-algebra $\ChS$ which admits the factorization
\begin{align}\label{eq:CSfact}
\ChS = \frg\otimes\big(\Kin_{\wedge^\bullet} \otimes_{\tau^{\Omega^\bullet}} \Scal \big) \ ,
\end{align}
because the colour-stripping is not twisted in the commutative case.
Instead of the double copy \eqref{eq:DCdgLieadstar} of the adjoint scalar theory with itself that we considered in \S\ref{sub:DCad}, we can take its double copy with commutative Chern--Simons theory for any gauge algebra $\frg$, which involves the kinematical vector space $\Kin_{\wedge^\bullet}$ and leads to the double copy prescription of \cref{eq:NCCSfact}. 

We denote this double copy operation symbolically by
\begin{align}\label{eq:ASCS}
\text{$\fru_\star(1)$ Chern--Simons} \ = \ \text{Adjoint Scalar} \ \otimes \ \text{Ordinary Chern--Simons}
\end{align}
with reducible fields $A_\mu^{\fru_\star(1)} = \phi\otimes A_\mu^\CS$. The twisted tensor product construction of \cref{eq:NCCSfact} combines the kinematic numerators of the adjoint scalar theory, based on the Lie algebra $\fru_\star(1)$, with those of commutative Chern--Simons theory, based on the Lie algebra $\mathfrak{diff}_{\text{vol}}(\FR^{1,2})$.

Alternatively, again with the understanding of $\fru_\star(1)$ as a kinematical Lie algebra analogously to \S\ref{sub:adjoint_scalar_theory_as_a_strict_linf}, the twisted factorisation \eqref{eq:CSfact1} for $N=1$ is given by 
\begin{align}\label{eq:NCCSfact1}
\widehat{\ChS_\star(1)} = \Kin_{\wedge^0} \otimes_{\tau^{\fru_\star(1)}}\big(\Kin_{\wedge^\bullet}\otimes_{\tau^{\Omega^\bullet}} \Scal\big) \ .
\end{align}
Comparing \cref{eq:CSfact,eq:NCCSfact1}, and recalling the colour-kinematics duality of \S\ref{sub:CKNCCS}, we conclude now that the noncommutative $\sU(1)$ Chern--Simons theory is a double copy of the commutative Chern--Simons theory for any gauge algebra $\frg$ with the adjoint scalar theory. This perspective is very natural, as the propagators in the commutative and noncommutative gauge theories are the same while the three-point interaction vertex is copied as 
\begin{align}
-\ii\,g\, \epsilon_{\mu\nu\rho} \, f^{abc} \longrightarrow -\tfrac{\ii\,\kappa}2\, \epsilon_{\mu\nu\rho} \, F(k,p,q)
\end{align}
in terms of the structure constants of the Lie algebras $\frg$ and $\fru_\star(1)$. As the twisted tensor product operations commute, i.e.~formally \smash{$\Kin_{\wedge^\bullet} \otimes_{\tau^{\Omega^\bullet}} \Kin_{\wedge^0}\otimes_{\tau^{\fru_\star(1)}} = \Kin_{\wedge^0} \otimes_{\tau^{\fru_\star(1)}} \Kin_{\wedge^\bullet}\otimes_{\tau^{\Omega^\bullet}}$}, the two double copy operations lead to the same theory \eqref{eq:ASCS}, as expected on heuristic grounds.

\paragraph{Diffeomorphism invariance.}

The rank one limit of the brackets \eqref{eq:BGCSproduct} is given by
\begin{equation}
\begin{split}
    \llbracket J(w_1) , J(w_2)\rrbracket^{\star}_{\CS} = [\varepsilon_{p_{w_1}}, \varepsilon_{p_{w_2}}]_{\mathfrak{diff}(\FR^{1,2})}  \otimes 2\, \ii\, \sin \big(\tfrac{t}{2}\, p_{w_1} \cdot \theta\, p_{w_2}\big) \, e_{p_{w_3}} \ \delta(p_{w_{1}} + p_{w_{2}} - p_{w_{3}}) \ .
\end{split}
\end{equation}
This infinite-dimensional Lie algebra is generated by the divergence-free vector fields
\begin{equation}
    \varepsilon_\mu (p) = \e^{\,\ii\, p\cdot x} \, \Pi_{\mu \nu}(p)\, \partial^\nu
\end{equation}
for a transverse projection tensor $\Pi_{\mu \nu}(p)$ with $\Pi_{\mu \nu}(p)\, p^\mu = 0$.
Their deformed Lie bracket is
\begin{equation}
\begin{split}
    [\varepsilon_\mu(p_1) \stackrel{\star}{,} \varepsilon_\nu(p_2)]_{\fru(1)} = \int\,\dd^3p_3 \ 
    {\tt F}_{\mu \nu}{}^\rho(p_1, p_2, p_3) \ \varepsilon_\rho(p_3)  \ ,
\end{split}
\end{equation}
where
\begin{equation}
\begin{split}
    {\tt F}_{\mu \nu }{}^\rho(p_1, p_2, p_3) &=2\, \ii\, \sin\big(\tfrac{t}{2}\, p_{1} \cdot \theta\, p_{2}\big)\, \Pi_{\mu \mu'}(p_1) \, \epsilon^{\mu' \nu' \rho} \, \Pi_{\nu' \nu}(p_2) \  \delta(p_{{1}} + p_{{2}} - p_{{3}}) \ .
\end{split}
\end{equation}
This is the twisted tensor product of the Lie algebra of infinitesimal volume-preserving diffeomorphisms of $\FR^{1,2}$ with the abelian Lie algebra $\fru(1)$.
In the semi-classical limit, it is the Poisson--Lie algebra of symplectic diffeomorphisms of $\FR^{1,2}$.

In the commutative case, a double copy theory which is mapped from a gauge theory should be invariant under diffeomorphisms.
Here we find that $\mathfrak{u}_\star(1)$ Chern--Simons theory, interpreted as a double copied theory, realises  a subalgebra of $\mathfrak{diff}_{\text{vol}}(\mathbbm{R}^{1,2})$ consisting of deformed symplectic diffeomorphisms of spacetime, in the sense of~\cite{Lizzi:2001nd}.
This vindicates our interpretation of the $\mathfrak{u}_\star(1)$ Chern--Simons theory as a genuine double copy, but with reduced diffeomorphism symmetry at the amplitude level. 

\section{Noncommutative Yang--Mills theory} 
\label{sec:klt_with_b_field}

\subsection{The $L_\infty$-structure of noncommutative Yang--Mills theory} 
\label{sub:noncommutative_yang_mills}

Consider the quadratic Lie algebra $\big(\mathfrak{u}(N), [- , -]_{\mathfrak{u}(N)}, \mathrm{Tr}_{\mathfrak{u}(N)} \big)$ with the usual normalisation of the trace of generators $\mathrm{Tr}_{\fru(N)}(T^a\, T^b) = \delta^{ab}$. We work on $d$-dimensional Minkowski spacetime $\FR^{1,d-1}$, with Hodge duality operator denoted $\ast_{\textrm {\tiny H}} : \Omega^{p}(\mathbbm{R}^{1,d-1}) \longrightarrow \Omega^{d-p}(\mathbbm{R}^{1,d-1})$. The corresponding codifferential $\delta: \Omega^{p}(\mathbbm{R}^{1,d-1}) \longrightarrow \Omega^{p-1}(\mathbbm{R}^{1,d-1})$ is given by $\delta = (-1)^{d\,p+1}\,\ast_{\textrm {\tiny H}}\, \mathrm{d}\, \ast_{\textrm {\tiny H}}$.

The classical action functional for standard $\sU(N)$ noncommutative Yang--Mills theory on $\mathbbm{R}^{1,d-1}$ is given by 
\begin{equation}\label{eq:ncym classical action}
    S^\star_{\textrm{\tiny YM}}[A] = \frac{1}{2}\, \int\, \mathrm{Tr}_{\mathfrak{u}(N)}\big(F_A^\star \wedge_\star \ast_{\textrm{\tiny H}}\, F_A^\star\big) \ ,
\end{equation}
with gauge field $A \in \Omega^1(\mathbbm{R}^{1,d-1}, \mathfrak{u}(N))$ and noncommutative field strength $F_A^\star \in \Omega^2(\mathbbm{R}^{1,d-1}, \mathfrak{u}(N))$ given by \cref{eq:fieldstrength}, where $g$ is the Yang--Mills coupling constant. The noncommutative Yang--Mills functional is invariant under infinitesimal star-gauge transformations $\delta^\star_c A = \mathrm{d} c + g\, [c \stackrel{\star}{,} A]_{\mathfrak{u}(N)}$ with $c \in \Omega^0(\mathbbm{R}^{1,d-1}, \mathfrak{u}(N))$.

This noncommutative gauge theory is organised into a cyclic $L_\infty$-algebra 
\begin{equation}\label{eq:NCYMLinfty}
    \mathfrak{YM}_\star = \Big(\mathsf{Ch}(\mathfrak{YM}_\star)\,,\,\mu_2^{\star \ym}\,,\, \mu_3^{\star \ym}\,,\, \langle -,- \rangle_{\ym}^\star \Big) \ .
\end{equation}
The underlying cochain complex $\Ch(\mathfrak{YM_\star}) $ is given by
\begin{subequations}\label{eq:ym bracekts}
\begin{align} \label{eq:ChYMstar}
\begin{split}
    \Omega^0(\mathbbm{R}^{1,d-1}, \mathfrak{u}(N)) \xrightarrow{ \ \dd \ }
    \Omega^1(\mathbbm{R}^{1,d-1},& \mathfrak{u}(N))[-1] \\ & \xrightarrow{ \ \delta\, \mathrm{d} \ } 
    \Omega^1(\mathbbm{R}^{1,d-1}, \mathfrak{u}(N))[-2] \xrightarrow{ \ \delta \ }
    \Omega^0(\mathbbm{R}^{1,d-1}, \mathfrak{u}(N))[-3]    
     \ ,
\end{split}
\end{align}
which identifies the differential as $\mu_1^{\star\textrm{\tiny YM}}(c)=\dd c$ on ghosts $ c $ in $ \Omega^0(\mathbbm{R}^{1,d-1}, \mathfrak{u}(N))$, as $\mu_1^{\star\textrm{\tiny YM}}(A)=\delta\,\dd A$ on gauge fields $A $ in $ \Omega^1(\mathbbm{R}^{1,d-1}, \mathfrak{u}(N))[-1]$, and as $\mu_1^{\star\textrm{\tiny YM}}(A^+)=\delta A^+$ on antifields  $A^+ $ in $ \Omega^1(\mathbbm{R}^{1,d-1}, \mathfrak{u}(N))[-2]$. 

Together with the ghost antifields $c^+ \in \Omega^0(\mathbbm{R}^{1,d-1}, \mathfrak{u}(N))[-3]$,  the non-zero $2$-brackets are given by
\begin{equation} \label{eq:nonstrictmu2}
\begin{split}    
    \mu_2^{\star \textrm{\tiny YM}}(c_1, c_2) =  g\,[c_1 \stackrel{\star}{,} c_2]_{\mathfrak{u}(N)}  \ , &\
    \mu_2^{\star \textrm{\tiny YM}}(c,A) =  g\,[c \stackrel{\star}{,} A]_{\mathfrak{u}(N)}\ , \
    \mu_2^{\star \textrm{\tiny YM}}(c, A^+) = g\,[c \stackrel{\star}{,} A^+]_{\mathfrak{u}(N)}  \ , \\[4pt]
    \mu_2^{\star \textrm{\tiny YM}}(c, c^+) = g\,[c \stackrel{\star}{,} c^+]_{\mathfrak{u}(N)}\ , & \
    \mu_2^{\star \textrm{\tiny YM}}(A, A^+) = g\,[A \stackrel{\star}{,} A^+]_{\mathfrak{u}(N)}\ ,  \\[4pt]
    \mu_2^{\star \textrm{\tiny YM}}(A_1, A_2) =  g\,\big(\delta\,[A_1 &\stackrel{\star}{,} A_2]_{\mathfrak{u}(N)} 
    + \ast_{\textrm{\tiny H}}\,[A_1 \stackrel{\star}{,} \ast_{\textrm{\tiny H}}\, \mathrm{d} A_2]_{\mathfrak{u}(N)}
    + \ast_{\textrm{\tiny H}}\,[\ast_{\textrm{\tiny H}}\, \mathrm{d} A_1 \stackrel{\star}{,} A_2]_{\mathfrak{u}(N)}\big) \ .
\end{split}
\end{equation}
The single non-zero higher bracket $\mu_3^{\star \textrm{\tiny YM}} : \Omega^1(\mathbbm{R}^{1,d-1}, \mathfrak{u}(N)))[-1]^{\otimes 3} \longrightarrow \Omega^1(\mathbbm{R}^{1,d-1}, \mathfrak{u}(N))[-3]$ acts as
\begin{equation}
\begin{split}\label{eq:ym 3bracket}
    \mu_3^{\star \textrm{\tiny YM}} (A_1, A_2, A_3) &=  g^2\, \big(
    \ast_{\textrm{\tiny H}} [A_1 \stackrel{\star}{,} \ast_{\textrm{\tiny H}}\, [A_2 \stackrel{\star}{,} A_3]_{\mathfrak{u}(N)}]_{\mathfrak{u}(N)}
    + \ast_{\textrm{\tiny H}}\, [A_2 \stackrel{\star}{,} \ast_{\textrm{\tiny H}}\, [A_3 \stackrel{\star}{,} A_1]_{\mathfrak{u}(N)}]_{\mathfrak{u}(N)}\\
    & \hspace{6cm}+ \ast_{\textrm{\tiny H}}\, [A_3 \stackrel{\star}{,} \ast_{\textrm{\tiny H}}\, [A_1 \stackrel{\star}{,} A_2]_{\mathfrak{u}(N)}]_{\mathfrak{u}(N)}\big) \ .
\end{split}
\end{equation}

The cyclic structure of $L_\infty$-degree $-3$ is given by the Hodge inner product of differential forms in the same exterior degree:
\begin{equation}
    \langle \alpha, \alpha^+ \rangle_{\textrm{\tiny YM}}^\star = \int\, \mathrm{Tr}_{\mathfrak{u}(N)}( \alpha \wedge_\star \ast_{\textrm{\tiny H}}\,\alpha^+) = \int\, \mathrm{Tr}_{\mathfrak{u}(N)}( \alpha \wedge \ast_{\textrm{\tiny H}}\,\alpha^+)\ ,
\end{equation}
for $\alpha\in\{c,A\}$.
\end{subequations}

Applying the Maurer--Cartan theory from \S\ref{sub:preliminaries} for the cyclic $L_\infty$-algebra \eqref{eq:NCYMLinfty}, the Maurer--Cartan curvature $f_A^{\star\ym} \in  \Omega^1(\mathbbm{R}^{1,d-1}, \mathfrak{u}(N))[-2]$ from \cref{eq:Maurer Cartan equation} reads
\begin{equation}\label{eq:MCNCYM eom}
    f^{\star\ym}_A = \mu_1^{\star \textrm{\tiny YM}}(A) + \tfrac{1}{2!}\, \mu_2^{\star \textrm{\tiny YM}}(A,A) + \tfrac{1}{3!}\, \mu_3^{\star \textrm{\tiny YM}}(A,A,A) =  -\ast_{\textrm{\tiny H}}\, \nabla_A^\star \ast_{\textrm{\tiny H}} F^\star_A  \ ,
\end{equation}
where $\nabla_A^\star : \Omega^p(\mathbbm{R}^{1,d-1}, \mathfrak{u}(N)) \longrightarrow \Omega^{p+1}(\mathbbm{R}^{1,d-1}, \mathfrak{u}(N))$ is the covariant derivative given by \cref{eq:covariantderivative}. 
The Maurer--Cartan equation $f^{\star\ym}_A = 0$ is therefore equivalent to the noncommutative Yang--Mills equations $\nabla^\star_A \ast_{\textrm{\tiny H}} F^\star_A = 0$. 

The Maurer--Cartan--Bianchi identity in  $\Omega^0(\mathbbm{R}^{1,d-1}, \mathfrak{u}(N))[-3]$  is the Noether identity for the star-gauge symmetry $\delta_c^{\star}A = \mu_1^{\star\textrm{\tiny YM}}(c) + \mu_2^{\star\textrm{\tiny YM}}(c,A)$ of noncommutative Yang--Mills theory, which is satisfied off-shell. It expresses the fact that the Maurer--Cartan curvature $f_A^{\star\ym}$ is covariantly constant for the Maurer--Cartan covariant derivative
\begin{align}
\begin{split}
\dd_A^{\star\ym}f_A^{\star\ym} &= \mu_1^{\star\ym}(f_A^{\star\ym}) + \mu_2^{\star\ym}(A,f_A^{\star\ym}) \\[4pt]
&= -\ast_{\textrm{\tiny H}}\,\nabla^\star_A \ast_{\textrm{\tiny H}} f^{\star\ym}_A = \ast_{\textrm{\tiny H}}\, (\nabla^\star_A)^2 \ast_{\textrm{\tiny H}} F^\star_A = \ast_{\hodge}\, [F_A^\star\ds\ast_\hodge\, F_A^\star]_{\fru(N)} =  0 \ ,
\end{split}
\end{align}
which follows directly from the symmetry properties of the four-form $[F_A^\star\ds\ast_\hodge F_A^\star]_{\fru(N)}$ in the Lie algebra $\fru_\star(N)$. 

Finally, the Maurer--Cartan functional
\begin{equation}
\begin{split}\label{eq:NCYM full action}
    S^\star_{\textrm{\tiny YM}}[A] &= \tfrac{1}{2}\, \big\langle A ,\mu_1^{\star\ym}(A) \big\rangle^\star_{\textrm{\tiny YM}}
    + \tfrac{1}{3!} \, \big\langle
    A,\mu_2^{\star\ym}(A,A)  \big\rangle^\star_{\textrm{\tiny YM}}
     + \tfrac{1}{4!}\, \big\langle 
    A,\mu_3^{\star\ym}(A,A,A)\big\rangle^{\star}_{\textrm{\tiny YM}}
\end{split}
\end{equation}
recovers the noncommutative Yang--Mills functional from \cref{eq:ncym classical action}.

\paragraph{Batalin--Vilkovisky formalism.}

Following the same steps as in \cref{sub:noncommutative_chern_simons}, the superspace extension of the noncommutative Yang--Mills functional \eqref{eq:ncym classical action} yields the BV action functional\footnote{See e.g.~\cite{Martin:2020ddo} for the BRST extension including the antighost of the ghost field $c$ and the Nakanishi--Lautrup auxiliary field.}
\begin{align}
S_{\textrm{\tiny BV}}^\star[\mbf{ A}] = \int \, \Tr_{\fru(N)}\Big(\frac12\,F_A^\star\wedge_\star\ast_{\textrm{\tiny H}}\, F^\star_A + A^+\wedge_\star\ast_{\textrm{\tiny H}}\, \nabla_A^\star\, c + \frac g2\,c^+\wedge_\star\ast_{\textrm{\tiny H}}\, [c\ds c]_{\fru(N)} \Big) \ .
\end{align}
In contrast to the noncommutative Chern--Simons theory of \cref{sec:cs}, the noncommutative deformation here corresponds to a non-trivial local BRST cohomology class and so it is non-trivial~\cite{Barnich:2002tz,Barnich:2003wq}: the $L_\infty$-algebra $\YM_{\star}$ is \emph{not} quasi-isomorphic to the $L_\infty$-algebra $\YM$ of ordinary Yang--Mills theory. This explains why the double copy construction of \cref{sec:cs} was so straightforward; for noncommutative Yang--Mills theory we have to work harder.

\paragraph{Colour ordering and decomposition.} 
\label{par:colour_ordering}

A common way to organise $n$-point amplitude calculations in gauge theories is by  summing over inequivalent orderings of colour factors, which yields  reduction formulas from $n!$ down to $(n-1)!$ inequivalent amplitudes. Similarly to the biadjoint scalar theory from \cref{sub:biadjoint}, tree-level scattering amplitudes of gluons are organised in terms of gauge-invariant partial amplitudes and colour structures as~\cite{Mangano:1990by}
\begin{align}
\CCA^{\textrm{\tiny YM}}_n(p,\zeta,a) = \sum_{\sigma\in S_n/\RZ_n} \,  \mathrm{Tr}_{\fru(N)}(T^{a_{\sigma(1)}} \cdots T^{ a_{\sigma(n)}}) \ \CA^{\textrm{\tiny YM}}_n\big( \sigma(1), \dots, \sigma(n)\big) \ .
\end{align}
Each Yang--Mills partial amplitude appears twice due to the ordering properties
\begin{align} \label{eq:YMorderingprops}
\CA^{\textrm{\tiny YM}}_n(1,\dots,n-1,n) = \CA^{\textrm{\tiny YM}}_n(n,1,\dots,n-1) =(-1)^n \, \CA^{\textrm{\tiny YM}}_n(n,n-1,\dots,1) \ ,
\end{align}
which enables one to restrict the sum to planar orderings as
\begin{align}
\CCA^{\textrm{\tiny YM}}_n(p,\zeta,a) = \sum_{\sigma\in S_n/\RZ_n\rtimes\RZ_2} \,  C(\sigma) \ \CA^{\textrm{\tiny YM}}_n\big( \sigma(1), \dots, \sigma(n) \big) \ .
\end{align}

In~\cite{Raju:2009yx,Huang:2010fc} it is shown that noncommutative gauge theories still respect the colour decomposition.
As we have seen already in \cref{eq:colour ordering structure}, the structure constants $F^{abc}(p_1,p_2,p_3)$ of $\fru_\star(N)$ can be decomposed to give the three-point vertex for noncommutative Yang--Mills theory in the form
\begin{subequations}
\begin{equation}
\begin{split}
\begin{tikzpicture}[scale = 0.8, line width=0.2 pt]
{\small
    \draw[gluon] (-140:1.8)node[below]{$\nu,b$}--(0,0);
    \draw[gluon] (140:1.8)node[above]{$\mu,a$}--(0,0);
    \draw[gluon] (0:1.8)node[right]{$\rho, c$} --(0,0);
    \draw[->, shift={(-0.25,0.6)}] (140:.5) node[above]{$p_1$} -- (0,0);
    \draw[->, shift={(-0.25,-0.6)}] (-140:.5) node[below]{$p_2$} -- (0,0);
    \draw[->, shift={(0.29,0.3)}] (0:.5) node[above]{$p_3$} -- (0,0);
    } \normalsize
    \node at (9.7, 0) {$= \ \mathrm{i}\,g\, \big(\e^{- \frac{\ii\,t}{2}\, p_1 \cdot \theta\, p_2}\, \mathrm{Tr}_{\fru(N)}(T^a\, T^b\, T^c) - 
    \e^{- \frac{\ii\,t}{2}\, p_2 \cdot \theta\, p_1}\, \mathrm{Tr}_{\fru(N)}(T^b\, T^a\, T^c)
    \big)$};
    \node at (11, -1) {$\times \big(  (p_1 - p_2)^\rho\, \eta^{\mu \nu} + (p_2-p_3)^\mu\, \eta^{\nu \rho} + (p_3 -p_1)^\nu\, \eta^{\rho \mu} \big) \ , $};
\end{tikzpicture}
\end{split}
\end{equation}
%
where for simplicity we suppress momentum conserving delta-functions.
Note that the two summands are considered as parts of different colour ordered amplitudes.
Similarly, the four-point vertex can be decomposed into a sum over inequivalent orderings with phase factors dressing the summands as
\begin{equation}
\begin{split}
\begin{tikzpicture}[scale=0.8, line width = 0.2 pt]
{\small
    \draw[gluon] (-135:1.8)node[left]{$\nu,b$}--(0,0);
    \draw[gluon] (135:1.8)node[left]{$\mu,a$}--(0,0);
    \draw[gluon] (-45:1.8)node[right]{$\rho, c$} --(0,0);
    \draw[gluon] (45:1.8)node[right]{$\sigma, d$} --(0,0);
    \draw[->, shift={(-0.25,0.7)}] (140:.5) node[above]{$p_1$} -- (0,0);
    \draw[->, shift={(-0.25,-0.7)}] (-140:.5) node[below]{$p_2$} -- (0,0);
    \draw[->, shift={(0.29,-0.7)}] (-45:.5) node[below]{$p_3$} -- (0,0);
    \draw[->, shift={(0.29,0.7)}] (45:.5) node[above]{$p_4$} -- (0,0);
    \node at (8,1.5) {$ \qquad g^2\,\big(
    - \e^{ - \frac{\ii\,t}{2}\, (p_1 \cdot \theta\, p_2 + p_1 \cdot \theta\, p_3 + p_2 \cdot \theta\, p_3)}\, \mathrm{Tr}_{\fru(N)}(T^a\, T^b\, T^c\, T^d) $};
    \node at (8.8,0.7) {$
    + \, \e^{ \frac{\,\ii\,t}{2}\,( p_1 \cdot \theta\, p_2 + p_1 \cdot \theta\, p_4 + p_2 \cdot \theta\, p_4)}\, \mathrm{Tr}_{\fru(N)}(T^a\, T^b\, T^d\, T^c) 
    $};
    \node at (8.1,0) {$ = \hspace{1cm}
    + \, \e^{ - \frac{\ii\,t}{2}\,( p_2 \cdot \theta\, p_1 + p_2 \cdot \theta\, p_3 + p_1 \cdot \theta\, p_3)}\, \mathrm{Tr}_{\fru(N)}(T^b\, T^a\, T^c\, T^d) $};
    \node at (8.95,-0.7) {$
    - \, \e^{ \,\frac{\ii\,t}{2}\,( p_2 \cdot \theta\, p_1 + p_2 \cdot \theta\, p_4 + p_1 \cdot \theta\, p_4)}\, \mathrm{Tr}_{\fru(N)}(T^b\, T^a\, T^d\, T^c) 
    \big)$};
    \node at (9, -1.6) {$ \times \, ( \eta^{\mu \sigma}\, \eta^{\nu \rho} - \eta^{\mu \rho}\, \eta^{\nu \sigma}) \ + \  \Big({\scriptsize\begin{matrix} \nu\longleftrightarrow\rho \\[-0.5ex] b\longleftrightarrow c\\[-0.5ex] p_2\longleftrightarrow p_3 \end{matrix}}\Big)   \ + \  \Big({\scriptsize\begin{matrix} \rho\longleftrightarrow\sigma \\[-0.5ex] c\longleftrightarrow d\\[-0.5ex] p_3\longleftrightarrow p_4 \end{matrix}}\Big) \ . $};
    } \normalsize
\end{tikzpicture}
\end{split}
\end{equation}
\end{subequations}

Since tree-level amplitudes involve only planar diagrams, colour ordered amplitudes in noncommutative Yang--Mills theory factorise with the appropriate phase factor \cite{Raju:2009yx,Huang:2010fc}.
Given an ordered word $w=k_1\cdots k_{n-1}k_n \in \CCW _n $ labelling colour, polarisations and momenta, for rank $N>1$ one finds that the noncommutative $n$-point Yang--Mills partial amplitudes $\CA_n^{\star\ym}(w)$ are related to the \emph{commutative} $n$-point partial amplitudes $\CA_n^\ym(w)$ through the simple relation
\begin{align}
    \CA^{\star\ym}_n(w) &= \varTheta_n(w) \ \CA_n^{\ym}(w) \ ,
\end{align}
with the overall momentum-dependent phase factor
\begin{align}
    \varTheta_n(w) &= \exp\bigg(- \frac{\ii\,t}{2}\, \sum_{i<j}\, p_{k_i} \cdot\theta\, p_{k_j}\bigg) \ .
    \label{eq:phase factor}
\end{align}
Due to antisymmetry of the bivector $\theta$, the phase is cyclically invariant: $\varTheta_n(w) = \varTheta_n(w^\circlearrowleft)$ where $w^\circlearrowleft:=k_nk_1\cdots k_{n-1}\in\CCW_n$, and symmetric under Kleiss--Kuijf reflection:
$\varTheta_n(w) = \varTheta_n( \bar w)^{-1}$ where $\bar w =k_nk_{n-1}\cdots k_1\in\CCW_n $. The ordering properties \eqref{eq:YMorderingprops} then imply
\begin{align}
 \CA^{\star\ym}_n(w) =  \CA^{\star\ym}_n(w^\circlearrowleft) = (-1)^n \, \varTheta_n( \bar w)^{-2} \ \CA^{\star\ym}_n(\bar w) \ .
\end{align}

This argument will be corroborated later on when we embed the gauge theory as a low energy limit of open string theory with a constant background $B$-field.
It suggests that a twisted homotopy factorisation is possible for noncommutative Yang--Mills theory.

\subsection{Strictification and twisted homotopy factorisation} 
\label{sub:strictification_of_ym}

\paragraph{Strictification in the second order formalism.} 
\label{par:strictification_}

Just like ordinary Yang--Mills theory, because of the non-trivial $3$-bracket in \cref{eq:ym 3bracket}, the $L_\infty$-structure of noncommutative Yang--Mills theory is \emph{not} strict.
  The twisted homotopy factorisation, and hence the homotopy double copy construction, only makes sense for strict $L_\infty$-algebras. This means finding a perturbatively equivalent theory with no vertices of order higher than three. 

If one further wishes to double copy this theory with itself, a cubic Lagrangian whose Feynman diagrams produce tree-level gluon amplitudes in
twisted colour-kinematics dual form must be sought; in the commutative case this is always possible~\cite{Reiterer:2019dys}.
This is done order by order in the multiplicity, by finding quasi-isomorphisms between the original theory and a strictified theory (organised by a strict $L_\infty$-algebra) that has manifest colour-kinematics duality. It extends the underlying cochain complex $\Ch(\YM_\star)$ to include auxiliary fields that enforce both the strict $L_\infty$-structure and the colour-kinematics duality, such that putting the auxiliary fields on-shell yields the higher order vertex. 

We generalise the arguments of \cite{Borsten:2021hua} to construct a strictification of noncommutative Yang--Mills theory that produces twisted colour-kinematics dual gluon amplitudes up to multiplicity four. For this, we introduce an auxiliary field $G \in \Omega^{1,2}(\mathbbm{R}^{1,d-1}, \mathfrak{u}(N))$, where 
\begin{align}
\Omega^{1,2}(\mathbbm{R}^{1,d-1}) := \Omega^{1}(\mathbbm{R}^{1,d-1}) \otimes_{\Omega^{0}(\mathbbm{R}^{1,d-1})}\Omega^{2}(\mathbbm{R}^{1,d-1}) \ .
\end{align}
Written in component form, the cubic noncommutative Yang--Mills functional we consider is
\begin{align} \label{eq:actionYM2}
\begin{split}
S_{\ym_2}^\star[A,G] &= \frac12\,\int\,\dd^dx \ \Tr_{\fru(N)}\big(A_\mu\star\square\, A^\mu  -2\,g\, A_\nu\star\partial_\mu[A^\mu \stackrel{\star}{,} A^\nu]_{\fru(N)} \\
& \hspace{4cm} + \sqrt2\,g 
    \, \partial_{\bar\alpha} ([A_\mu \stackrel{\star}{,} A_\nu]_{\fru(N)}) \star  G^{\bar\alpha \mu \nu} 
    - G_{\bar\alpha \mu \nu} \star \square\, G^{\bar\alpha \mu \nu} \big) \ .
\end{split}
\end{align}

The resulting noncommutative gauge theory is perturbatively equivalent to the original theory after integration over the auxiliary field $G$, whose equation of motion is
\begin{align} \label{eq:eomG}
\square\, G_{\bar\alpha \mu \nu} = \tfrac{1}{2\, \sqrt{2}}\, g\, \partial_{\bar\alpha} [A_\mu \stackrel{\star}{,} A_\nu]_{\fru(N)} \ .
\end{align}
Substituting \cref{eq:eomG} into \cref{eq:actionYM2} gives back the noncommutative Yang--Mills functional \eqref{eq:NCYM full action}, because of the ${\rm ad}(\fru_\star(N))$-invariance of the cyclic structure.
This defines an $L_\infty$-quasi-isomorphism \smash{$\psi^{\star\ym_2}:\YM_\star\longrightarrow\YM_{\star 2}$} whose non-vanishing components on gauge fields is given by
\begin{align}
\psi_1^{\star\ym_2}(A) = A \qquad \mbox{and} \qquad \psi_2^{\star\ym_2}(A_1,A_2) = \frac g{2\,\sqrt2 \ \square} \ \dd\otimes[A_1\ds A_2]_{\fru(N)} \ ,
\end{align}
so that $G = \psi^{\star\ym_2}_2(A,A)$. 

The new cyclic strict $L_\infty$-algebra is
\begin{align}
 \mathfrak{YM}_{\star 2} = \Big(\mathsf{Ch}(\mathfrak{YM}_{\star 2}) \,,\, \mu_2^{\star \ym_2} \,,\, \langle -,- \rangle_{\ym_2}^\star \Big) \ .
\end{align}
In the following we focus only on its subspaces in degrees~$1$ and~$2$ for illustration, and drop the antighost as well as various auxiliary fields including the Nakanishi--Lautrup field. Its cochain complex $\mathsf{Ch}(\YM_{\star2}) $ extends the four-term Yang--Mills cochain complex \eqref{eq:ChYMstar} as 
\begin{equation}\label{eq:ChYM2}
  \begin{tikzcd}[row sep=0ex]
     \cdots \arrow{r} &   \Omega^1(\mathbbm{R}^{1,d-1}, \mathfrak{u}(N))[-1] \arrow["\delta\,\dd"]{rr}& & \Omega^1(\mathbbm{R}^{1,d-1}, \mathfrak{u}(N))[-2] \arrow{r} & \cdots \\
      &  \oplus & & \oplus & \\
       \cdots \arrow{r} & \Omega^{1,2}(\mathbbm{R}^{1,d-1},\mathfrak{u}(N))[-1] \arrow["-\delta\,\dd\,\otimes\,\ident"]{rr}& & \Omega^{1,2}(\mathbbm{R}^{1,d-1}, \mathfrak{u}(N))[-2] \arrow{r} & \cdots
    \end{tikzcd} 
\end{equation}
with $2$-bracket modified to 
\begin{equation}\label{eq:strict 2-bracket deg 1}
\begin{split}
&  \mu_2^{\star\ym_2}\Big({\small \bigg(\begin{matrix} A_1 \\ G_1\end{matrix}\bigg) , \bigg(\begin{matrix} A_2 \\ G_2\end{matrix}\bigg) } \normalsize \Big) \\[4pt]
& \hspace{1cm} = 
    \begin{pmatrix}
    \mu_2^{\star \ym}(A_1, A_2) + \sqrt2\, g \, \ast_{\textrm {\tiny H}}\big([A_1 \stackrel{\star}{,} \ast_{\textrm {\tiny H}}(\delta\otimes\ident)\, G_2]_{\fru(N)} + [A_2 \stackrel{\star}{,} \ast_{\textrm {\tiny H}}(\delta\otimes\ident)\, G_1]_{\fru(N)} \big) \\[1ex]  \sqrt{2}\,g\, \mathrm{d}\otimes [A_1 \stackrel{\star}{,} A_2]_{\fru(N)} 
    \end{pmatrix} \ .
\end{split}
\end{equation}
The cyclic structure is extended by
\begin{equation}
\begin{split}\label{eq:auxiliary structure}
\langle G, G^+  \rangle_{\ym_2}^\star = \int \, \dd^dx \ \mathrm{Tr}_{\fru(N)}\big(G_{\bar\alpha \mu \nu}\star G^{+ \bar\alpha \mu \nu}\big) \ ,
\end{split}
\end{equation}
for $G\in\Omega^{1,2}(\mathbbm{R}^{1,d-1}, \mathfrak{u}(N))[-1]$ and $G^+\in\Omega^{1,2}(\mathbbm{R}^{1,d-1}, \mathfrak{u}(N))[-2]$. 

We will first factorise the kinematic dependence  from the differential form part, much like we did in \cref{eq:CS as double copy}. For this, we introduce a graded vector space organising the differential forms for this theory through
\begin{equation}
{\small
\mathfrak{Kin}_{\ym_2} \ := \ 
\cdots  \ \oplus \  \begin{matrix} \midwedge^1(\mathbbm{R}^{1,d-1})^* \\[1ex] \oplus \\[1ex] \midwedge^{1}(\mathbbm{R}^{1,d-1})^* \otimes \midwedge^2(\mathbbm{R}^{1,d-1})^* \end{matrix} \  \oplus \  \begin{matrix} \midwedge^1(\mathbbm{R}^{1,d-1})^*[-1] \\[1ex] \oplus \\[1ex] \midwedge^{1}(\mathbbm{R}^{1,d-1})^* \otimes \midwedge^2(\mathbbm{R}^{1,d-1})^*[-1] \end{matrix} \  \oplus \  \cdots \ . } \normalsize
\end{equation}
We write $e^\mu$ for the basis of covectors on $\FR^{1,d-1}$ in degree~$0$ and $e^{+\mu}$ in degree~$1$. We will also abbreviate $e^{\bar\alpha\mu\nu}:=e^{\bar\alpha}\otimes(e^\mu\wedge e^\nu)$ and $e^{+\bar\alpha\mu\nu}:=e^{+\bar\alpha}\otimes(e^{+\mu}\wedge e^{+\nu})$, similarly to our conventions from~\cref{sub:DCCS}. This space carries the non-vanishing inner products
\begin{align}
\langle\!\langle e^\mu,e^{+\nu}\rangle\!\rangle := \eta^{\mu\nu} \qquad \mbox{and} \qquad \langle\!\langle e^{\bar\alpha\mu\nu},e^{+\bar\beta\rho\lambda}\rangle\!\rangle := \tfrac12\,\eta^{\bar\alpha\bar\beta}\, ( \eta^{\mu\rho}\,\eta^{\nu\lambda} - \eta^{\mu\lambda}\,\eta^{\nu\rho}) \ .
\end{align}

Following~\cite{Borsten:2021hua}, we introduce a twist datum \smash{$\tau^{\ym_2}=\big(\tau_1^{\ym_2},\tau_2^{\ym_2}\big)$} for $\fru(N)\otimes L$, where $L$ is the graded vector space \eqref{eq:scalarvectorspace}. Its non-trivial actions on basis vectors  in $\mathfrak{Kin}_{\ym_2}$ are given by
\begin{align}\label{eq:tauYM2}
\begin{split}
    \tau_1^{\ym_2}(e^\mu) &= e^{+\mu}\otimes\mathbbm{1} \quad , \quad \tau_1^{\ym_2}(e^{\bar\alpha \mu \nu})= - e^{+\bar\alpha \mu \nu} \otimes \mathbbm{1} \ , \\[4pt]
    \tau_2^{\ym_2}(e^\mu, e^\nu) &= 3\, \big( e^{+\mu} \otimes (\partial^\nu \otimes \mathbbm{1} + \mathbbm{1} \otimes \partial^\nu) - e^{+\nu} \otimes (\partial^\mu \otimes \mathbbm{1} + \mathbbm{1} \otimes \partial^\mu)\big) \\
    & \quad \, + \sqrt{2} \, {e}^{+\bar\alpha \mu \nu} \otimes ( \partial_{\bar\alpha} \otimes \mathbbm{1} +  \mathbbm{1} \otimes \partial_{\bar\alpha}) \ , \\[4pt]
    \tau_2^{\ym_2}(e^\mu, e^{\bar\alpha \nu \rho}) &= - \tfrac{1}{\sqrt2}\, \big(\eta^{\mu \nu}\, e^{+\rho} \otimes \mathbbm{1} \otimes \partial^{\bar\alpha} - \eta^{\mu \rho}\, e^{+\nu} \otimes \mathbbm{1} \otimes \partial^{\bar\alpha}\big) \ , \\[4pt]
    \tau_2^{\ym_2}(e^{\bar\alpha \nu \rho}, e^\mu) &=  \tfrac{1}{\sqrt2}\, \big(\eta^{\mu \nu}\, e^{+\rho} \otimes  \partial^{\bar\alpha} \otimes \mathbbm{1} - \eta^{\mu \rho}\, e^{+\nu}  \otimes \partial^{\bar\alpha} \otimes \mathbbm{1}\big) \ .
\end{split}
\end{align}

Let $\mathfrak{C}_{\mathfrak{u}_\star(N)}$ be the cyclic strict $C_\infty$-algebra which was introduced in \S\ref{sub:factNCbiadjoint} for the twisted homotopy factorisation of the star-commutator. It can be used to factorise the star-commutators of coordinate functions $A_\mu,G_{\alpha\mu\nu}:\FR^{1,d-1}\longrightarrow\fru(N)$ of the gauge fields and the auxiliary fields. In this way we obtain the twisted homotopy factorisation
\begin{equation}
    \mathfrak{YM}_{\star 2} =  \mathfrak{Kin}_{\ym_2} \otimes_{\tau^{\ym_2}} \mathfrak{C}_{\mathfrak{u}_\star(N)} \ .
\end{equation}

To see this, note that, after twisting against the $C_\infty$-algebra $\mathfrak{C}_{\mathfrak{u}_\star(N)}$, the first line of \cref{eq:tauYM2} easily reproduces the differential \smash{$\mu_1^{\star\ym_2}(A,G)=(\delta\,\dd A,-(\delta\,\dd\otimes\ident)G)$} of \cref{eq:ChYM2} on gauge fields and auxiliary fields. On gauge fields $A_1, A_2 \in \Omega^1(\mathbbm{R}^{1,d-1}, \mathfrak{u}(N))$, the third equality reproduces the non-strict noncommutative Yang--Mills $2$-bracket $\mu_2^{\star \ym}(A_1, A_2)$ from \cref{eq:nonstrictmu2}, as well as the last term proportional to $ \mathrm{d}\otimes [A_1 \stackrel{\star}{,} A_2]_{\fru(N)}$ of the strict $2$-bracket on degree~$1$ fields in \cref{eq:strict 2-bracket deg 1} which lives in \smash{$\Omega^1(\mathbbm{R}^{1,d-1}) \otimes_{\Omega^0(\FR^{1,d-1})} \Omega^2(\mathbbm{R}^{1,d-1}) \otimes \mathfrak{u}(N)$}:
\begin{align}
\begin{split}
    \mu_2^{\tau^{\ym_2}}(e^\mu \otimes A_{1 \mu} \,,\, e^\nu \otimes A_{2 \nu}) &= 
    3\, g\,e^{+\mu} \otimes \big([\partial^\nu A_{1 \mu} \stackrel{\star}{,} A_{2 \nu}]_{\fru(N)}  + [A_{1 \mu} \stackrel{\star}{,} \partial^\nu A_{2 \nu}]_{\fru(N)} \big)\\
    & \quad \, -3\,g\, e^{+\nu} \otimes \big([\partial^\mu A_{1 \mu} \stackrel{\star}{,} A_{2 \nu}]_{\fru(N)}
    + [A_{1 \mu} \stackrel{\star}{,} \partial^\mu A_{2 \nu}]_{\fru(N)} \big) \\
    & \quad \, + \sqrt2\,g \, e^{+\bar\alpha \mu \nu} \otimes
    \big( [\partial_{\bar\alpha} A_{1 \mu} \stackrel{\star}{,} A_{2 \nu}]_{\fru(N)} + [A_{1 \mu} \stackrel{\star}{,} \partial_{\bar\alpha} A_{2 \nu}]_{\fru(N)}\big) \ .
\end{split}
\end{align}
Similarly, the last two equalities of the twist data from \cref{eq:tauYM2} reproduce the mixed terms among $A_i$ and $G_i$ in \cref{eq:strict 2-bracket deg 1}. The cyclic structure $\langle-,-\rangle_{\ym_2}$ is easily seen to be reproduced by $\langle\!\langle-,-\rangle\!\rangle\otimes \Tr_{\fru_\star(N)}$.

Finally, the $C_\infty$-algebra part of the theory has already been factorised in \cref{eq:CCCbiadstarfact} using the twist \eqref{eq:twist with lie derivatives} applied to coordinate functions in $\Omega^0(\FR^{1,d-1},\fru(N))$. Hence we have constructed a twisted colour-kinematics dual factorisation
\begin{equation} \label{eq:YM2fact1}
    \mathfrak{YM}_{\star 2} =  \mathfrak{Kin}_{\ym_2} \otimes_{\tau^{\ym_2}} \big(\fru(N)\otimes_{\tau^{\fru_\star(N)}}\Scal\big) \ .
\end{equation}
This factorisation allows one to compute double copies up to multiplicity four scattering~\cite{Borsten:2021hua} with any theory, regardless of whether its strictification has manifest colour-kinematics duality.

Alternatively, by simply extracting the colour structure first analogously to \cref{eq:CSfact1}, we find that the factorisation \eqref{eq:YM2fact1} commutes with the twist data defined by \cref{eq:tauYM2}. Thus we can write the equivalent twisted homotopy factorisation
\begin{equation}
    \mathfrak{YM}_{\star 2} = \fru(N)\otimes_{\tau^{\fru_\star(N)}}  \big(\mathfrak{Kin}_{\ym_2} \otimes_{\tau^{\ym_2}} \Scal\big) \ .
\end{equation}


\paragraph{First order formalism.} 
\label{par:noncommutative_yang_mills_in_the_first_order_formalism}

In the foregoing discussion we set $d=4$, because a much more tractable way to strictify noncommutative Yang--Mills theory in four dimensions is to rewrite it in the first order formalism. Following~\cite{Benaoum:1999ca}, this reformulates the theory as a non-topological deformation of noncommutative $BF$-theory with only cubic vertices; we will organise it below into a cyclic strict $L_\infty$-algebra akin to that of the noncommutative Chern--Simons theory of \S\ref{sub:noncommutative_chern_simons}.
This strictification is \emph{not} manifestly twisted colour-kinematics dual, and so it cannot be used to double copy itself. However, in this form the theory also admits a twisted homotopy factorisation, which can be used to double copy noncommutative Yang--Mills theory with any other theory that has manifest colour-kinematics duality and compute arbitrary $n$-point scattering amplitudes.

In four spacetime dimensions, the Hodge duality operator defines a complex structure on the vector space $\Omega^{2}(\mathbbm{R}^{1,3}, \mathfrak{u}(N))$, i.e. an endomorphism $\ast_{\textrm {\tiny H}} \!:\! \Omega^{2}(\mathbbm{R}^{1,3}, \mathfrak{u}(N))\! \longrightarrow\! \Omega^{2}(\mathbbm{R}^{1,3}, \mathfrak{u}(N))$, acting solely on the differential form part, such that $\ast_{\textrm {\tiny H}}^2 = -\ident$. The cubic noncommutative Yang--Mills functional for 
an auxiliary $\fru(N)$-valued two-form field $B\in \Omega^2(\mathbbm{R}^{1,3}, \mathfrak{u}(N))$ and a gauge field $A \in \Omega^1(\mathbbm{R}^{1,3}, \mathfrak{u}(N))$ is given by
\begin{equation}\label{eq:ncym1 classical action}
    S^\star_{\ym_1}[A, B] = \int \, \mathrm{Tr}_{\mathfrak{u}(N)}\Big(\frac{1}{2}\, B \wedge_\star \ast_{\textrm {\tiny H}}\,B - B\wedge_\star F^\star_A  \Big) \ ,
\end{equation}
where as usual the corresponding noncommutative field strength is $F_A^\star = \mathrm{d} A + \frac{g}{2}\, [A \stackrel{\star}{,} A]_{\mathfrak{u}(N)}$ in $ \Omega^2(\mathbbm{R}^{1,3}, \mathfrak{u}(N))$. Dropping the metric-dependent term defines noncommutative $BF$-theory, whose restriction to self-dual fields is the action functional of noncommutative self-dual Yang--Mills theory.

The equations of motion for $B$ and $A$ are found by using $\mathrm{ad}(\fru(N))$-invariance of the trace. They respectively read as
\begin{equation} 
    F_A^\star = \ast_{\textrm {\tiny H}}\,B \qquad \textrm{and} 
    \qquad \nabla_A^\star\, B = 0 \ ,
\end{equation}
where $\nabla_A^\star:\Omega^p(\FR^{1,3},\fru(N))\longrightarrow\Omega^{p+1}(\FR^{1,3},\fru(N))$ is the star-gauge covariant derivative \eqref{eq:covariantderivative}.
Integrating out $B$ by imposing the on-shell condition in \cref{eq:ncym1 classical action}, we recover the noncommutative Yang--Mills functional \eqref{eq:ncym classical action}.
For a gauge parameter $c \in \Omega^0(\mathbbm{R}^{1,3}, \mathfrak{u}(N))$, the action functional \eqref{eq:ncym1 classical action} is invariant under the star-gauge transformations $\delta_c A= \mathrm{d} c + g\, [c \stackrel{\star}{,} A]_{\mathfrak{u}(N)}$ and $\delta_c B =g\,[c \stackrel{\star}{,} B]_{\mathfrak{u}(N)}$.

Noncommutative Yang--Mills theory in the first order formalism is organised by the cyclic \textit{strict} $L_\infty$-algebra 
\begin{equation}
    \mathfrak{YM}_{\star 1} = \Big(\mathsf{Ch}(\mathfrak{YM}_{\star 1}) \,,\, \mu_2^{\star \ym_1} \,,\, \langle -,- \rangle_{\ym_1}^\star \Big) \ ,
\end{equation}
which is quasi-isomorphic to the cyclic $L_\infty$-algebra \eqref{eq:NCYMLinfty} through an injective $L_\infty$-morphism $\psi^{\star\ym_2}:\YM_\star \lhook\joinrel\longrightarrow \YM_{\star 1}$.
The BV cochain complex $\mathsf{Ch}(\mathfrak{YM}_{\star 1})$ associated to this $L_\infty$-algebra is
\begin{equation} \label{eq:BF YM complex}
\small
    \begin{tikzcd}[row sep=4ex]
    \Omega^0(\mathbbm{R}^{1,3}, \mathfrak{u}(N)) \arrow["\mathrm{d}"]{r}
    &
    \Omega^1(\mathbbm{R}^{1,3}, \mathfrak{u}(N))[-1] \arrow["\mathrm{d}"]{r} \arrow[phantom, "\oplus"]{d} 
    & \Omega^2(\mathbbm{R}^{1,3}, \mathfrak{u}(N))[-2] \arrow[phantom, "\oplus"]{d}
    &
    \\
    &\Omega^2(\mathbbm{R}^{1,3}, \mathfrak{u}(N))[-1] \arrow["\mathrm{d}"]{r} \arrow["-\ast_{\textrm {\tiny H}}"]{ur} 
    & \Omega^3(\mathbbm{R}^{1,3}, \mathfrak{u}(N))[-2] \arrow["\mathrm{d}"]{r}
    & \Omega^4(\mathbbm{R}^{1,3}, \mathfrak{u}(N))[-3]
    \end{tikzcd}
\normalsize
\end{equation}
which identifies the differential $\mu_1^{\star \ym_1}$ of degree~$1$ acting on homogeneous elements as
\begin{subequations}\label{eq:ncym1 brackets}
\begin{equation}
\begin{split}
    \mu_1^{\star \ym_1}(c) = (\mathrm{d}c,0) \quad , \quad \mu_1^{\star \ym_1} (A ,  B) =  (\mathrm{d} A - \ast_{\textrm {\tiny H}}\, B, \mathrm{d}B) \quad , \quad    \mu_1^{\star \ym_1}(B^+,A^+) = \mathrm{d}A^+ 
\end{split} \ .
\end{equation}

The non-zero $2$-brackets $\mu_2^{\star \ym_1} $ are given by
\begin{equation}
\begin{split}    
    \mu_2^{\star \ym_1}(c_1, c_2) = g\,[c_1 \stackrel{\star}{,} c_2]_{\fru(N)} \qquad & , \qquad
    \mu_2^{\star \ym_1}\Big(c, {\small \bigg(\begin{matrix} A \\ B\end{matrix}\bigg)} \normalsize \Big) =  g\,\begin{pmatrix} [c \stackrel{\star}{,} A]_{\fru(N)} \\[1ex] [c \stackrel{\star}{,} B]_{\fru(N)} \end{pmatrix} \ , \\[4pt]
    \mu_2^{\star \ym_1}\Big(c, {\small \bigg(\begin{matrix} B^+ \\ A^+\end{matrix} \bigg)} \normalsize \Big) =g\,\begin{pmatrix} [c \stackrel{\star}{,} B^+]_{\fru(N)} \\[1ex] [c \stackrel{\star}{,} A^+]_{\fru(N)} \end{pmatrix} \qquad & , \qquad \mu_2^{\star \ym_1}(c, c^+) = g\,[c \stackrel{\star}{,} c^+]_{\fru(N)} \ , \\[4pt]
    \mu_2^{\star \ym_1}\Big({\small \bigg(\begin{matrix} A_1 \\ B_{1}\end{matrix}\bigg),\bigg(\begin{matrix} A_2 \\ B_{2}\end{matrix}\bigg) } \normalsize\Big) &= g\,\begin{pmatrix}  [A_1 \stackrel{\star}{,} A_2]_{\fru(N)} \\[1ex] 
    [A_1 \stackrel{\star}{,} B_{2} ]_{\fru(N)} + [A_2 \stackrel{\star}{,} B_{1}]_{\fru(N)}\end{pmatrix} \ , \\[4pt]
    \mu_2^{\star \ym_1}\Big({\small \bigg(\begin{matrix} A \\ B\end{matrix}\bigg) , \bigg(\begin{matrix} B^+ \\ A^+\end{matrix}\bigg) } \normalsize\Big) &= g\,\big([A \stackrel{\star}{,} A^+]_{\fru(N)} + [B \stackrel{\star}{,} B^+]_{\fru(N)}\big) \ .
\end{split}
\end{equation}

Finally, as in \S\ref{sub:noncommutative_chern_simons}, the cyclic structure $\langle -,- \rangle^\star_{\ym_1}$ pairs forms of complementary degrees:
\begin{equation}
\begin{split}\label{eq:ym1 traces}
    \langle \alpha, \alpha^+ \rangle_{\ym_1}^\star := \int \, \mathrm{Tr}_{\mathfrak{u}(N)}(\alpha \wedge_\star \alpha^+)  = \int \, \Tr_{\fru(N)}(\alpha \wedge \alpha^+) \ ,
\end{split}
\end{equation}
for $\alpha\in\{c,A,B\}$.
\end{subequations}

\paragraph{Factorisation in the first order formalism.} 
\label{par:factorisation_of_ustar_ncym}

We shall now demonstrate the homotopy factorisation of first order noncommutative Yang--Mills theory, which is similar to that of noncommutative Chern--Simons theory. 
We first factorise the colour algebra using the twisting map from \cref{sub:factNCCS}, now applied to differential forms on $\FR^{1,3}$.
It reproduces the star-commutator on $\Omega^\bullet(\mathbbm{R}^{1,3}, \mathfrak{u}(N))$ as a twisted tensor product with \smash{$\tau^{\mathfrak{u}_\star(N)}_2$}, and similarly the differential on $\Omega^\bullet(\mathbbm{R}^{1,3}, \mathfrak{u}(N))$ as a twisted tensor product with \smash{$\tau_1^{\mathfrak{u}_\star(N)}$}.

This yields the factorisation
\begin{equation}
    \mathfrak{YM}_{\star 1} = \mathfrak{u}(N) \otimes_{\tau^{\mathfrak{u}_\star(N)}} \mathfrak{C}_{\ym_1} \ ,
\end{equation}
where $\mathfrak{C}_{\ym_1}$ is the colour-stripped $C_\infty$-algebra with underlying cochain complex
\begin{equation}\label{eq:ch of ncym1}
    \mathsf{Ch}(\mathfrak{C}_{\ym_1}) = \left(
    \begin{tikzcd}[row sep=4ex]
    \Omega^0(\mathbbm{R}^{1,3}) \arrow["\mathrm{d}"]{r}
    &
    \Omega^1(\mathbbm{R}^{1,3})[-1] \arrow["\mathrm{d}"]{r} \arrow[phantom, "\oplus"]{d} 
    & \Omega^2(\mathbbm{R}^{1,3})[-2] \arrow[phantom, "\oplus"]{d}
    &
    \\
    &\Omega^2(\mathbbm{R}^{1,3})[-1] \arrow["\mathrm{d}"]{r} \arrow["-\ast_{\textrm {\tiny H}}"]{ur} 
    & \Omega^3(\mathbbm{R}^{1,3})[-2] \arrow["\mathrm{d}"]{r}
    & \Omega^4(\mathbbm{R}^{1,3})[-3]
    \end{tikzcd}
    \right) \ .
\end{equation}
Its brackets are given by colour-stripping the $L_\infty$-structure of \cref{eq:ncym1 brackets}, so the differentials \smash{$m_1^{\ym_1}$} are of the same form, while the $2$-brackets \smash{$m_2^{\ym_1}$} are simply given by exterior products as
\begin{equation}\label{eq:ncym1 fact brackets}
\begin{split}
    m_2^{ \ym_1}(c_1, c_2) = g\, c_1\,c_2 \ \ , \ & \
    m_2^{ \ym_1}\Big(c, {\small \bigg(\begin{matrix} A \\ B\end{matrix}\bigg)} \normalsize \Big) =  g\,\bigg(\begin{matrix} c \, A \\ c \, B  \end{matrix}\bigg) \ \ , \ \
    m_2^{ \ym_1}\Big(c, {\small \bigg(\begin{matrix} B^+ \\ A^+\end{matrix} \bigg)} \normalsize \Big) =g\,\bigg(\begin{matrix} c\, B^+ \\ c\, A^+\end{matrix}\bigg) \\[4pt]
     m_2^{ \ym_1}(c, c^+) = g\,c\,c^+ \quad  , \ & \quad
    m_2^{ \ym_1}\Big({\small \bigg(\begin{matrix} A_1 \\ B_{1}\end{matrix}\bigg),\bigg(\begin{matrix} A_2 \\ B_{2}\end{matrix}\bigg) } \normalsize\Big) = g\,\bigg(\begin{matrix} A_1\wedge A_2 \\ 
  A_1 \wedge B_{2}  + A_2 \wedge B_{1} \end{matrix}\bigg) \  , \\[4pt]
    m_2^{ \ym_1}\Big({\small \bigg(\begin{matrix} A \\ B\end{matrix}\bigg) , \bigg(\begin{matrix} B^+ \\ A^+\end{matrix}\bigg) } \normalsize\Big) &= g\,\big( A \wedge A^+ + B \wedge B^+\big) \ . 
\end{split}
\end{equation}

We further factorise this $C_\infty$-algebra as a twisted tensor product
\begin{align}
\mathfrak{C}_{\ym_1} = \mathfrak{Kin}_{\ym_1} \otimes_{\tau^{\ym_1}} \mathfrak{Scal} \ ,
\end{align}
where the kinematical vector space is identified as
\begin{equation}
    \mathfrak{Kin}_{\ym_1} \ := \ 
    \midwedge^0(\mathbbm{R}^{1,3})^*[1] \quad \oplus \quad 
    \begin{matrix}
    \midwedge^1(\mathbbm{R}^{1,3})^* \\[1ex] \oplus \\[1ex] \midwedge^2(\mathbbm{R}^{1,3})^* \end{matrix} \quad \oplus \quad
    \begin{matrix}
    \midwedge^2(\mathbbm{R}^{1,3})^*[-1] \\[1ex] \oplus \\[1ex] \midwedge^3(\mathbbm{R}^{1,3})^*[-1] \end{matrix} \quad \oplus \quad
    \midwedge^4(\mathbbm{R}^{1,3})^*[-2] \ .
\end{equation}
For a basis of covectors $e^\mu$ on $\FR^{1,3}$, we abbreviate $e^{\mu\nu\cdots}:=e^\mu\wedge e^\nu\wedge\cdots$; we distinguish the bases of $\midwedge^2(\mathbbm{R}^{1,3})^*$ in degrees~$0$ and~$1$ by denoting them respectively as $e^{\mu\nu}$ and $e^{+\mu\nu}$. As in \cref{sub:factNCCS}, this vector space is equipped with a pairing given by wedging forms in complementary degrees and applying the Hodge duality operator $\Tr_{\wedge^\bullet}:\midwedge^4(\FR^{1,3})^*\longrightarrow\FR$ to the resulting top form on $\FR^{1,3}$, which on basis elements is given by
\begin{equation}
\begin{split}
    \Tr_{\wedge^\bullet}(e^{\mu\nu\rho\sigma}) = \epsilon^{\mu \nu \rho \sigma} \ ,
\end{split}
\end{equation}
where $\epsilon^{\mu\nu\rho\sigma}$ is the Levi--Civita symbol in four dimensions with $\epsilon^{0123}=1$.

The twist datum $\tau^{\ym_1}=\big(\tau_1^{\ym_1},\tau_2^{\ym_1}\big)$ is defined as follows. The non-zero values of the twisting map \smash{$\tau^{\ym_1}_1: \mathfrak{Kin}_{\ym_1}  \longrightarrow \mathfrak{Kin}_{\ym_1} \otimes \mathsf{End}(L)$} for the graded vector space \eqref{eq:scalarvectorspace} are given by
\begin{subequations}
\begin{align}
\begin{split}
    \tau^{\ym_1}_1(1) = e^\mu \otimes \frac{\partial_\mu}{\square} & \quad ,  \quad
    \tau_1^{\ym_1}(e^\mu) = e^{+\mu \nu} \otimes \frac{\partial_\nu}{\square} \quad , \quad \tau_1^{\ym_1}(e^{\mu \nu \rho}) = e^{\mu \nu \rho \sigma} \otimes \frac{\partial_\sigma}{\square} \ ,
    \\[4pt]
    & \tau_1^{\ym_1}(e^{\mu \nu})  = -e^{+\rho\sigma} \otimes \epsilon_{\rho\sigma}{}^{\mu\nu}\,\frac{1}{\square} + e^{\mu \nu \rho} \otimes \frac{\partial_\rho}{\square} \ .
\end{split}
\end{align}
This recovers the colour-stripped differential: $m_1^{\tau^{\ym_1}} = m_1^{\ym_1}$. For example, on degree~$1$ fields $(A, B)$ the differential decomposes into 
\begin{align}
\begin{split}
    m_1^{\tau^{\ym_1}}(e^\mu \otimes A_\mu + e^{\nu \rho} \otimes B_{\nu \rho} ) 
    &=
    e^{+\mu\nu} \otimes 
    \mu_1^{\rm Scal}\big(\tfrac{\partial_\nu}\square\,A_\mu
    \big)\\
    & \quad \, 
    - e^{+\rho\sigma}\otimes \mu_1^{\rm Scal}\big(\tfrac{1}\square\,\epsilon_{\rho\sigma}{}^{\mu\nu}\,B_{\mu\nu}\big) + e^{\nu\rho\sigma}\otimes \mu_1^{\rm Scal}\big(\tfrac{\partial_\sigma}\square\,B_{\nu\rho}\big) \\[4pt]
&= e^{+\sigma \mu} \otimes (\partial_\sigma A_\mu -\epsilon_{\sigma\mu}{}^{\nu\rho}\, B_{\nu\rho}) + e^{\mu \nu \rho}\otimes \partial _\mu B_{\nu \rho} \\[4pt]
& = ( \mathrm{d}A - \ast_{\textrm {\tiny H}}\, B , \mathrm{d}B) \ .
\end{split}
\end{align}

The colour-stripped $2$-bracket $m_2^{\tau^{\ym_1}} = m_2^{\ym_1}$ is recovered using the following twisting map \smash{$\tau_2^{\ym_1} : \mathfrak{Kin}_{\ym_1} \otimes \mathfrak{Kin}_{\ym_1}  \longrightarrow \mathfrak{Kin}_{\ym_1} \otimes \mathsf{End}(L) \otimes\mathsf{End}(L) $}. Any bracket involving a degree $0$ element is twisted in the same way, recovering the first four brackets of \cref{eq:ncym1 fact brackets} from
\begin{equation}
\begin{split}
    \tau_2^{\ym_1}(1,v) = v \otimes \mathbbm{1} \otimes \mathbbm{1} \ ,
\end{split}
\end{equation}
for all $v\in \mathfrak{Kin}_{\ym_1}$.
Brackets between two degree $1$ fields as well as brackets between a degree $1$ field and a degree $2$ field are recovered respectively by the non-vanishing values
\begin{equation} \label{eq:tau2YM1}
\begin{split}
    \tau_2^{\ym_1}(e^\mu, e^\nu) = e^{+\mu \nu} \otimes \mathbbm{1} \otimes \mathbbm{1} \qquad , & \qquad
    \tau_2^{\ym_1}(e^\mu, e^{\nu \rho}) = e^{\mu \nu \rho} \otimes \mathbbm{1} \otimes \mathbbm{1} \ , \\[4pt]
    \tau_2^{\ym_1}(e^{\mu \nu}, e^\rho) = - e^{\mu \nu \rho} \otimes \mathbbm{1} \otimes \mathbbm{1} \qquad , & \qquad
    \tau_2^{\ym_1}(e^\mu, e^{\nu \rho \sigma}) = e^{\mu \nu \rho \sigma} \otimes \mathbbm{1} \otimes \mathbbm{1} \ , \\[4pt]
    \tau_2^{\ym_1} (e^{\mu \nu}, e^{+\rho \sigma}) = e^{\mu \nu \rho \sigma} \otimes \mathbbm{1} \otimes \mathbbm{1} \qquad , & \qquad
    \tau_2^{\ym_1} (e^{+\mu \nu}, e^{\rho \sigma}) = - e^{\mu \nu \rho \sigma} \otimes \mathbbm{1} \otimes \mathbbm{1}  \ .
\end{split}
\end{equation}
For example, $2$-brackets between fields in degrees $1$ and $2$ are generated by the twisted bracket using the non-zero twist data  from \cref{eq:tau2YM1} to reduce to top forms  in $\midwedge^4 (\mathbbm{R}^{1,3})^*$:
\begin{equation}
\begin{split}
& m_2^{\tau^{\ym_1}}\big( e^\mu \otimes A_\mu + e^{\nu\rho} \otimes B_{\nu \rho}\,,\,
    e^{+\alpha \sigma} \otimes B^+_{\alpha\sigma} + e^{\beta \lambda \kappa} \otimes A^+_{\beta\lambda \kappa}\big) \\[4pt]
 & \hspace{5cm}  =  e^{\mu\beta\lambda\kappa} \, \otimes
    \mu_2^{\rm Scal}(A_\mu,A^+_{\beta\lambda\kappa})  + e^{\nu\rho\alpha\sigma} \,
    \otimes
    \mu_2^{\rm Scal}(B_{\nu\rho},B^+_{\alpha \sigma})
    \\[4pt]
    & \hspace{5cm} = g\,e^{\mu \nu \rho \sigma}\otimes \big(A_\mu \, A^+_{\nu \rho \sigma} + B_{\mu \nu}\, B_{ \rho \sigma}^+\big) \\[4pt]
    & \hspace{5cm} = g\,\big(A \wedge A^+ + B \wedge B^+\big) \ .
\end{split}
\end{equation}
\end{subequations}

Finally, as in \cref{sub:factNCCS} the cyclic structure \eqref{eq:ym1 traces} is reproduced by $\Tr_{\wedge^\bullet}\otimes\langle-,-\rangle_{\rm Scal}$, and altogether we have shown that the twisted homotopy factorisation of the cyclic strict $L_\infty$-algebra underlying noncommutative Yang--Mills theory in the first order formalism is given by
\begin{equation}\label{eq:ncym1 linf factorisation final}
    \mathfrak{YM}_{\star 1} = \mathfrak{u}(N) \otimes_{\tau^{\mathfrak{u}_\star(N)}} \big(\mathfrak{Kin}_{\ym_1} \otimes_{\tau^{\ym_1}} \mathfrak{Scal}\big) \ .
\end{equation}
As in \cref{sub:factNCCS}, both factorisations \eqref{eq:YM2fact1} and \eqref{eq:ncym1 linf factorisation final} identify the binoncommutative biadjoint scalar theory of \S\ref{sub:the_other_single_copy} as the zeroth copy of noncommutative Yang--Mills theory.
\subsection{Noncommutative gauge theories on D-branes} 
\label{sub:modified_k}

We will now aim to understand the structural features of noncommutative Yang--Mills amplitudes that we have discussed as a natural consequence of the embedding of the noncommutative gauge theory into string theory as a low-energy limit of open strings in constant $B$-field backgrounds. 

\paragraph{Open strings in Kalb--Ramond fields.} 
\label{par:open_string_theory_in_the_presence_of_a_kalb_ramond_field}
Consider the open string sigma-model with fields $X=(X^\mu)$ mapping from a Euclidean worldsheet $\Sigma$ with boundary conditions corresponding to a stack of D$p$-branes filling flat space $\FR^{1,p}$ in a closed string background with metric $g$, Kalb--Ramond two-form $b$ and dilaton $\phi$, all of which are assumed to be constant.  The D-brane worldvolume supports an abelian gauge field $A$ of constant curvature $F$. This combines with the Kalb--Ramond field to the gauge-invariant Born--Infeld field strength $B:=b+F$, which can be regarded as an electromagnetic field on the D-branes that the open string endpoints are charged with respect to. We assume that $A$ is a sum of gauge fields restricted to each connected component of the boundary $\partial\Sigma$, so that different branes can support independent $\sU(1)$ gauge fluxes; this allows for open strings to end on D-branes with \emph{different} overall background $B$-fields. For simplicity we suppose that the gauge flux $F$ has the same rank $r\leq p+1$ as $b$, and that $g_{\mu\nu}=0$ whenever $\mu\in\{0,1,\dots,r-1\}$ and $\nu\notin\{0,1,\dots,r-1\}$. 

At tree-level in open string perturbation theory, the worldsheet $\Sigma$ is a disk, or the conformally equivalent complex upper-half plane $\mathbbm{H}^+\subset\FC$. In the boundary conformal field theory on $\mathbbm{H}^+$, the presence of the $B$-field modifies the Neumann boundary conditions in the longitudinal directions to the D$p$-branes to the mixed boundary conditions
\begin{align}
(\partial_z- \partial_{\bar z})\,g\, X+2\pi\,\alpha'\,(\partial_z+\partial_{\bar z})\,B\,X\,\big|_{\partial\,\mathbbm{H}^+} = 0 \ ,
\end{align}
where ${\rm Im}(z)\geq0$ and $\alpha'$ is the string Regge slope. The bulk propagator  on the worldsheet $\mathbbm{H}^+$ is then~\cite{Abouelsaood:1986gd}
\begin{align} \label{eq:uhpprop}
 \langle 
    X^\mu(z,\overline{z}) \, X^\nu(z',\overline{z}^{\,\prime})
    \rangle_{\mathbbm{H}^+} = -2\pi\,\alpha'\,\big(g^{\mu\nu} \log|z-z'| + D^{\mu\nu} \log|z-\overline{z}^{\,\prime}| + c^{\mu\nu} \big) \ ,
\end{align}
with the projector
\begin{align}
D = -\frac1g + \frac2{g+2\pi\,\alpha'\,B}
\end{align}
and $c^{\mu\nu}$ are arbitrary integration constants. 

This Green's function is defined on the double cover $\FC$ of the upper-half plane $\mathbbm{H}^+$ by worldsheet parity $\Omega:z\longmapsto\bar z$ and is obtained using the method of images: starting from the free Green's function $-2\pi\,\alpha'\,g^{\mu\nu}\log|z-z'|^2$ for the two-dimensional Laplace equation, one enforces the mixed boundary conditions by adding to it the contribution of an image charge symmetric with respect to reflection through the real line $z=\bar z$, which corresponds to worldsheet parity. 
The propagator \eqref{eq:uhpprop} is single-valued if the logarithmic branch cut is placed in the complex lower-half plane $\mathbbm{H}^-\subset\FC$. 

The effective target space geometry seen by the open strings ending on the D$p$-branes consists of the open string metric $G$ and the Poisson bivector $\theta$. They are related to the parameters $(g,b)$ of the closed string background  through the open-closed relations~\cite{Seiberg:1999vs}
\begin{equation}
\begin{split}
    G = g - (2 \pi\, \alpha')^2\,B\,\frac1g\,B \qquad \mbox{and} \qquad
    \theta = -(2 \pi\, \alpha')^2\, \frac{1}{g+ 2 \pi\, \alpha'\, B}\, B\, \frac{1}{g - 2 \pi\, \alpha'\, B}  \ .
\end{split}
\end{equation}
This parametrization allows us to write the string propagator as
\begin{equation}\label{eq:uhp correlator}
\begin{split}
    \langle 
    X^\mu(z,\overline{z}) \, X^\nu(z',\overline{z}^{\,\prime})
    \rangle_{\mathbbm{H}^+} &= 
    -2\pi\, \alpha'\, \bigg(
    g^{\mu \nu} \log|z - z'| - g^{\mu \nu} \log|z - \overline{z}^{\,\prime}|
    + G^{\mu \nu} \log|z - \overline{z}^{\,\prime}|^2  \\
    &\hspace{3cm}
    - \frac{\ii\,\theta^{\mu \nu}}{2 \pi\, \alpha'} 
    \log \Big(\frac{z - \overline{z}^{\,\prime}}{\overline{z} - z'} \Big) + c^{\mu\nu}
    \bigg) \ .
\end{split}
\end{equation}

On the boundary $\partial\,\mathbbm{H}^+=\FR$,  the choice of constant counterterm $c^{\mu\nu}=-\frac\ii2 \,\theta^{\mu\nu}$ reduces the propagator to the correlation function~\cite{Seiberg:1999vs}
\begin{equation}\label{eq:boundary prop}
    \langle 
    X^\mu(\tau) \, X^\nu(\tau')
    \rangle_{\partial \mathbbm{H}^+} = 
    - \alpha'\,
    G^{\mu \nu} \log(\tau - \tau')^2 
    + \tfrac\ii2\,\theta^{\mu \nu} \, \mathrm{sgn}(\tau - \tau') \ .
\end{equation}
This implies that the operator product expansion of tachyon vertex operators in the limit $\tau\to\tau'$ with $\tau>\tau'$ is given by
\begin{align}\label{eq:tachyon vertex effect}
    \e^{\,\ii\,p\cdot X}(\tau) \, \e^{\,\ii\,q\cdot X}(\tau') = |\tau-\tau'|^{\alpha'\,p\cdot G\,q} \ \exp\big(\tfrac\ii2\,p\cdot\theta\,q\big) \ \e^{\,\ii\,(p+q)\cdot X}(\tau') + \cdots \ ,
\end{align}
where as usual the ellipses denote less singular terms as $\tau\to\tau'$.
The term in the boundary propagator \eqref{eq:boundary prop} involving $\theta$ is a piecewise-constant function of $\tau$ and $\tau'$, so it does not contribute to correlation functions of $\tau$-derivatives of $X$.

Consider open strings with $\sU(N)$ Chan--Paton factors, and the tree-level scattering of $n$ gluons of momenta $p_i$, polarizations $\zeta_i$, and Chan--Paton wavefunctions $\lambda_i = T^{a_i}$, with $i=1,\dots,n$.
The scattering amplitude for an ordering of colour factors $(a_1, \ldots, a_n)$ and insertion points $( \tau_1, \ldots, \tau_n )$ on $\partial\,\mathbbm{H}^+$ is given by
\begin{align}\label{eq:partial open string amplitude}
\begin{split}
    \CM_n^{\rm open}(p,\zeta,a)_{G,\theta} 
    &= \Tr_{\fru(N)}(T^{a_1}\cdots T^{a_n}) \\
    & \qquad \times \int_{(\partial\,\mathbbm{H}^+)^{\times n}} \, \dd\mu(\tau_1,\dots,\tau_n) \ \Big\langle\prod_{i=1}^n \, \zeta_i\cdot\frac{\dd X}{\dd\tau} \ \e^{\,\ii\, p_i\cdot X}(\tau_i) \Big\rangle_{G,\theta} \ ,
    \end{split}
\end{align}
where the subscripts $_{G,\theta}$ indicate that we evaluate correlation functions and amplitudes as functions of the open string parameters $(G,\theta)$. The term multiplying the trace is the colour-stripped open string amplitude $\mathcal{A}_n^{\rm open}(1,\dots,n)_{G, \theta}$.
The vertex operators are inserted on $\partial\,\mathbbm{H}^+$ in a definite cyclic order, and the measure $\dd\mu$ refers to the integral over the positions $\tau_i$ modulo the action of~$\sSL(2,\FR)$ by boundary conformal transformations. 

The only $\theta$-dependence of the amplitude \eqref{eq:partial open string amplitude} is in the phase factor $\exp\big(\frac\ii2 \, \sum_{i<j} \, p_i\cdot\theta\,p_j\big)$ that comes from the expectation value of products of the tachyon vertex operators \smash{$\e^{\,\ii\,p_i\cdot X(\tau_i)}$} in \cref{eq:tachyon vertex effect}. 
This factor arises from the correlation function inside the $\dd\mu$ integral, and it is a piecewise-constant function of $\tau_1,\dots,\tau_n$ that depends only on their cyclic ordering. 
Since the cyclic ordering is kept fixed in evaluating the integral that gives the scattering amplitude, this factor multiplies an otherwise $\theta$-independent amplitude. Rescaling $\theta\to t\,\theta$ for a parameter $t\in\FR$, the colour-stripped amplitude $\mathcal{A}_n^{\rm open}(1,\dots,n)_{G,t\,\theta}$ thus factors as
\begin{equation}\label{eq: q dependence factored out}
\begin{split}
    \CA_n^{\rm open} (1,\dots,n)_{G,t\,\theta} &= \exp\Big(-\frac {\ii\,t}{2} \, \sum_{i<j} \, p_i\cdot\theta\, p_j\Big) \ \mathcal{A}_n^{\rm open}(1,\dots,n)_{G, t = 0}
\end{split}\ .
\end{equation}
In the limit $B\to0$, the flat space string amplitudes are recovered.

We may interpret these equations in terms of an ordering of momenta $(p_1, \ldots, p_n)$ viewed as the word $w = 1\cdots n \in \CCW _n $. The phase factor appearing in \cref{eq: q dependence factored out}  may then be denoted as $\varTheta_n(w)$. This is the same as the phase \eqref{eq:phase factor} factoring tree-level amplitudes in noncommutative Yang--Mills theory.
The full amplitude is given by summing over the $(n-3)!$ inequivalent orderings.
Since all the data of partial amplitudes respects an ordering $w \in \CCW _n $, all terms in \cref{eq: q dependence factored out} can be written in terms of $w$.
%

\paragraph{Seiberg--Witten limit.} 
\label{par:seiberg_witten_limit}

To make the link with noncommutative gauge theories, we take the Seiberg--Witten limit \cite{Seiberg:1999vs}.
This scales the closed string metric to zero, while keeping both the open string metric $G$ and Poisson bivector $\theta$ fixed. It can be achieved by taking the usual low-energy limit $\alpha'\to0$, but now correlated with the limit $g_{\mu\nu}\sim(\alpha')^2\to0$, or equivalently with the limit $B_{\mu\nu}\sim(\alpha')^{-2}\to\infty$. In either of these scaling limits, the boundary propagator \eqref{eq:boundary prop} becomes
\begin{equation}
    \langle X^\mu(\tau)\, X^\nu(\tau') \rangle^\sw_{\partial \mathbbm{H}^+} = \tfrac\ii2 \,  \theta^{\mu \nu} \, \mathrm{sgn}(\tau - \tau') \ .
\end{equation} 

In the Seiberg--Witten limit, only the phase  in the operator product expansion \eqref{eq:tachyon vertex effect} remains as an overall pre-factor. 
This has the effect of replacing ordinary multiplication of wavefunctions by the Moyal--Weyl star-product which quantizes the Poisson structure $\theta$.
The closed string propagator \eqref{eq:uhp correlator} diverges in this limit, as can also be seen at the level of the sigma-model action functional where the bulk kinetic term disappears and the worldsheet theory becomes a topological field theory on the boundary with a degenerate phase space.
BV quantisation of this topological string theory in the first order formalism reproduces star-products of fields in correlation functions of boundary observables~\cite{Kontsevich:1997vb,Cattaneo:1999fm}. The tree-level S-matrix is generated by the spacetime effective action functional of noncommutative Yang--Mills theory on $\FR^{1,p}$~\cite{Seiberg:1999vs}, with the constant metric $G$ replacing the $d=p+1$ Minkowski metric $\eta$ everywhere in \S\ref{sub:noncommutative_yang_mills}. 

Application of the Seiberg--Witten limit to \cref{eq: q dependence factored out} relates colour ordered tree-level scattering amplitudes in $\sU(N)$ noncommutative Yang--Mills theory with $N>1$ to those of the corresponding commutative theory through the same phase factor \eqref{eq:phase factor}:
\begin{equation}
    \CA^{\star\ym}_n(w)_{G} = \varTheta_n(w) \ \CA_n^{\ym}(w)_{G} \ ,
\end{equation}
for $w \in \CCW _n $. Here the subscript ${}_G$ emphasises that all kinematic invariants used in the computation of amplitudes through this relation are evaluated with respect to the constant open string metric. 

\subsection{Bern--Carrasco--Johansson relations} 
\label{par:bcj_relations_for_noncommutative_ym}
The Bern--Carrasco--Johansson (BCJ) relations\cite{Bern:2008qj} are linear relations, whose coefficients are rational functions of Mandelstam variables, between tree-level partial amplitudes with different cyclic orderings that differ by the insertion position of a single gluon. They further reduce the number of independent subamplitudes at multiplicity $n$ to $(n-3)!$.
After their advent it was realised that a natural explanation for them appears as the low-energy limit of monodromy relations in open string theory, see e.g.~\cite{Stieberger:2009hq}.

BCJ relations for noncommutative Yang--Mills amplitudes are obtained in \cite{Huang:2010fc}, building on the BCFW recursion formulas discussed in~\cite{Raju:2009yx}. This requires some care, as the noncommutative amplitudes contain essential singularities in the complex plane.
As we now explicitly demonstrate, these relations can also be derived from the same set of monodromy equations for open string scattering amplitudes in a constant Kalb--Ramond field that were found in~\cite{Boels:2010bv}.

For an ordering of punctures $w=k_1\cdots k_n \in \CCW _n $, the tachyonic part of the colour-stripped open string amplitude in \cref{eq:partial open string amplitude} is given by an iterated integral of the corresponding Koba--Nielsen factor 
\begin{equation}
\begin{split}
    \mathcal{A}_n^{\rm open}(w)_{G, t\,\theta} = \varTheta_n(w) \  \int_{\mathbbm{D}(w)} \  \prod_{i=1}^n\, \mathrm{d} \tau_{k_i} \ \prod_{i<j}\, \big(\tau_{k_i} - \tau_{k_j}\big)^{\alpha'\, p_{k_i}\cdot G\,p_{k_j}}
\end{split}
\end{equation}
over the domain $\mathbbm{D}(w) = \{ (\tau_{k_1},\dots,\tau_{k_n}) \in \mathbbm{R}^n\ |\ \tau_{k_1} < \cdots < \tau_{k_{n}} \}$.
For illustration we choose the trivial ordering $w = 1\cdots n$.
We may then choose to single out $\tau_{1}$ and integrate over it along the boundary of the worldsheet $\FR$.
The integrand is singular at $\tau_{1} = \tau_{i}$ for $i > 1$, where it has poles, while analytic everywhere else. Using this information, we analytically continue the integral over $\tau_1$ to an integral over a suitable closed contour in the complex plane.

For each pole $\tau_i$, the integrand picks up a monodromy factor $\e^{-2\pi\,\ii\, \alpha' \, p_1\cdot G\,\,p_i}$ while the noncommutativity phase picks up a factor $\e^{\,\ii\,t\, p_1 \cdot \theta\, p_i}$.
In addition, integration around each chamber $\tau_i < \tau_1 < \tau_{i+1}$ gives the partial amplitude $\mathcal{A}_n^{\rm open}(2,\dots. i, 1, i{+}1, \dots, n)_{G,t\,\theta}$.
Introducing the massless Mandelstam invariants $s_{k_ik_j} := 2\,p_{k_i} \cdot  G\,p_{k_j}$, and deforming the contour to infinity, it follows that the result of the contour integration yields
\begin{align}
\begin{split}
& \mathcal{A}_n^\textrm{open}(1,\dots, n)_{G,t\,\theta} \\
& \hspace{2cm} + \varTheta_n(1\cdots n) \,  \sum_{i=2}^{n-1} \, \frac{\e^{-\pi\,\ii\, \alpha' \,( s_{12}+\cdots+s_{1i})}}{ \varTheta_n(2\cdots i\,1\,i{+}1\cdots n)} \ \mathcal{A}_n^\textrm{open}(2,\dots,i,1,i{+}1,\dots,n)_{G,t\,\theta}
= 0 \ .
    \end{split}
\end{align}

Now we take the Seiberg--Witten limit, in which the noncommutativity factor survives the low-energy limit $\alpha'\to0$.
In this way, we obtain the Ward identities
\begin{equation}
   \sum_{i=1}^{n-1} \, \varTheta_n(n\cdots i{+}1\,1\,i\cdots 21)\ \CA^{\star\ym}_n(1,2,\dots,i,1,i{+}1,\cdots n)_G  = 0 
\end{equation}
as well as the BCJ relations
\begin{equation}
   \sum_{i=2}^{n-1} \, \frac{s_{12}+\cdots +s_{1i}}{\varTheta_n(12\cdots i\,1\,i{+}1\cdots n)} \ \CA^{\star\ym}_n(1,2,\dots,i,1,i{+}1,\cdots n)_G  = 0 
\end{equation}
for noncommutative Yang--Mills theory.

\subsection{Kawai--Lewellen--Tye relations} 
\label{sub:noncommutative_gauge_theories_from_open_string_theory}

The factorisations discussed in \cref{sub:strictification_of_ym} can be used to construct a variety of double copy theories from noncommutative Yang--Mills theory. As we saw already in \cref{sub:DCCS}, these dual theories do not undergo any noncommutative deformation and coincide with their classical double copies. In particular, the homotopy double copy of noncommutative Yang--Mills theory with itself in the second order formalism yields ordinary perturbative gravity in $d$ dimensions, or more exactly $\CN=0$ supergravity which involves a two-form and a dilaton in addition to the graviton~\cite{Borsten:2021hua,Bern:2019prr}. Again we demonstrate that this is a natural consequence of properties of open string amplitudes in constant $B$-field backgrounds.

\paragraph{Kawai--Lewellen--Tye relations with $\boldsymbol{B}$-fields.} 
\label{par:similarity_transformation_of_the_klt}

The Kawai--Lewellen--Tye (KLT) relations~\cite{Kawai:1985xq} are an explicit realisation of open-closed worldsheet duality for the tachyonic parts of string amplitudes.
A first topological argument comes from the observation that one can glue two oppositely oriented copies of a disk along their boundaries to form a Riemann sphere.
For open string amplitudes, punctures are inserted along the boundaries of the disks, so  the gluing between amplitudes is a function only of kinematic invariants and the ordering of vertices along the two boundaries. The KLT relations have been understood more recently from several perspectives: through a more computationally efficient paradigm~\cite{Mizera:2016jhj,Mizera:2017cqs}, through an ultraviolet completion in $\alpha'$ of biadjoint scalar theory called Z-theory~\cite{Mafra:2016mcc}, and through intersection theory on the moduli space of $n$-punctured Riemann spheres~\cite{Mizera:2016jhj,Mizera:2019gea}. The one-loop extension of the tree-level KLT relations, involving torus and annulus amplitudes, is found in~\cite{Stieberger:2022lss}.

Here we consider the tachyonic part $\mathcal{A}^{\rm open}_n(w)_G$ of the open string amplitude for a given ordering of punctures (and associated momenta) $w \in \CCW _{n} $, and kinematical invariants calculated with respect to the open string metric $G$.
A closed form of the KLT relation is given in terms of a \textit{momentum kernel} $ \boldsymbol{S}_n(w|w')_G$ linking the colour-stripped partial amplitudes to tree-level closed string amplitudes through
\begin{equation}\label{eq:string klt}
    \CCA_{n,G}^{\rm closed}(p) = \left(-\frac{\ii}{4}\right)^{n-3} \
    \sum_{w, w' \in \CCW _{n-3} } \,
    \mathcal{A}_{n}^{\rm open}(w)_G \ {\boldsymbol{S}}_n(w | w')_G \ \mathcal{A}^{\rm open}_n(w')_G \ .
\end{equation}
The reduced number of word letters here is due to the $\mathsf{SL}(2, \mathbbm{C})$ conformal invariance of the worldsheet theory, which enables one to fix three insertion points;  following the standard choice we fix $z_1 = 0,$ $ z_{n-1} = 1$ and $ z_n= \infty$.
The $n$-point momentum kernel $\boldsymbol{S}_n(w|w')_G$ is not unique: it is derived from an iterated integral over $n-3$ complex variables, so there are $n-3$ equivalent contour integrals to choose from.

One possible choice is found in \cite{Bjerrum-Bohr:2010pnr}. For $w=k_1\cdots k_n$ and fixed ordering of the right insertion $w' = 1\cdots n$, define  the operation
\begin{equation}
    \mathtt{H}(k_i, k_j)_{w'} = \begin{cases}
    \ 1 \quad \textrm{if the ordering of } k_i, k_j \textrm{ is opposite in } w' \ , 
    \\ \ 0 \quad \textrm{otherwise} \ .
    \end{cases}
\end{equation}
By conformal invariance, three points are fixed with labels $\{ 1, n-1, n \}$, so any summation over words in equations below is understood as leaving these letters fixed; for example, when writing $w \in \CCW  _n$ we mean $w = 1\, w^\circ\, n{-}1\,n$ for some $ w^\circ \in \CCW  _{n-3}$. We then set
\begin{equation}\label{eq:momentum kernel}
    \boldsymbol{S}_n(w| 1\cdots n)_G = \left(\frac{2}{\pi\, \alpha'}\right)^{n-3} \
    \prod_{i=2}^{n-2}\, \sin \pi\, \alpha'\,\Big(p_1 \cdot G\,p_{k_i}  + \textstyle{\sum\limits_{1\leq j<l\leq i}}\, \mathtt{H}(k_j, k_l)_{1 \cdots n} \ p_{k_j} \cdot G\,p_{k_l}\Big) \ ,
\end{equation}
and the full formula for the momentum kernel is simply obtained by permuting the right letters of $w'\in\CCW _{n-3} $ in the algebra of ordered words on the letters $\{ 2, \ldots, n-2 \}$. For example, unpacking \cref{eq:momentum kernel} we find the momentum kernel
\begin{align}
\boldsymbol{S}_5(12345|32145)_G = \big(\tfrac2{\pi\,\alpha'}\big)^2 \, \sin\big(\pi\,\alpha'\,p_1\cdot G\,p_2\big) \sin\big(\pi\,\alpha'\,p_3\cdot G\,(p_1+p_2)\big) 
\end{align}
involved in the gluing of multiplicity five string amplitudes.

The colour-stripped open string amplitudes appearing in the KLT formula \eqref{eq:string klt} combine in an analytic way to compute closed string amplitudes. Using \cref{eq: q dependence factored out} they can be written in terms of open string amplitudes with a Kalb--Ramond background. This scales separately the left and right partial amplitudes with respective phase factors, and leaves the KLT formula invariant under redefinition of the momentum kernel according to\footnote{Recall from \cref{sub:modified_k} that a situation with $\theta\neq\bar\theta$ can arise if we allow for distinct worldvolume gauge fluxes on different stacks of D-branes, which give rise to distinct Born--Infeld field strengths $B\neq\bar B$ in a fixed closed string background $(g,b)$. Even when $\theta=\bar\theta$, the notation aids in distinguishing the left-moving and right-moving open string sectors involved in the gluing of amplitudes.}
\begin{equation}
\boldsymbol{S}_n(w|w')_{G, t\,\theta,\bar t\,\bar\theta}  :=   \varTheta_n(\bar w) \ \boldsymbol{S}_n(w|w')_G \ \bar\varTheta_n(\bar w') \ .
\end{equation}
We interpret this trivial modification as saying that, for a constant $B$-field, it is possible to glue open string amplitudes in a manner consistent with the fact that the Kalb--Ramond field is invisible to closed strings on a Riemann sphere. We thus propose the $B$-field modified KLT relations
\begin{align}\label{eq:klt with theta}
    \CCA_{n,G}^{\rm closed}(p) &= \left(-\frac{\ii}{4}\right)^{n-3} \ \sum_{w, w' \in \CCW _{n-3} } \,
    \mathcal{A}_n^{\rm open}(w)_{G, t\,\theta} \ \boldsymbol{S}_n(w| w')_{G, t\,\theta,\bar t\,\bar\theta} \ \mathcal{A}_n^{\rm open}(w')_{G, \bar t\,\bar\theta} \ ,
\end{align}
leaving unchanged the closed string scattering amplitudes.

\paragraph{Double copy relations.} 

In the $\alpha'\rightarrow0$ limit, the closed string amplitudes taken with respect to the open string metric $G$ reduce to pure gravity amplitudes $\mathscr{A}_{n,G}^{\textrm{\tiny GR}}(p)$ on the background $\FR^{1,p}$ with metric $G$.
Therefore the Seiberg--Witten limit of the KLT relations \eqref{eq:klt with theta} gives the double copy relations
\begin{align}\label{eq:field klt with theta}
    \mathscr{A}_{n,G}^{\textrm{\tiny GR}}(p) &= \left(-\frac{\ii}{4}\right)^{n-3} \ \sum_{w, w' \in \CCW  _{n-3}} \,
    \mathcal{A}_n^{\star\ym}(w)_{G} \ \boldsymbol{S}^{\star\bar\star\ym}_n(w| w')_{G} \ \mathcal{A}_n^{\bar\star\ym}(w')_{G} \ .
\end{align}
In terms of the massless Mandelstam invariants $s_{ij} = 2\, p_i \cdot G\,p_j$ computed with respect to the spacetime metric $G$, the new field theory KLT kernel is
\begin{equation}\label{eq:nc momentum kernel field}
    \boldsymbol{S}^{\star\bar\star\ym}_n(w| w')_{G} =  \varTheta_n(\bar{w}) \ \bar\varTheta_n(\bar{w}') \ 
    \prod_{i=2}^{n-2}\, \Big( s_{1 k_i}  + \sum_{1\leq j<l\leq i}\, \mathtt{H}(k_j, k_l)_{w'} \ s_{k_jk_l} \Big) \ ,
\end{equation}
giving $\boldsymbol{S}^{\star\bar\star\ym}_n(w | w')_{G} = \varTheta_n(\bar w) \, \boldsymbol{S}^\ym_n(w | w')_G \, \bar\varTheta_n(\bar w')$, the momentum kernel for gluing two copies of noncommutative gauge theories with respect to the original momentum kernel. This provides the explicit double copy construction of noncommutative Yang--Mills theory with itself to ordinary perturbative gravity: an ordinary graviton can also be regarded as composed of two \emph{noncommutative} gluons.

A well-known feature of the field theory KLT kernel, originally derived in \cite{Cachazo:2013iea}, is that it can be expressed as the inverse matrix of double colour ordered biadjoint amplitudes $\CA^{\textrm{\tiny BAS}}_n(w|w')_G$ that were discussed in \cref{sub:biadjoint}, here evaluated in the flat background metric $G$. This partial amplitude is the amplitude of a $\phi^3$-theory restricted to an ordering of external momenta, and it is related to the commutative momentum kernel as $\CA^{\textrm{\tiny BAS}}_n(w|w')_G = \boldsymbol{S}^\ym_n(w|w')_G^{-1}$. In other words, tree amplitudes of the  zeroth copy uniquely determine the double copy kernel. This connection has been more recently understood through different perspectives: the inverse of the string theory KLT kernel can be related to the doubly ordered tree amplitudes of the Z-theory prescription~\cite[\S2.2]{Mizera:2017cqs}, while a simpler argument in the field theory limit is found in~\cite{Frost:2021qju}.

In the noncommutative field theory, this relation applied to the modified momentum kernel \eqref{eq:nc momentum kernel field} involves the sub-amplitudes
\begin{align}
\CA^{\star\bar\star\textrm{\tiny BAS}}_n(w|w')_G = \varTheta_n(w) \ \CA^{\textrm{\tiny BAS}}_n(w|w')_G \ \bar\varTheta_n(w') 
\end{align}
of the binoncommutative biadjoint scalar theory which we defined in \cref{sub:the_other_single_copy}: the interaction vertex \eqref{eq:vertex double deformed} keeps track of the two copies of noncommutative phase factors that are required by the noncommutative KLT kernel.
Thus the momentum kernel associated to the double copy of noncommutative gauge theory is sourced by the binoncommutative biadjoint scalar theory through
\begin{equation}
    \boldsymbol{S}_n^{\star\bar\star\ym}(w | w')_{G} = \CA^{\star\bar\star\textrm{\tiny BAS}}_n(w|w')_G^{-1} \ .
\end{equation}
Crucially, the noncommutative corrections preserve the rank of the matrix of double colour ordered amplitudes, which is $(n-3)!$, hence they satisfy the minimal rank condition and result in an admissible KLT kernel~\cite{CEHJP21}.
This further vindicates our understanding that double copies of noncommutative gauge theories are the same as the double copies of the corresponding commutative gauge theories.

\begin{remark}
In the commutative case, the (non-strict) Yang--Mills $L_\infty$ algebra $\YM$ can be viewed as a tensor product $\fru(N)\otimes\frC_\YM$ of the gauge algebra with a kinematical $C_\infty$-algebra~\cite{Zeitlin:2008cc}. In a quasi-isomorphic description of $\YM$ inspired by open string field theory, this was used by~\cite{Bonezzi:2022yuh} to show that a subspace of the $C_\infty$-algebra \smash{$\frC_\YM^{\textrm{\tiny L}} \otimes \frC_\YM^{\textrm{\tiny R}}$}, corresponding to states satisfying the level-matching
constraints of closed string theory, induces the cubic truncation of the $L_\infty$-algebra of double field theory. Following the prescription of the present paper, the same construction using our notion of twisted colour-stripping also gives ordinary double field theory as a doubling of noncommutative Yang--Mills theory, which results in the same $C_\infty$-algebra $\frC_\YM$.
\end{remark}
\subsection{$\mathfrak{u}_\star(1)$ Yang--Mills theory as a double copy} 
\label{sub:u_star_1_noncommutative_yang_mills_as_a_double_copy}

Most of what we have said so far in this section only applies to rank $N>1$, and it is natural to ask what is the fate of noncommutative $\sU(1)$ Yang--Mills theory from the double copy perspective. This is an interacting theory with non-trivial amplitudes that cannot be simply related to its commutative counterpart, which is the non-interacting Maxwell theory; scattering amplitudes in this theory are studied in e.g.~\cite{Latas:2020nji,Trampetic:2021awu}. Instead, we will interpret it analogously to what we did in \cref{sub: u1CS as DC}, by viewing $\mathfrak{u}(1) \simeq \mathbbm{R} =: \mathfrak{Kin}_{\wedge^0}$ as the kinematic vector space underlying the adjoint scalar theory introduced in~\cref{sub:adjoint_scalar_theory_as_a_strict_linf}. 

Consider the rank one limit of the factorisation of noncommutative Yang--Mills theory in the first order formalism from \cref{eq:ncym1 linf factorisation final}, whose $L_\infty$-algebra is now elusively denoted as
\begin{equation}\label{eq:ncym1_1_fact}
    \widehat{\mathfrak{YM}_{\star 1}(1)} =   \mathfrak{Kin}_{\wedge^0}\otimes_{\tau^{\mathfrak{u}_\star(1)}} \big(\mathfrak{Kin}_{\ym_1} \otimes_{\tau^{\ym_1}} \mathfrak{Scal}\big) \ .
\end{equation}
The underlying cochain complex in this factorization is again given by \cref{eq:ch of ncym1}.
Comparing \cref{eq:adfact,eq:ncym1_1_fact}, and recalling the colour-kinematics duality of \cref{sub:adjointCKduality}, we conclude that noncommutative $\mathsf{U}(1)$ Yang--Mills theory in the first order formalism is a double copy of the adjoint scalar theory of \cref{sub:adjoint_scalar_theory_as_a_strict_linf} with commutative Yang--Mills theory in the first order formalism, for any colour algebra $\frg$.
Symbolically, this double copy relation reads as
\begin{align}\label{eq:ASYM}
\text{$\fru_\star(1)$ Yang--Mills} \ = \ \text{Adjoint Scalar} \ \otimes \ \text{Ordinary Yang--Mills} \ .
\end{align}
This renders an interpretation of rank one noncommutative Yang--Mills theory as a gravitational theory, despite the absence of dynamical spin two fields, thus harvesting old anticipations discussed in \cref{sec:introduction}. 

\subsection{Noncommutative self-dual Yang--Mills theory}

We can make the double copy relations of this section somewhat more precise in the self-dual sector of Yang--Mills theory in $d=4$ dimensions, which we studied already in \cref{sub:topgravity}. Noncommutative self-dual Yang--Mills theory was originally studied in~\cite{Nekrasov:1998ss}. In this sector the first order formalism of \cref{sub:strictification_of_ym} is  a $BF$-theory involving self-dual two-forms which describes (noncommutative) self-dual Yang--Mills theory. As shown by~\cite{Lechtenfeld:2000nm}, this theory (in the Leznov gauge) can be obtained from quantizing open $\CN=2$ strings in a constant background $B$-field in the Seiberg--Witten zero-slope scaling limit. For rank $N>1$, our twisted homotopy factorization is compatible with the twisted form of colour-kinematics duality that provides a double copy map of noncommutative $\sU(N)$ self-dual Yang--Mills theory with itself to \emph{ordinary} self-dual gravity (perturbed around the open string metric $G$), along similar lines as discussed in \cref{sub:topgravity}.

Restricting both sides of \cref{eq:ASYM} to the self-dual sector in $d=4$ dimensions provides a  further corroboration of this double copy relation as a map to a gravitational theory. As we discussed in \cref{sub:topgravity}, self-dual $\fru_\star(1)$ Yang--Mills theory is the same theory as noncommutative self-dual gravity whose equation of motion is the deformed Pleba\'nski equation \eqref{eq:NCgrav}. There we also exhibited the explicit double copy construction of noncommutative self-dual gravity from the adjoint scalar theory with ordinary self-dual Yang--Mills theory. The connections of noncommutative self-dual gravity and Yang--Mills theory to Lorentz-invariant chiral higher-spin theories, discussed in~\cite{Monteiro:2022xwq}, hints at possible interpretations of the double copy relation \eqref{eq:ASYM} beyond the self-dual sector.

\section{Final remarks} 
\label{sec:final_remarks}

In this paper we have applied the homotopy algebraic formalism of~\cite{Borsten:2021hua} to the study of noncommutative gauge theories descending from the low-energy limit of open string theories with a stack of D-branes in constant $B$-fields.
We resolve the issue of colour and kinematic degrees of freedom mixing, which naively obstruct factorisation and colour-kinematics duality, by introducing a \textit{twisted} form of factorisation and colour-kinematics duality.
We then construct the double copy of noncommutative gauge theories in this framework and show that it coincides with the commutative limit.
This apparently trivial result is substantiated by the well-known fact that noncommutativity is an open string effect which leads us to introducing a modified KLT relation; nevertheless the homotopy algebraic techniques used in this work are themselves interesting and may lead to further insights into the structure of noncommutative theories. 
We also introduced a corresponding zeroth copy for noncommutative field theories, and checked that it plays the role of the inverse momentum kernel in our modified field theory KLT relation.

We illustrated our arguments by studying noncommutative deformations of Chern--Simons theory in three dimensions as well as of Yang--Mills theory in both first and second order formalisms.
We interpreted the special case of the rank one limits of noncommutative gauge theories as double copies themselves with the purely noncommutative adjoint scalar theory. 
We also have reviewed applications of this formalism in the study of the self-dual sector of Yang--Mills theory and gravity by studying the semi-classical limits of a new binoncommutative scalar theory. 
The diverse relations between theories are succinctly summarised in \cref{fig:web}.

Let us address here one final question: is there a way to modify our picture in such a way that noncommutative Yang--Mills theory can be double copied to a noncommutative theory of gravity? We assert that the answer is affirmative. As is well-known, noncommutative gravity involves a twisted form of diffeomorphism invariance, see e.g.~\cite{Szabo:2006wx} for a review. This twisted symmetry does not fit nicely into the standard $L_\infty$-algebra formalism; see~\cite{Szabo:2022edp} for an explicit exposition. This in itself formally hints as to why the standard noncommutative gauge theories, which \emph{are} organised systematically by $L_\infty$-algebras~\cite{Blumenhagen:2018kwq,Giotopoulos:2021ieg}, do not homotopy double copy to noncommutative gravity. However, we believe that this obstacle can be overcome by passing from $L_\infty$-algebras to \emph{braided} $L_\infty$-algebras, which organise noncommutative field theories with \emph{braided} gauge symmetries~\cite{DimitrijevicCiric:2021jea,Giotopoulos:2021ieg}; this formalism was in fact initially developed with a homotopy algebraic approach to noncommutative gravity in mind. 

By working with homotopy algebras in a symmetric monoidal category with non-trivial braiding, one can avoid the twisted factorizations and colour-kinematics duality that we had to introduce to stay in the usual categories of homotopy algebras, and instead work with the standard concepts, albeit in a braided setting. By a suitable extension of the homotopy double copy prescription to braided homotopy algebras, the double copy theory in this case would be encoded by a braided $L_\infty$-algebra and so would have braided diffeomorphism symmetry. Hence it should describe noncommutative gravity. We plan to address this interesting perspective and its consequences in future work.


\appendix

\section{Homotopical techniques in quantum field theory} 
\label{sec:linf}

In this appendix we review the relevant aspects, used throughout the main text, of $L_\infty$-algebras and their uses in purely algebraic computations of correlation functions for perturbative quantum field theories. 

\subsection{Primer on quantum field theory} 
\label{sub:primer_in_qft}

We begin by briefly recalling standard textbook material on the computation of correlation functions and scattering amplitudes in quantum field theory. 
A free scalar field theory on  Minkowski spacetime $\mathbbm{R}^{1,d-1}$ is defined by the kernel of a hyperbolic operator $P$ acting on the Schwartz space $\mathcal{S}(\mathbbm{R}^{1,d-1})$ of smooth functions with rapidly decreasing derivatives of all orders at infinity.
The Minkowski--Fourier transform is an automorphism of this space.

For example, the massive Klein--Gordon operator $P = \square + m^2 - \ii\, \epsilon$ is invertible on the Schwartz space with
\begin{equation}\label{eq:Greenid}
    \big(\square_x + m^2 - \ii\, \epsilon\big)\, G_{\mathrm{F}}(x,y) = \ii\, \delta(x-y) \ .
\end{equation}
The $\ii\, \epsilon$-prescription is used to denote a choice of contour integral prescription which uniquely fixes the Feynman propagator $G_{\mathrm{F}}$. 
It also moves states away from the mass-shell, dispelling the need to project zero-modes out, as the Feynman propagator is not defined on non-trivial fields in $\ker(P)$.
Thus
\begin{equation}\label{eq:scalar field propagator consistent}
    P \circ (-\ii\, G_\mathrm{F}) = (-\ii\, G_\mathrm{F}) \circ P = \mathbbm{1}
\end{equation}
and we regard $-\ii\, G_\mathrm{F}$ as the partial inverse of $P$ on the subspace of off-shell fields.

The representation of the Feynman propagator in momentum space as
\begin{equation}
    G_{\mathrm{F}}(x,y) = -{\ii} \, \int\, \frac{\mathrm{d}^d p}{(2 \pi)^d} \ \frac{\e^{-\ii \,p \cdot (x-y)}}{ p^2 - m^2 + \ii\, \epsilon} 
\end{equation}
solves the Green's identity \eqref{eq:Greenid} for all Schwartz functions and makes it a tempered distribution. We usually drop the explicit $\ii\,\epsilon$-prescription to simplify notation.

Given an action functional $S[\phi]$ for a field $\phi \in \mathcal{S}(\mathbbm{R}^{1,d-1})$, we construct correlation functions of the quantum field theory from the partition function for the theory coupled to an external source field $J \in \mathcal{S}(\mathbbm{R}^{1,d-1})$:
\begin{equation}
    Z[J] = \int_{\mathcal{S}(\mathbbm{R}^{1,d-1})} \, \mathscr{D}\phi \ \exp\Big(\,\ii\, S[\phi] + \ii \, \int\, \dd^dx \ J\, \phi\Big) \ .
\end{equation}
By separating the action functional into free and interacting parts as $S := S_{\rm free} + S_{\rm int}$, the partition function has the functional derivative representation
\begin{equation}\label{eq:partition func give Feynmann}
    Z[J] = \exp\Big(\,\ii \, S_{\rm int}\Big[\frac{\delta}{\ii\,\delta J}\Big]\Big) \, Z_{\rm free}[J]\ ,
\end{equation}
where $Z_{\rm free}[J]$ is the partition function for the free field theory coupled to $J$, which is evaluated by a functional Gaussian integration.

The off-shell $n$-point correlators $G_n(x_1, \ldots, x_n)$, or Green's functions, are defined by further functional differentiation with respect to $J \in \mathcal{S}(\mathbbm{R}^{1,d-1})$:
\begin{equation}\label{eq:correlation from partition}
    G_n(x_1, \ldots, x_n) 
    = \left. \frac{1}{\ii^n}\,\frac{\delta^n}{\delta J(x_1) \cdots \delta J(x_n)} \, \frac{Z[J]}{Z[0]}\, \right \vert_{J = 0} \ .
\end{equation}
Both the partition function and the correlators have a Feynman diagram expansion in terms of unamputated graphs, which include propagators on external lines.

To find on-shell correlators, or scattering amplitudes, one appeals to the Lehman--Symanzik--Zimmermann (LSZ) reduction theorem.
This requires extending the field content to include states that are asymptotically on-shell. These are solutions of the free field equations of motion which are of Schwartz-type for any fixed time-slice of Minkowski space, the vector space of which we denote by $\ker_ \mathcal{S}(P)$.
An early geometric treatment of this issue in the context of Chern--Simons perturbation theory was given in~\cite{Axelrod:1991vq}, while a more modern treatment in terms of $L_\infty$-algebras is found in \cite{Arvanitakis:2019ald,Macrelli:2019afx}.
We continue the Feynman propagator ${G}_\mathrm{F}$ on $\mathcal{S}(\FR^{1,d-1})$ trivially to $\widetilde{G}_\mathrm{F}$ on the full field space $ \mathcal{S}(\FR^{1,d-1})\oplus\ker_ \mathcal{S}(P)$ by setting
\begin{equation}
    \widetilde{G}_\mathrm{F} \big \vert_{\mathcal{S}(\mathbbm{R}^{1,d-1})} =  G_\mathrm{F} \qquad 
    \textrm{and} \qquad \ker(\widetilde{ G}_\mathrm{F}) = \ker_{\mathcal{S}}(P) \ .
\end{equation}

Scattering amplitudes are obtained by applying the kinetic operator $P_x$ at position $x\in\FR^{1,d-1}$ to a correlation function $G_n(x_1, \ldots, x_n)$.
In a scalar field theory with on-shell normalised wavefunctions $\e^{\,\ii\, p_k \cdot x_k}$, the $n$-point S-matrix element is given by
\begin{equation}\label{eq:LSZ reduction}
    S_n(p_1, \ldots, p_n) := \ii^n \, \prod_{k=1}^n \ \int\,  \mathrm{d}^dx_k \ \e^{\,\ii\, p_k \cdot x_k} \, P_{x_k}\, G_n(x_1, \ldots, x_n) \ .
\end{equation}
This procedure amputates the external legs of a diagram with the correct pole structure and appends the correct wavefunctions, giving the S-matrix element as the multiple on-shell residue of the Fourier transformed Green's function.
Only external states lying in $\ker_{\mathcal{S}}(P_{x_k}) = \ker_\mathcal{S}(\widetilde{G}_{\mathrm{F}\,x_k})$ contribute to the scattering amplitude.

For theories with fermions and/or gauge bosons, the in or out states to add depends on polarisation, spin and/or colour degrees of freedom.
The generalization of the LSZ formula for such quantum field theories is straightforward.

\subsection{Homotopy algebra methods} 
\label{sub:preliminaries}

We next review  material on homotopy Lie ($L_\infty$-)algebras, and related homotopy commutative algebras, focusing on those features of direct relevance to the present paper. Unless otherwise explicitly stated, all vector spaces are defined over the field $\FR$ of real numbers.

\paragraph{\texorpdfstring{$\boldsymbol{L_\infty}$}\ -algebras.} 
\label{par:linf_algebra}
{
    An $L_\infty$-algebra $\mathfrak{L}=(L,\{\mu_n\})$ consists of a $\mathbbm{Z}$-graded vector space $L = \bigoplus_{k\in\RZ}\, L^k$ together with a (possibly infinite) sequence of brackets $\{\mu_n\}$ which are multilinear maps of homogeneous degree $|\mu_n| = 2 - n$ for $n \geq 1$:
    \begin{equation}
    \begin{split}
        \mu_n : L^{\otimes n} \longrightarrow L \ , \quad
        v_1 \otimes \cdots \otimes v_n \longmapsto \mu_n(v_1, \ldots, v_n) \ .
    \end{split}
    \end{equation}
They respect \textit{strong graded antisymmetry}:
    \begin{equation}
        \mu_n( \ldots, v, v', \ldots) = - (-1)^{|v|\,|v'|}\, \mu_n(\ldots, v',v,\ldots) \ ,
    \end{equation}
for homogeneous elements $v,v'\in L$ of degrees $|v|$ and $|v'|$, respectively.
In particular, on degree 1 elements $v,v' \in L^1$, the 2-bracket is symmetric: $\mu_2(v,v') = \mu_2(v',v)$.
    
As opposed to graded antisymmetry, the Jacobi identity is only imposed in a weak sense.
    The brackets fulfil (possibly infinitely many) higher Jacobi identities $\mathcal{J}_n = 0$ for $n\geq1$, where $\CJ_n:L^{\otimes n}\longrightarrow L$ is the map of degree $|\CJ_n| = 3-n$ defined by
    \begin{equation}\label{eq:Jacobi}
    \begin{split}
        \mathcal{J}_n(v_1, \ldots, v_n) :=
        \sum_{j+k = n}\, (-1)^{k} \ \sum_{\sigma \in \mathrm{Sh}(j;n)} \ &
        \chi(\sigma; |v_1|, \ldots, |v_n|) \\
        & \quad \times
        \mu_{k+1}\big( 
        \mu_j(v_{\sigma(1)}, \ldots, v_{\sigma(j)}), v_{\sigma(j+1)} , \ldots, v_{\sigma(n)}
        \big) \ . 
        \end{split}
    \end{equation}
The sum runs over permutations of degree $n$ in the shuffle group $\mathrm{Sh}(j;n) \subset S_n$ that represents how two decks of $j$ and $n-j$ cards get shuffled once together:
    \begin{equation}\label{eq:shuffle group}
        \mathrm{Sh}(j; n) = \big\{ 
        \sigma \in S_n \ \big\vert \
        \sigma(1) < \cdots < \sigma(j) \  \textrm{ and } \ \sigma(j+1) < \cdots < \sigma(n)
        \big\} \ .
    \end{equation}
The Koszul sign factor $\chi$ takes into account antisymmetry and grading degree, enabling one to express  graded antisymmetry of higher brackets as
    \begin{equation}
        \mu_n( v_{\sigma(1)}, \ldots, v_{\sigma(n)}) = \chi(\sigma; |v_1|, \ldots, |v_n|) \ \mu_n( v_1 , \ldots , v_n) \ .
    \end{equation}
    
Let us look in detail at the identities $\CJ_n=0$ for $n=1,2,3$. The first Jacobi identity imposes the requirement that the map $\mu_1$ squares to zero, making $\Ch(\frL):= (L,\mu_1)$ into a cochain complex, whose cohomology is an $\FR$-module denoted by $\sfH^\bullet(\frL)$. The second Jacobi identity is a compatibility condition which states that the differential $\mu_1$ is a graded derivation of the 2-bracket $\mu_2$, i.e. $\mu_2:L\otimes L\longrightarrow L$ is a cochain map of degree~$0$. By precomposing with permutations $\sigma \in S_3$ acting on vector labels $(1,2,3)$, 
    the third Jacobi identity is given by
    \begin{align}
    \begin{split}
        \mathcal{J}_3 &= \mu_1 \circ \mu_3 - \mu_2 \circ(\mu_2 \otimes \mathbbm{1}) \circ \big(
        \mathbbm{1}^{\otimes 3} + {(123)} - {(23)}
        \big) \\
        & \qquad \, + \mu_3 \circ \big(\mu_1 \otimes \mathbbm{1}^{\otimes2}\big) \circ 
        \big(
        \mathbbm{1}^{\otimes 3} + {(132)} - {(12)}
        \big) \ .
        \end{split}
    \end{align}
    This is the usual Jacobi identity for a graded Lie algebra up to terms involving $\mu_1$ and $\mu_3$, i.e. the Jacobi identity is controlled by a cochain map $\mu_3:L\otimes L\otimes L\longrightarrow L$ of degree~$-1$.
    This weak but controlled definition is what is meant by a (strong) homotopy algebra:
    The algebraic structure is respected up to a differential. In particular, the cohomology $\sfH^\bullet(\frL)$ with the induced maps  is a graded Lie algebra.
    
 \paragraph{Cyclic structures.}
    
    A \textit{cyclic} $L_\infty$-algebra  $\frL = (L, \{ \mu_n \},\langle-,-\rangle)$ is an $L_\infty$-algebra together with a non-degenerate bilinear form of degree $-3$:
    \begin{equation}
        \langle -,- \rangle : L \otimes L \longrightarrow \mathbbm{R} \ ,
    \end{equation}
    which obeys the graded cyclic condition
    \begin{equation}
        \langle v_1, \mu_n(v_2, \ldots, v_{n+1}) \rangle = (-1)^{n + n\,( |v_1| + |v_{n+1}|) + |v_{n+1}|\, ( |v_1|+\cdots+|v_n|)} \ \langle v_{n+1}, \mu_n(v_1, \ldots, v_n) \rangle \ ,
    \end{equation}
    for all $n\geq 1$ and for homogeneous elements $v_i \in L$ of degree $|v_i|$.
}

%
\paragraph{Maurer--Cartan theory.} 
\label{par:mc}
{
    A Maurer--Cartan element $A \in L^1$ is an element of vanishing Maurer--Cartan curvature  $f_A \in L^2$:
    \begin{equation}\label{eq:Maurer Cartan equation}
        f_A := \sum_{n = 1}^\infty\, \frac{1}{n!} \  \mu_n(A^{\otimes n}) = 0 \ .
    \end{equation}
    This is called the Maurer--Cartan equation. It is identified as the equation of motion for gauge fields in a generalised gauge theory organised by a local $L_\infty$-algebra $\frL=(L,\{\mu_n\})$.
    
The Maurer--Cartan curvature $f_A \in L^2$ satisfies the identity
    \begin{equation}\label{eq:second noether}
        \mathrm{d}_A f_A := \mu_1(f_A) +  \sum_{n=1}^\infty\, \frac{(-1)^n}{n!} \ \mu_{n+1}(f_A, A^{\otimes n}) = 0 \ .
    \end{equation}
This is satisfied off-shell, i.e. when $f_A \neq 0$. It follows from a useful convergent series formula (see e.g.~\cite[Appendix~A]{Jurco:2018sby}) which enables one to recast it in terms of homotopy Jacobi identities \eqref{eq:Jacobi} in $L^3$ as 
    \begin{equation}
        \mathrm{d}_A f_A= \sum_{n=1}^\infty\, \frac{1}{n!} \ \mathcal{J}_n(A^{\otimes n}) =0 \ .
    \end{equation}
    This is a generalisation of the Bianchi identity and is called the Maurer--Cartan--Bianchi identity. In a generalized gauge field theory it is identified as Noether's second identity. 
    
    The Maurer--Cartan functional is an action functional  for a Maurer--Cartan element $A \in L^1$ on a cyclic $L_\infty$-algebra $\frL = (L, \{\mu_n\}, \langle -,- \rangle)$. It is given by
    \begin{equation}\label{eq:MC functional}
        S[A] := \sum_{n = 1}^\infty\, \frac{1}{(n+1)!} \ \langle A, \mu_n(A^{\otimes n}) \rangle \ .
    \end{equation}
The choice of degree $-3$ for the cyclic structure is chosen precisely so that $\langle A, \mu_n(A^{\otimes n}) \rangle$ is valued in $\FR$ (sitting in degree~$0$) for gauge fields $A \in L^1$. Via the Leibniz rule for $\delta$ together with cyclicity of the inner product, the variational principle $\delta S[A]=0$ for arbitrary field variations $\delta A\in L^1$ is equivalent to the Maurer--Cartan equation \eqref{eq:Maurer Cartan equation}, since
\begin{align}
\delta S[A] = \sum_{n = 1}^\infty\, \frac{1}{(n+1)!} \ \delta\langle A, \mu_n(A^{\otimes n}) \rangle = \langle\delta A,f_A\rangle \ .
\end{align}
    
    Gauge variations are encoded in elements $c$ of the vector space $L^0$.
    The Maurer--Cartan action functional $S[A]$ is invariant under the gauge transformation
    \begin{equation}
        \delta_c A = \mu_1(c) + \sum_{n=1}^\infty\, \frac{1}{n!} \ \mu_{n+1}(A^{\otimes n}, c) \ .
    \end{equation}
To see this, one uses cyclicity of the inner product to rewrite the gauge variation of $S[A]$ for arbitrary gauge parameters $c\in L^0$ in terms of the Noether identity \eqref{eq:second noether}:
    \begin{equation}
    \begin{split}
        \delta_c S[A] &= \langle \delta_c A , f_A \rangle = \langle c, \mathrm{d}_A f_A \rangle = 0 \ .
    \end{split}
    \end{equation}
    
\paragraph{Batalin--Vilkovisky theory.}

We apply the Batalin--Vilkovisky (BV) formalism to a generalised gauge theory with local cyclic $L_\infty$-algebra $ \frL=(L, \{ \mu_n\}, \langle -,- \rangle)$.
For this, we consider superfields $\mbf A \in \mathsf{Fun}(L[1])\otimes L$, where generally $L[p]$ for $p\in\RZ$ is the graded vector space obtained from $L$ by shifting the degrees of its homogeneous subspaces by $-p$ units: $L[p]^k := L^{k+p}$. The brackets $\mu_n$ and cyclic structure $\langle-,-\rangle$ are extended in the obvious way to operations that make $\mathsf{Fun}(L[1])\otimes L$ into a cyclic $L_\infty$-algebra (see e.g.~\cite{Jurco:2018sby}); by an abuse of notation, we continue to denote these extended operations with the same symbols.
    Applying homotopy Maurer--Cartan theory to this extended cyclic $L_\infty$-algebra, we retrieve the full BV action functional
    \begin{align}
    \begin{split}
        S_{\textrm{\tiny BV}}[\mbf A] = \sum_{n = 1}^\infty\, \frac{1}{(n+1)!} \ \langle \mbf A, \mu_n( \mbf A^{\otimes n}) \rangle \ ,
    \end{split}
    \end{align}
for $\mbf A$ a degree~$1$ superfield, also known as a contracted coordinate function. 
    
A useful way of writing contracted coordinate functions on the BV field space is as follows. Let $\{\mathtt{e}^I\}\subset L$ be a basis for the local cyclic $L_\infty$-algebra $\frL$, where $I$ is a DeWitt multi-index representing colour and kinematic degrees of freedom. Then degree~$1$ superfields may be decomposed as $\mbf A = \mathtt{e}^I\otimes\mathtt{A}_I$, where $\{\mathtt{A}_I\}$ includes the BRST ghosts $c\in L^0$ and the gauge fields $A\in L^1$, as well as their antifields $c^+\in L^3$ and $A^+\in L^2$, together with ghosts-for-ghosts in negative degrees and corresponding antifields in $L^k$ for $k\geq4$. 

In the quantum field theory, correlation functions are determined from the gauge-fixed partition function sourced by a degree~$2$ external superfield $\mbf J \in \mathsf{Fun}(L[1])\otimes L$:
    \begin{equation}
        Z[\mbf J] = \int_{\CL} \, \mathscr{D} \mathtt{A}_\CL \ \exp\big(\,\ii\, S_{\textrm{\tiny BV}}[\mbf A] + \ii\, \langle \mbf J, \mbf A \rangle\big)\big|_{\CL} \ ,
    \end{equation}
where $\CL$ is a Lagrangian submanifold of the BV field space $L$ with respect to the natural symplectic structure induced by the cyclic inner product, which is parametrized by fields $ \mathtt{A}_\CL$. 
Analogously to \cref{eq:partition func give Feynmann}, the Feynman expansion for the partition function of a Maurer--Cartan theory in the superfield formalism is given by 
\begin{equation}
\begin{split}\label{eq:partition funciton in hmc}
    Z[\mbf J] = \exp\bigg(\,\ii \ {\textstyle{\sum\limits_{n = 2}^\infty} \, \frac{1}{(n+1)!} \ \Big\langle\frac{\delta}{\ii\,\delta \mathtt{J}}, \mu_n  \Big(\big(\frac{\delta}{\ii\,\delta \mathtt{J}}\big)^{\otimes n}\Big)\Big\rangle} \bigg) \,
    \int_\CL \, \mathscr{D} \mathtt{A}_\CL \ \exp\big(\, \ii\, \langle \mbf A,\mu_1(\mbf A) \rangle + \ii\, \langle \mbf J, \mbf A \rangle \big)\big|_\CL  \ .
\end{split}
\end{equation}

By decomposing $\mbf J = \mathtt{e}_I\otimes\mathtt{J}^I$, where $\{\mathtt{e}_I\}\subset L^*\simeq L[3]$ is a dual basis to $\{\mathtt{e}_I\}$ with respect to the cyclic structure, the gauge-fixed $n$-point correlator is then
    \begin{equation}\label{eq:correlation from partition superfield}
  G_n[\mathtt{A}_{I_1},\dots,\mathtt{A}_{I_n}]\big|_\CL =   \left.   \frac{1}{\ii^n} \, \frac{\delta^n}{\delta \mathtt{J}^{I_1} \cdots \delta \mathtt{J}^{I_n}} \, \frac{Z[\mbf J]}{Z[\mbf 0]} \, \right \vert_{\mbf J = \mbf 0} \ .
    \end{equation}
Physically meaningful information is extracted from this correlator upon contracting the fields, which is done using the cyclic structure of the $L_\infty$-algebra. 
}
%

\paragraph{Morphisms of \texorpdfstring{$\boldsymbol{L_\infty}$}\ -algebras.} 
\label{par:morphism_of_linf _algebras}
{
 Let us now describe the category of (cyclic) $L_\infty$-algebras. Let $\frL=(L,\{\mu_n\})$ and $\frL'=(L',\{\mu_n'\})$ be $L_\infty$-algebras.     An $L_\infty$-morphism $\psi : \frL \longrightarrow \frL'$ is a sequence $\{\psi_n\}$ of multilinear graded antisymmetric maps $\psi_n : L^{\otimes n} \longrightarrow L'$ of degree $|\psi_n| = 1-n$ for $n\geq1$ such that all maps are compatible with all brackets. 
 
    Informally, for  each fixed partition $n=j+k$ and $n=k_1+\cdots +k_l$ of $n$ tensor powers of the graded vector space $L$, one can regard compatibility in terms of the diagram
    \begin{equation}
        \begin{tikzcd}
            L^{\otimes n} \arrow["\mu_j\otimes\mathbbm{1}^{\otimes k}"]{rr} 
            \arrow["\psi_{k_1} \otimes \cdots\otimes\psi_{k_l}"']{dd}
            && 
            L^{\otimes k+1} \arrow["\psi_{k+1}"]{dd}
            \\ \\
             L'^{\,\otimes l} \arrow["\mu_l'"]{rr}
            &&
            L'
        \end{tikzcd}
    \end{equation}
The sum of all possible combinations of diagrams commute and give the formula
    \begin{equation}
    \begin{split}\label{eq:Linf morphism}
        &\sum_{j+k = n}\, (-1)^{k} \ \sum_{\sigma \in \mathrm{Sh}(j;n)} \,
        \chi(\sigma; |v_1|, \ldots, |v_n|) \
        \psi_{k+1}\big( 
        \mu_{j}(v_{\sigma(1)}, \ldots, v_{\sigma(j)})
        , v_{\sigma(j+1)}, \ldots, v_{\sigma(n)}
        \big) \\[4pt]
        & \quad = 
        \sum_{l=1}^n \, \frac{1}{l!} \ 
        \sum_{k_1 + \cdots + k_l = n} \ 
        \sum_{\sigma \in \mathrm{Sh}(k_1, \ldots,k_{l-1};n)}\,
        \chi(\sigma;|v_1|,\dots,|v_n|) \ (-1)^{\zeta_{k_1,\dots,k_l}(\sigma;|v_1|,\dots,|v_n|)} \\
        & \hspace{5cm} \times
        \mu'_l\big( \psi_{k_1}(v_{\sigma(1)},\ldots, v_{\sigma(k_1)}), \ldots, 
        \psi_{k_l}(v_{\sigma(n - k_l + 1)},\ldots, v_{\sigma(n)})\big) \ ,
    \end{split}
    \end{equation}
    for all $n\geq1$, where 
\begin{align}
\zeta_{k_1,\dots,k_l}(\sigma;|v_1|,\dots,|v_n|) = \sum_{1\leq i<j\leq l} \, k_i\,k_j + \sum_{i=1}^{l-1} \, (l-i)\,k_i + \sum_{i=2}^l\, (1-k_i) \ \sum_{k=1}^{k_1+\cdots+k_{l-1}} \, |v_{\sigma(k)}| \ .
\end{align}
Note that the left-hand side of \cref{eq:Linf morphism} is formally equivalent to the higher Jacobi identity \eqref{eq:Jacobi} if one replaces $\psi_{k+1}$ with $\mu_{k+1}$.

A \emph{strict} $L_\infty$-algebra is an $L_\infty$-algebra of the form $\frL=(L,\mu_1,\mu_2)$, i.e. such that $\mu_n=0$ for all $n\geq3$. A strict $L_\infty$-algebra is the same thing as a differential graded (dg-)Lie algebra, but regarded as an object in the category of $L_\infty$-algebras. This perspective allows for more flexibility and stronger properties, as morphisms between dg-Lie algebras are more restrictive than $L_\infty$-morphisms.
    
If $\frL$ and $\frL'$ are moreover endowed with cyclic structures $\langle-,-\rangle_L$ and $\langle-,-\rangle_{L'}$, respectively, then   an $L_\infty$-morphism $\psi: \frL \longrightarrow \frL'$ extends to a morphism of cyclic $L_\infty$-algebras if
    \begin{equation}
        \langle \psi_1(v_1)\,,\, \psi_1(v_2) \rangle_{L'} = \langle v_1, v_2 \rangle_L
    \end{equation}
    and 
    \begin{equation}
        \sum_{j+k=n} \, \langle \psi_j(v_1, \ldots, v_j) \,,\, \psi_k(v_{j+1}, \ldots, v_n) \rangle_{L'} = 0 \ ,
    \end{equation}
    for $n\geq3$. For strict $L_\infty$-algebras, a cyclic inner product is the same thing as a non-degenerate invariant bilinear form on a dg-Lie algebra.

    Recall that to any $L_\infty$-algebra $\frL=(L, \{ \mu_n\})$ there is an associated cochain complex given by $\Ch(\frL) = (L,\mu_1)$ with cohomology $\sfH^\bullet(\frL)$.
An \textit{$L_\infty$-quasi-isomorphism} $\psi:\frL\longrightarrow\frL'$ is an $L_\infty$-morphism whose induced map on cohomology \smash{$\psi_*:\sfH^\bullet(\frL) \longrightarrow \sfH^{\bullet}(\frL')$} is an isomorphism. In contrast to dg-Lie algebras, quasi-isomorphism is an equivalence relation on the category of $L_\infty$-algebras. 

Under an $L_\infty$-morphism $\psi : \frL\longrightarrow\frL'$, the Maurer--Cartan theory for $\frL$ is mapped to the Maurer--Cartan theory for $\frL'$ according to
    \begin{equation}
    \begin{split}\label{eq:MC element under morphism}
        A \longmapsto A' &= \sum_{n = 1}^\infty\, \frac{1}{n!} \ \psi_n(A^{\otimes n}) =  \sum_{n = 1}^\infty\, \frac{1}{n!} \ \mu'_n(A'^{\,\otimes n}) \ , \\[4pt]
        f_A \longmapsto f_{A'} &= \psi_1(f_A) +  \sum_{n = 1}^\infty\, \frac{(-1)^n}{n!} \ \psi_{n+1}( f_A,A^{\otimes n}) \ , \\[4pt]
        c \longmapsto c' &= \psi_1(c) + \sum_{n = 1}^\infty\, \frac{1}{n!} \  \psi_{n+1}(A^{\otimes n}, c) \ .
    \end{split}
    \end{equation}
Thus Maurer--Cartan elements $A \in L^1$ are sent to Maurer--Cartan elements $A' \in L'^{\,1}$. In particular, an $L_\infty$-quasi-isomorphism $\psi$ induces an isomorphism between the moduli spaces of Maurer--Cartan elements in $L$ and $L'$ modulo gauge transformations. In the language of field theory, this means that quasi-isomorphic theories are classically equivalent.
}

\paragraph{Minimal models and recursion relations.} 
{
Any $L_\infty$-algebra $\frL=(L,\{\mu_n\})$ has an associated \textit{minimal model} $\frL^\circ=(\sfH^\bullet(\frL),\{\mu_n^\circ\})$, which is an $L_\infty$-algebra based on the graded cohomology $\sfH^\bullet(\frL)$ with respect to the differential $\mu_1$. To close the $L_\infty$-structure of the minimal model, infinitely-many higher brackets need to be introduced using \emph{homotopy transfer} of $L_\infty$-algebras, which achieves the process of integrating out fields in the path integral. These have the interpretation of effective currents generating scattering amplitudes in quantum field theory, as we will discuss later on.

As originally shown in \cite{Kajiura:2001ng,Kajiura:2003ax}, any $L_\infty$-algebra $\frL$ is isomorphic to a direct sum of its minimal model and an $L_\infty$-algebra with trivial cohomology as well as trivial higher brackets. The $L_\infty$-algebras $\frL$ and $\frL^\circ$ are homotopy equivalent due to the existence of a deformation retract given by a map $\sfh : L \longrightarrow L$ of degree $-1$, called a contracting homotopy, and a choice of section \smash{$\sfe:\sfH^\bullet(\frL)\lhook\joinrel\longrightarrow L$}  of degree~$0$ which fit into the diagram
    \begin{equation}
        \begin{tikzcd}
            \sfH^\bullet(\frL) \arrow[hook, shift right=.5ex, "\sfe"', above]{rr} 
            && 
            L \arrow[twoheadrightarrow, shift right=1.5ex,"\sfp"']{ll}
            \arrow[loop right, distance=2em, start anchor={[yshift=1ex]east},
             end anchor={[yshift=-1ex]east}]{}{\sfh}
        \end{tikzcd}
        \label{eq:minimal model}
    \end{equation}
for the natural projection \smash{$\sfp: L\, -\!\!\!\twoheadrightarrow \sfH^\bullet(\frL)$} of degree~$0$, and yield the Hodge--Kodaira decomposition
    \begin{equation}\label{eq:abstract hodge kodaira}
        \mathbbm{1}_L = \sfe \circ \sfp + \sfh \circ \mu_1 + \mu_1 \circ \sfh \ .
    \end{equation}
 Thus the condition that the degree 0 maps $\sfh\circ\mu_1$ and $\mu_1\circ\sfh$ are projectors is relaxed in general.
 
 One can construct a recursion formula \cite{Macrelli:2019afx} for the quasi-isomorphism $\psi_n : \sfH^\bullet(\frL)^{\otimes n}  \longrightarrow L$ by considering Maurer--Cartan elements $v_1^\circ,\dots,v_n^\circ \in \sfH^1(\frL)$.
 The original proof of quasi-isomorphism between homotopy algebras and their minimal models, given in \cite{Kajiura:2003ax} for the case of homotopy associative ($A_\infty$-)algebras, does not require the triple of maps $(\sfp,\sfh\circ\mu_1,\mu_1\circ\sfh)$ to be projections.
However, given the Hodge--Kodaira decomposition \eqref{eq:abstract hodge kodaira}, one can redefine the contracting homotopy $\sfh$ such that $\sfh \circ \sfh = 0$, and $\mu_1 \circ\sfh$ as well as $\sfh\circ \mu_1$ do become projectors. Thus \cref{eq:abstract hodge kodaira} is replaced by
    \begin{equation}\label{eq:HKmodified}
    \begin{split}
        \mathbbm{1}_L &= \sfe \circ\sf p + \sfh \circ \mu_1 + \mu_1 \circ \sfh \  , \quad  
        \sfh = \sfh \circ \mu_1 \circ \sfh \qquad \mbox{and} \qquad 
        \mu_1 = \mu_1 \circ \sfh \circ \mu_1 \ .
    \end{split}
    \end{equation}
    
    \begin{subequations}
    The recursion relations are given by
    \begin{equation}
    \begin{split}
    \psi_1(v^\circ_1) &= \sfe(v_1^\circ)
    \end{split}
    \end{equation}
    and
    \begin{equation}
    \begin{split}
    & \psi_n(v^\circ_1, \ldots , v^\circ_n) \\[4pt]
    & \quad = - \sum_{j=2}^n \,  \frac{1}{j!} \
    \sum_{k_1 + \cdots + k_j = n} \
    \sum_{\sigma \in \mathrm{Sh}(k_1, \ldots, k_{j-1};n)} \,
    \chi(\sigma;|v_i^\circ|,\dots,|v_n^\circ|) \ (-1)^{ \zeta_{k_1,\dots,k_j}(\sigma;|v_1^\circ|,\dots,|v_n^\circ|)}  \\
    &\hspace{4cm} \times (\sfh \circ \mu_j) 
    \big(\psi_{k_1}(v_{\sigma(1)}^\circ, \ldots, v^\circ_{\sigma(k_1)}), \ldots, \psi_{k_{j}}( v^\circ_{\sigma(i - k_{j}+1)}, \ldots, v^\circ_{\sigma(i)})\big) \ .
    \end{split}\label{eq:minimal model recursion relations}
    \end{equation}
    Similarly, the brackets on the minimal model can be recursively computed as
    \begin{equation}
    \begin{split}
    \mu_1^\circ(v_1^\circ)&= 0
    \end{split}
    \end{equation}
    and
    \begin{equation}
    \begin{split} 
    & \mu_n^\circ(v_1^\circ, \ldots, v_n^\circ) \\[4pt]
    & \quad = \sum_{j=2}^n \, \frac{1}{j!} \
    \sum_{k_1 + \ldots + k_j = n} \
    \sum_{\sigma \in \mathrm{Sh}(k_1, \ldots, k_{j-1};n)} \,
    \chi(\sigma;|v_i^\circ|,\dots,|v_n^\circ|) \ (-1)^{ \zeta_{k_1,\dots,k_j}(\sigma;|v_1^\circ|,\dots,|v_n^\circ|)} \\
    & \hspace{4cm} \times (\sfp \circ \mu_j) 
    \big(\psi_{k_1}(v_{\sigma(1)}^\circ, \ldots, v^\circ_{\sigma(k_1)}), \ldots, \psi_{k_{j}}( v^\circ_{\sigma(i - k_{j} + 1)}, \ldots, v^\circ_{\sigma(i)})\big) \ .
    \end{split}\label{eq:min model brackets}
    \end{equation}
    \end{subequations}
    
In field theory, the differential $\mu_1:L^1\longrightarrow L^2$ generates the free field equations; its inverse is the Feynman propagator on the subspace of field space that excludes $\ker(\mu_1)$.
 The contracting homotopy $\sfh:L^2\longrightarrow L^1$ then has the interpretation of the Feynman propagator extended trivially to on-shell states, as we have seen in \S\ref{sub:primer_in_qft}, with gauge fixing imposed by setting $\sfh(v)=0$ for $v\in L^1$.
    If the $L_\infty$-algebra $\frL$ carries a cyclic structure $\langle-,-\rangle$, then its minimal model $\frL^\circ$ also has a cyclic structure given by pullback $\langle v_1^\circ,v_2^\circ\rangle_\circ := \langle\sfe(v^\circ_1),\sfe(v^\circ_2)\rangle$. 
    It should be stressed that quasi-isomorphic theories do not necessarily yield the same quantum theory even though they are equivalent at the classical level.
   We will see an explicit example of this later on, in which we compute tree-level scattering amplitudes using the minimal model.   
}

\paragraph{Twisted tensor products and $\boldsymbol{C_\infty}$-algebras.}{
    \begin{subequations}
    Let  $\frR = (R,m_1,m_2)$ be a differential graded algebra with underlying graded vector space $R = \bigoplus_{k\in\RZ}\, R^k$, or equivalently a strict homotopy algebra.
    The \textit{twisted tensor product} of the underlying cochain complex  $\Ch(\frR) = (R, m_1)$ with a graded vector space $V = \bigoplus_{k\in\RZ}\, V^k$ is the cochain complex $\Ch(V\otimes_\tau\frR) = (V \otimes R, m_1^\tau)$, with underlying graded vector space $V\otimes R=\bigoplus_{k\in\RZ}\,(V\otimes R)^k$ whose homogeneous subspaces are $(V \otimes R)^k = \bigoplus_{i+j=k}\, V^i \otimes R^j$.
    Using the standard Sweedler notation, the  twist map
    \begin{equation}
    \begin{split}
        \tau_1: V \longrightarrow V \otimes \mathsf{End}(R) \quad , \quad
        v \longmapsto \tau_1^{\textrm{\tiny(1)}}(v) \otimes \tau_1^{\textrm{\tiny(2)}}(v)
    \end{split}
    \end{equation}
    defines a \textit{twisted differential $m_1^\tau:V \otimes R\longrightarrow V \otimes R$} which is the linear degree 1 map  squaring to zero given by 
    \begin{equation}\label{eq:twisted differential def}
        m_1^\tau(v \otimes r) = (-1)^{|\tau_1^{\textrm{\tiny(1)}}(v)|} \ \tau^{\textrm{\tiny(1)}}_1(v) \otimes m_{1}\big(\tau^{\textrm{\tiny(2)}}_1(v)(r)\big) \ ,
    \end{equation}
    for $v\in V$ and $r\in R$.
    
    One can extend this construction to the full dg-algebra $\frR$ \cite{Borsten:2021hua}.
    Introducing a second twist map
    \begin{equation}
    \begin{split}
        \tau_2: V \otimes V &\longrightarrow V \otimes \mathsf{End}(R) \otimes \mathsf{End}(R) \ , \\
        v_1\otimes v_2 & \longmapsto \tau^{\textrm{\tiny(1)}}_2(v_1, v_2) \otimes \tau_2^{\textrm{\tiny(2)}}(v_1,v_2)
        \otimes  \tau_2^{\textrm{\tiny(3)}}(v_1,v_2) \ ,
    \end{split}
    \end{equation}
    we define a \emph{twisted $2$-bracket} $m_2^\tau:(V \otimes R)^{\otimes 2}\longrightarrow V\otimes R$ which is the bilinear degree 0 map given by
    \begin{equation}
    \begin{split}\label{eq:twisted 2-brackets}
        m_2^\tau(v_1 \otimes r_1, v_2 \otimes r_2) &= 
        (-1)^{|v_2|\,|r_1|} \
        \tau^{\textrm{\tiny(1)}}_2(v_1, v_2) \otimes 
        m_2\big(
        \tau_2^{\textrm{\tiny(2)}}(v_1,v_2) (r_1)\,
        , \, \tau_2^{\textrm{\tiny(3)}}(v_1,v_2) (r_2) \big) \ ,
    \end{split}
    \end{equation}
    for $v_1,v_2\in V$ and $r_1,r_2\in R$.
 Depending on the homotopical structure of $\frR$, an appropriate choice of twist datum $\tau:=(\tau_1, \tau_2)$ turns $V\otimes_\tau\frR:=(V \otimes R, m_1^\tau, m_2^\tau)$ into a differential graded algebra. 
 
This often involves the notion of a \emph{$C_\infty$-algebra}: a $C_\infty$-algebra is a commutative algebra in the category of $A_\infty$-algebras. In the main text we only use \emph{strict} $C_\infty$-algebras (often representing a kinematic algebra), which are equivalently described as differential graded commutative algebras. Concretely, a strict $C_\infty$-algebra $\frC = (C,m_1,m_2)$ is a cochain complex $\Ch(\frC) = (C,m_1)$ with a cochain map $m_2:C\otimes C\longrightarrow C$ of degree~$0$ which is  graded commutative and associative:
\begin{align}
m_2(c_1,c_2) = (-1)^{|c_1|\,|c_2|} \, m_2(c_2,c_1) \qquad \mbox{and} \qquad m_2\big(m_2(c_1,c_2),c_3\big) = m_2\big(c_1,m_2(c_2,c_3)\big) \ ,
\end{align}
for $c_1,c_2,c_3\in C$.

The twisted tensor product is a way to factorize strict homotopy algebras in a unique fashion, as necessitated by the double copy; one needs to check by hand that it defines a dg-algebra. In this paper we start from a known homotopy algebra, and factorise it in terms of a twist datum $\tau$ such that the given homotopical structure is recovered. 
In our examples, the twisted tensor product mixes the types of homotopy algebras: the twisted tensor product of an $L_\infty$-algebra $\frR$ yields a $C_\infty$-algebra, while conversely the twisted tensor product of a $C_\infty$-algebra $\frR$ is an $L_\infty$-algebra.
    \end{subequations}
}\label{def:Twisted factorisation}

\subsection{Scattering in the \texorpdfstring{$L_\infty$}\ -algebra formalism} 
\label{sub:scattering_in_the_textlinf_formalism}
%

\paragraph{Tree-level amplitudes.}

To connect the formalisms of \S\ref{sub:primer_in_qft} and \S\ref{sub:preliminaries}, we start from the observation of \cite{Arvanitakis:2019ald,Macrelli:2019afx}  that scattering amplitudes in a quantum field theory computed with the LSZ reduction formula \eqref{eq:LSZ reduction} can be interpreted in terms of the cyclic structure on the minimal model for the local $L_\infty$-algebra $\frL=(L,\{\mu_n\},\langle-,-\rangle)$ which organises the field theory.
Indeed, for the scattering of fields which are sections of a (trivial) vector bundle $E$ over Minkowski spacetime $\FR^{1,d-1}$, one considers the field content in $L^1 = {\Gamma}_{\mathcal{S}}(\mathbbm{R}^{1,d-1},E) \oplus \sfH^1(\frL)$, that is, interacting fields with the correct asymptotic behaviour and on-shell states at infinity.

For the minimal model of the BV theory, we trade the brackets $\{\mu_n\}_{n\geq 2}$ on $\mathsf{Fun}(L[1]) \otimes L$ for infinitely-many brackets $\{ \mu_n^{\circ}\}_{n\geq2}$ on $\mathsf{Fun}\big(\sfH^\bullet(\frL)[1]\big) \otimes \sfH^\bullet(\frL)$.
Since $\mu_1^\circ = 0$, the corresponding gauge-fixed  partition function \eqref{eq:partition funciton in hmc} is explicitly calculated through the Legendre transformation
\begin{equation}
    Z^\circ[\mbf J] = \exp\big(\,\ii \,  S^\circ[\mbf A^\circ] + \ii\, \langle \mbf J, \mbf A^\circ \rangle_\circ\big)\big|_{\sfp(\CL)} \ ,
\end{equation}
where $\mbf J=-f_{\mbf A^\circ}$ and
\begin{align}
S^\circ[\mbf A^\circ] := \sum_{n = 2}^\infty \, \frac{1}{(n+1)!} \ \langle \mbf A^\circ, \mu_n^{\circ}( \mbf A^{\circ \,\otimes n}) \rangle_\circ \ .
\end{align}
One recognises the argument of the exponential as the generating functional of connected tree diagrams, which is given in terms of the Maurer--Cartan functional $S^\circ[\mbf A^\circ]$ of the minimal model.

Using \cref{eq:correlation from partition superfield} for the minimal model partition function, it follows that the $n$-point correlation functions are given entirely in terms of the algebraic structure of the minimal model. That is, the Maurer--Cartan elements of the minimal model $\mbf A^\circ$ generate partial tree-level amplitudes  
\begin{equation} \label{eq:treelevelscattering}
    \CM_{n}(1, \ldots,n) = \langle \mbf A^\circ_1, \mu_{n-1}^\circ( \mbf A_2^\circ , \ldots, \mbf A^\circ_{n}) \rangle_\circ\big|_{\sfp(\CL)} \ ,
\end{equation}
with the full amplitude given by taking the sum over all planar ordered permutations of external particles.
In summary, the computation of tree-level amplitudes is reduced to the problem of solving the recursion relations in \cref{eq:minimal model recursion relations}. That the recursion relations of the minimal model yield tree-level amplitudes was already realised in early work on the subject~\cite{Kajiura:2003ax}. 

\paragraph{Perturbiner expansions.} 
\label{par:example}

The link between perturbiner expansions and $L_\infty$-recursion relations was established by~\cite{Lopez-Arcos:2019hvg}.
Working at the level of vector spaces, a \textit{perturbiner element} of a field theory with $L_\infty$-algebra $\frL=(L, \{ \mu_n\}, \langle   -,- \rangle )$ is a Maurer--Cartan element $A^\circ \in \sfH^1(\frL)$ of the minimal model $\frL^\circ=(\sfH^\bullet(\frL), \{ \mu_n^\circ \}, \langle  -,-  \rangle_\circ)$.
Since the quasi-isomorphisms from \cref{eq:minimal model recursion relations} map Maurer--Cartan elements to Maurer--Cartan elements, from a perturbiner element $A^\circ$ one reconstructs an on-shell interacting field of the full theory as
\begin{equation}\label{eq:perturbiner element}
     A = \sum_{n=1}^\infty\, \frac{1}{n!} \ \psi_n(A^{\circ\,\otimes n}) \ \in \ L^1 \ .
\end{equation}

The on-shell free fields $A^\circ$ in the minimal model source the interactions. Hence \cref{eq:perturbiner element} has the interpretation of an expansion in terms of plane waves, first introduced in \cite{Berends:1987me}.
Indeed, the Berends--Giele relations provide an alternative way to compute the quasi-isomorphism and thus the tree-level scattering amplitudes. This is discussed further below. 

The perturbiner expansion can also be extended to compute both  off-shell tree and loop correlation functions, see e.g.~\cite{Lee:2022aiu,Gomez:2022dzk}, as we tacitly do in several parts of the main text. Loop-level scattering amplitudes can be computed recursively from minimal models of quantum $A_\infty$-algebras using homological perturbation theory~\cite{Jurco:2019yfd}, generalising the tree-level Berend--Gieles recursion relations; 
for correlation functions of scalar field theories this corresponds to taking the trivial projection $\sfp=0$ in \cref{eq:HKmodified}~\cite{Okawa:2022sjf}.

\paragraph{Word combinatorics.} 
\label{par:word_combinatorics}

In explicit examples, one is faced with combinatorics on words, which we briefly overview here.
This arises because there is a homomorphism between the shuffle groups in the standard $L_\infty$-relations and tensor products of word algebras.

Let $M\subset\mathbbm{N}$ be a subset of positive integers of cardinality $n$. 
An \textit{ordering} $w$ of $M$ is a word of length $|w|=n$ with no repeated letters; equivalently, a word is an ordering if it is maximal in the set $M$.
Denote by $\mathscr{O}_M$ the set of orderings of $M$, and by $\CCW_M$ the $\mathbbm{R}$-vector space generated by orderings, with the inner product 
\begin{equation}
    (w_1,w_2) = \delta_{w_1,w_2} \ ,
\end{equation}
that is, two words $w_1$ and $w_2$ are orthogonal if they are distinct.
Let $\mathscr{L}_M$ be the set of multilinear Lie polynomials in $\CCW_M$, i.e. elements of the free Lie algebra $\CCF(M)$ on the set $M$ of maximal length:
\begin{equation}
    \mathscr{L}_M = \CCF(M) \cap \CCW_M \ ,
\end{equation}
in the free associative algebra on $M$.
For example, if $M = \{ 1,2,3,4 \}$, then $[ [ [1,2],4],3] \in \mathscr{L}_M$, whereas~$[1,2] \notin \mathscr{L}_M$.

As we often work on the set $M = \{ 1, \ldots , n \}$, in the following we will replace the subscripts $_M$ with $_n$ on all sets defined above. 
For example, $\mathscr{L}_n$ is the set of multilinear Lie polynomials of length $n$.
Given a word $w = k_1 \cdots k_n \in \CCW_n$, we denote its transpose by $\bar w = k_n \cdots k_1$, with $|w| = |\bar w|=n$.

The left bracketing operation on words of length $n$ is the map
\begin{equation}
\begin{split}\label{eq:left bracketing map}
    \ell : \CCW_n &\longrightarrow \mathscr{L}_n \ , \\
    k_1k_2k_3 \cdots k_n &\longmapsto [ [\cdots [[k_1, k_2], k_3],\ldots],k_n ] = (-1)^{n-1} \ \mathrm{ad}_{k_n} \circ \cdots \circ \mathrm{ad}_{k_3}\circ\mathrm{ad}_{k_2}(k_1) \ .
\end{split}
\end{equation}
For example, $\ell(123) = [[1,2],3] = 123 - 213 - 312 + 321 \in \mathscr{L}_3$.\footnote{One can similarly define a right bracketing map $
    r(k_1 \cdots k_{n-1}k_n) = \mathrm{ad}_{k_1} \circ \cdots \circ \mathrm{ad}_{k_{n-1}}(k_n) = (-1)^{n-1} \ \ell( \overline{k_1 \ldots k_n})$,
but it will not be used here.}
For non-empty words $w,w' \in \CCW_n$, one has
\begin{equation}
\begin{split}
    \ell\big(w\, \ell(w')\big) = [\ell(w), \ell(w')] \qquad \mbox{and} \qquad 
    \ell^2(w) &= |w| \, \ell(w) \ .
\end{split}
\end{equation}
Conversely, by the Dynkin--Specht--Wever lemma, if a word $w$ satisfies $\ell(w) = |w| \, w$ then $w$ is a Lie polynomial. 
Any Lie polynomial $\varGamma \in \mathscr{L}_n$ can be expanded as 
\begin{equation}
    \varGamma = \sum_{w \in \mathscr{L}_{n-1}}\, (1w, \varGamma) \ \ell(1w) \ ,
\end{equation}
as the bases $1w$ and $\ell(1w)$ for $\mathscr{L}_n^*$ and $ \mathscr{L}_n$ respectively with $w\in\mathscr{O}_{n-1}$ are dual:
\begin{equation}
    (1w, \ell(1w')) = \delta_{ww'} \ .
\end{equation}

The shuffle product $\shuffle$ on words $w$ and $w'$ of lengths $n$ and $n'$, respectively, is defined inductively from the empty word $\varnothing$:
\begin{equation}
    w \shuffle \varnothing = \varnothing \shuffle w = w \qquad \mbox{and} \qquad w\,a \shuffle w'\,a' = (w \shuffle w'\,a')\,a + (w\,a \shuffle w')\,a' \ ,
\end{equation}
for non-empty letters $a,a' \in \mathbbm{N}$.
Shuffled words are orthogonal to Lie polynomials, so
\begin{equation}
    \ell(w \shuffle w') = 0 \ .
\end{equation}

Following~\cite{Bridges:2019siz}, we define the binary tree map $b: \CCW_n \longrightarrow \mathscr{L}_n$ recursively by 
\begin{equation}
\begin{split}\label{eq:binary tree map}
    b(a) = a \qquad \mbox{and} \qquad
    b(w) = \frac{1}{s_w} \, \sum_{w = w_1\sqcup w_2}\, [b(w_1), b(w_2)] \ ,
\end{split}
\end{equation}
where $a\in\mathbb{N}$ and $s_w$ is the kinematical invariant Mandelstam variable associated with momenta labelled by the letters in $w$. The sum runs over deconcatenations of the word $w\in\CCW_n$ into two non-empty words $w_1 \in \CCW_i$ and $w_2 \in \CCW_{n-i}$.

To make contact with the $L_\infty$-algebra formalism, we consider the evaluation map $\mathrm{ev}$ between the shuffle group \eqref{eq:shuffle group} and word algebras. We view elements $\sigma \in \mathrm{Sh}(i;n)$ as words of length $n$ with image $\mathrm{im}(\sigma) = (w_1 , w_{2})$; notice that there exists a conjugate element $\bar{\sigma} \in \mathrm{Sh}(n-i; n)$ with image $\mathrm{im}(\bar{\sigma}) = (w_{2}, w_1)$. This defines the map
\begin{equation}
\begin{split}\label{eq:evaluation map}
    \mathrm{ev}: \mathrm{Sh}(i;n) \longrightarrow \CCW_i \otimes \CCW_{n-i} \quad , \quad
    \sigma \longmapsto \mathrm{im}(\sigma) = w_1 \otimes w_{2} \ ,
\end{split}
\end{equation}
where here we consider the action of $\sigma$ on $M = \{ 1,\ldots, n \}$.


\paragraph{Biadjoint scalar theory.}

As an explicit example illustrating these techniques, we consider the commutative biadjoint scalar theory from \cref{sub:biadjoint}.
    Because of the homotopy factorisation \eqref{eq:ncbiadjointfact}, we can use the colour-stripped version of Berends--Giele recursion\cite{Berends:1987me,Mafra:2016ltu}.
    The idea is to consider a superposition of $n$ on-shell fields $\phi^\circ \in \sfH^1(\BAS)=\ker(\square)$ and expand the interacting field in a basis of the tensor product. We label colour by the index sets $\{ a_1, \ldots, a_n \}$ and $\{ \bar a_1, \ldots, \bar a_n \}$ for the Lie algebras $\frg$ and $\bar\frg$, respectively, and associated momenta by $\left\{ p_1, \ldots, p_n \right\}$. We take our expansion to be valued in the free Lie algebra $\mathscr{F}(\mathfrak{g} \otimes \bar{\frg})$ \cite{Frost:2020eoa}; this requires an extension of the $L_\infty$-algebra $\BAS$ to values in $\mathscr{F}(\mathfrak{g} \otimes \bar{\frg})$, which however we do not indicate explicitly in the notation. For a word $w=k_1\cdots k_n$ on $n$ letters, we denote $T^w:=T^{a_{k_1}}\cdots T^{a_{k_n}}$, and similarly for $\bar T^w$, as well as $p_w:=p_{k_1}+\cdots + p_{k_n}$.
       
    The biadjoint field  $\phi \in \Omega^0(\mathbbm{R}^{1,d-1}, \mathscr{F}(\mathfrak{g} \otimes \bar{\mathfrak{g}}))$ admits an expansion in terms of Lie words
    \begin{equation}\label{eq:freelie expansion}
        \phi(x) = \sum_{n = 1}^\infty \ \sum_{w,w' \in \CCW_n}\, \phi_{w|w'}^\circ \ \e^{\,\ii\, p_w \cdot x} \ T^w \otimes \bar{T}^{w'} \ , 
    \end{equation}
    where $\CCW_n$ is the algebra of words of length $n$, and  $\phi_{w|w'}^\circ$ is a function valued in double words modulo shuffles:
    \begin{equation}
        \phi^\circ_{w \shuffle \tilde w | w'} = \phi^\circ_{w | \tilde w \shuffle w'}\ .
    \end{equation}
    We call $\phi_{w|w'} \in \Omega^0(\mathbbm{R}^{1,d-1} , \mathscr{F}( \mathfrak{g} \otimes \bar{ \mathfrak{g}})^*)$ the \textit{Berends--Giele double current}.
    The currents are computed recursively starting from the one-particle states $\phi^\circ_{1| 1} \in \sfH^1(\BAS)$ satisfying the Maurer--Cartan equation of the minimal model $\BAS^\circ$.
    The first few terms can be found from the equation of motion $\square\, \phi = \frac{g}{2}\, \llbracket \phi, \phi\rrbracket_{\frg\otimes\bar\frg}$, giving
    \begin{equation}\label{eq:first term pert expansion}
        \phi^\circ_{12|12} = \phi^\circ_{21|{21}}= - \phi^\circ_{12|{21}} = -\phi^\circ_{21|{12}} = \frac{g}{s_{12}}
        \ .
    \end{equation}
    
    The Berends--Giele expansion must match with the $L_\infty$-recursion relations.
    Hence we identify \cref{eq:freelie expansion} with \cref{eq:perturbiner element}, where on the left-hand side we use the single-particle expansion $\phi^\circ(x) = \sum_{w,w'\in\CCW_n}\,\phi^\circ_{w|w'}\ \e^{\,\mathrm{i}\,p_w \cdot x}\ T^w \otimes \bar{T}^{w'}$. In this way we arrive at
    \begin{equation}\label{eq:perturbiner quasiiso correspondence}
        \psi_n( \phi^{\circ\,\otimes n}) = n! \, \sum_{w,w'\in\CCW_n}\, \phi_{w|w'}^\circ \ \e^{\,\ii\, p_w\cdot x} \ T^{w} \otimes \bar T^{w'} \ .
    \end{equation}
    
    This is checked by using \cref{eq:nc biad bracket} in \cref{eq:minimal model recursion relations} for the contracting homotopy $\sfh$ given by the Feynman propagator acting as
\begin{align}    
    G_{\rm F}(\mathrm{e}^{\,\mathrm{i}\, p \cdot x}) = -\frac{\ii}{p^2} \, \mathrm{e}^{\,\mathrm{i}\, p \cdot x} \ , 
\end{align}    
   so that the recursion relations are
    \begin{equation}\label{eq:strict quasi recursion}
        \psi_n(\phi^{\circ\,\otimes n}) = \frac{\ii\,g}{2} \, \sum_{i=1}^{n-1} \,
        \bigg(\begin{matrix} n \\ i\end{matrix}\bigg) \
        G_{\rm F}\big( \llbracket \psi_i(\phi^{\circ\,\otimes i}), \psi_{n-i}(\phi^{\circ\,\otimes n-i}) \rrbracket_{\frg\otimes\bar\frg}\big) \ .
    \end{equation}
    Direct computation of the $L_\infty$-recursion matches with the Berends--Giele expansion found in \cref{eq:first term pert expansion} up  to second order, when inserted back in \cref{eq:freelie expansion} to reproduce the factorised bracket:
    \begin{equation}
    \begin{split}
        \psi_1(\phi^\circ) &= \sfe(\phi^\circ) = \mathrm{e}^{\,\mathrm{i}\, p_1 \cdot x} \ T^{a_1} \otimes \bar{T}^{\bar a_1} \ , \\[4pt]
        \psi_2(\phi^\circ, \phi^\circ) &=  \ii\,g \, G_{\rm F}\big(\llbracket \phi^\circ,\phi^\circ\rrbracket_{\frg\otimes\bar\frg}\big) 
        = \frac{ g}{s_{12}} \, \mathrm{e}^{\,\mathrm{i}\,p_{12} \cdot x}
        \ [T^{a_1},T^{a_2}]_{\frg} \otimes [\bar{T}^{\bar a_1},\bar {T}^{\bar a_2}]_{\bar\frg} \ . 
    \end{split}
    \end{equation}

The tree-level scattering amplitudes \eqref{eq:treelevelscattering} between identical particles are given by the cyclic structure and the higher brackets of the minimal model 
\begin{align}
\BAS^\circ= \Big(\ker(\square)[-1] \xrightarrow{ \ 0 \ } \mathrm{coker}(\square)[-2] \ , \ 
\big\{\mu_{n}^{\circ\textrm{\tiny BAS}}\big\} \ , \ \langle-,-\rangle_{\circ\textrm{\tiny BAS}}\Big) \ .
\end{align}
For $\sfp$ the trivial projection onto on-shell states, the $L_\infty$-recursion in \cref{eq:min model brackets} is given by
    \begin{equation}\label{eq:strict quasi recursion 2}
        \mu^\circ_n(\phi^\circ_1, \ldots, \phi^\circ_n) = -\frac{g}{2} \, \sum_{i=1}^{n-1} \ \sum_{\sigma \in \mathrm{Sh}(i;n)} \,
        \sfp \circ \llbracket \psi_i(\phi^\circ_{\sigma(1)},\ldots,\phi^\circ_{\sigma(i)}) \,,\, \psi_{n-i}(\phi^\circ_{\sigma(i+1)}, \ldots,\phi^\circ_{\sigma(n)})\rrbracket_{\frg\otimes\bar\frg} \ .
    \end{equation}
    This has the first non-trivial bracket
\begin{align}    
    \mu_2^\circ(\phi^\circ_1, \phi^\circ_2)= -g \, \llbracket \phi^\circ_1, \phi^\circ_2\rrbracket_{\frg\otimes\bar\frg} \ \e^{\,\ii\,p_{12}\cdot x} \ .
    \end{align}
Using the cyclic structure \eqref{eq:bipairing}, the partial three-point amplitude is then given by
    \begin{equation}
        \mathcal{M}_3^{\textrm{\tiny BAS}}(1,2,3|1,2,3) = -  g \,
         f^{a_1a_2a_3} \, \bar{f}^{\bar a_1\bar a_2\bar a_3} \ (2 \pi)^d \, \delta(p_1 + p_2 + p_3) \ ,
    \end{equation}
    as expected.



\bibliographystyle{ourstyle.bst}
\bibliography{good_lib}

\end{document}